\documentclass[12pt,leqno]{article}
\usepackage{amsmath,amsthm,amsfonts,amssymb,color,mathrsfs,framed}
\usepackage{stmaryrd}

\DeclareMathAlphabet{\mathpzc}{OT1}{pzc}{m}{it}

\newtheorem{thm}{Theorem}[section]

\newtheorem{lem}[thm]{Lemma}
\newtheorem{prp}[thm]{Proposition}

\newtheorem{remark}[thm]{Remark}

\theoremstyle{definition}

\newcommand{\scr}[1]{\mathscr #1}
\definecolor{wco}{rgb}{0.5,0.2,0.3}

\numberwithin{equation}{section}

\def\u{u_{\omega}}
\def\G{\mathfrak{g}}
\def\R{\mathbb R}

\def\gg{\mathpzc{g_{\omega}}}
\def\gl{\mathpzc{g}}

\def\<{\langle}
\def\>{\rangle}

\def\D{\scr D}
\def\E{\mathbb E}

\def\c{\scr C}

\def\B{\scr B}
\def\e{\text{\rm{e}}}

\def\g{g_{\omega}}

\def\cu{{\rm curl}}
\def\P{\mathbb P}
\def\p{\mathcal{P}}

 \def\ee{\varepsilon}
\def\B{\mathbf B}
\def\S{\scr S}
\def\ee{\varepsilon}

\def\w{\boldsymbol{\mathsf{w}}}
\def\T{\mathbb T}
\def\t{\theta}
\def\d{\mathbf{d}}
\def\e{\mathbb E}
\def\x{\mathfrak{X}}
\def\a{\alpha_{\omega}}
\def\A{\tilde A}

\def\s{{\bf S}}
\def\G{\mathfrak{g}}

\begin{document}

\title{Stochastic variational principles for
dissipative equations with advected quantities}
\author{Xin Chen$^{1}$,\; Ana Bela Cruzeiro$^{2}$,\;
Tudor S. Ratiu$^{1,3}$}
\addtocounter{footnote}{1}
\footnotetext{School of Mathematics, Shanghai Jiao Tong
University, Minhang District, 800 Dong Chuan Road, 200240 Shanghai, China. Partially supported by
National Natural Science Foundation of China (No. 11501361).
\texttt{chenxin217@sjtu.edu.cn}
\addtocounter{footnote}{1} }
\footnotetext{Grupo de F\'isica-Matem\'atica and Departamento de
Matem\'atica, Instituto Superior
T\'ecnico, Univ. Lisboa,  Av. Rovisco Pais, 1049-001 Lisbon, Portugal. Partially supported by project
PTDC/MAT-STA/0975/2014 from the Portuguese F.C.T..
\texttt{abcruz@math.tecnico.ulisboa.pt}
\addtocounter{footnote}{1} }
\footnotetext{
Section de Math\'ematiques, Universit\'e de Gen\`eve, 2-4 rue du
Li\`evre, Case postale 64,
1211 Gen\`eve 4, Switzerland. Partially supported by the National Natural Science Foundation of China grant number 11871334 and by the
NCCR SwissMAP grant of the Swiss National
Science Foundation.
\texttt{ratiu@sjtu.edu.cn, tudor.ratiu@epfl.ch}
\addtocounter{footnote}{1} }

\date{}

\maketitle

\allowdisplaybreaks

This paper presents symmetry reduction for material stochastic
Lagrangian systems with advected quantities whose configuration
space is a Lie group. Such variational principles
yield deterministic as well as stochastic constrained
variational principles for dissipative equations of motion
in spatial representation. The general theory is presented
for the finite dimensional situation. In infinite dimensions
we obtain partial differential equations and stochastic partial
differential equations.
When the Lie group is, for example, a diffeomorphism group, the
general result is not directly applicable but the
setup and method
suggest rigorous proofs valid in infinite dimensions which lead
to similar results. We apply this technique
to the compressible Navier-Stokes equation and to
magnetohydrodynamics for charged viscous compressible fluids.
A stochastic Kelvin-Noether theorem is presented. We derive,
among others, the classical deterministic dissipative equations from purely
variational and stochastic principles, without any appeal to
thermodynamics.

\section{Introduction}

The goal of this paper is to develop a Lagrangian symmetry
reduction process for a large class of stochastic systems with
advected parameters. The general theory, which yields both deterministic
and stochastic constrained variational principles and deterministic,
as well as stochastic reduced equations of motion,
is developed for finite dimensional systems. The resulting abstract
equations then serve as a template for the
study of infinite dimensional stochastic systems, for which the
rigorous analysis has to be carried out separately. The examples of
the compressible Navier-Stokes equations and
dissipative compressible magnetohydrodynamics equations, as well as
their randomly perturbed counterparts and  are treated in detail.
 We recover with our method the classical dissipative fluid
and magnetohydrodynamic equations without any appeal to thermodynamical
considerations, except for the form of the internal energy density.
\medskip

The dynamics of many conservative physical
systems can be described geometrically taking advantage of
the intrinsic symmetries in their material description. These
symmetries induce Noether conserved quantities and allow for
the elimination of unknowns, producing an equivalent system
consisting of new equations of motion in spaces with less
variables and a non-autonomous ordinary differential equation,
called the reconstruction equation. This geometric
procedure is known as reduction, a method that is ubiquitous
in symplectic, Poisson, and Dirac geometry and has wide applications
in theoretical physics, quantum and continuum mechanics, control
theory, and various branches of engineering. For example, in
continuum mechanics, the passage from the material (Lagrangian)
to the spatial (Eulerian) or convective (body) description is
a reduction procedure. Of course, depending on
the problem, one of the three representations may be preferable.
However, it is often the case that insight from the other two 
representations, although apparently more intricate, leads to a
deeper understanding of the physical phenomenon under consideration
and is useful in the description of the dynamics.

A simple example in which the three descriptions
are useful and serve different purposes is free rigid body dynamics
(e.g., \cite[Section 15]{MaRa1999}).
If one is interested in the motion of the attitude matrix,
the material picture is appropriate. The classical free rigid
body dynamics result, obtained by applying Hamilton's standard
variational principle on the tangent bundle of the proper rotation
group $SO(3)$, states that the attitude matrix describes
a geodesic of a left invariant Riemannian metric on $SO(3)$,
characterized by the mass distribution of the body. However,
as shown already by Euler, the equations of motion simplify
considerably in the convective (or body) picture because the
total energy of the rotating body, which in this case is just
kinetic energy, is invariant under left translations. The convective
description takes place on the Lie algebra $\mathfrak{so}(3)$ of $SO(3)$ 
and is given by the classical Euler equations for a free rigid body,
after implementing the Lie algebra isomorphism of $\mathbb{R}^3$ with
$\mathfrak{so}(3)$ given by the cross product operation. Finally, 
the spatial description comes into play,
because the spatial angular momentum is conserved during the motion
and is hence used in the description of the rigid body motion.

The present paper uses exclusively Lagrangian mechanics, where
variational principles play a fundamental role since they produce
the equations of motion. In continuum mechanics, the variational
principle used in the material description is the standard Hamilton
principle producing curves in the configuration space of the
problem that are critical points of the action functional.
However, in the spatial and convective representations, if the
configuration space of the problem is a Lie group, the
induced variational principle requires the use of constrained
variations, a fundamental result of Poincar\'e \cite{Poincare1901};
the resulting equations of motion are called today the
\textit{Euler-Poincar\'e equations} (\cite{MaSc1993},
\cite[Section 13.5]{MaRa1999}, \cite{CeMaPeRa2003}). 
These equations have been vastly
extended to include the motion of advected quantities
(\cite{CeMaHoRa1998,HMR, HoMaRa2002})
as well as affine (\cite{GBRa2008}) and
non-commutative versions  thereof that naturally appear
in models of complex materials with internal structure
(\cite{GBRa2009, GBRa2011a}) and whose geometric description has
led to the solution of a long-standing controversy in the
nematodynamics of liquid crystals (\cite{GBRaTr2012, GBRaTr2013}).
Euler-Poincar\'e equations have also very important generalizations to 
problems whose configuration space is an
arbitrary manifold and the Lagrangian is invariant under a
Lie group action (\cite{CeMaRa2001a}) as well as its extension
to higher order Lagrangians
(\cite{GBHoRa2011, GBHoMeRaVi2012a,GBHoMeRaVi2012b}).
Lagrange-Poincar\'e equations turn out to model the motion
of spin  systems (\cite{GBHoRa2009}), long molecules
(\cite{ElGBHoPuRa2010, GBHoPuRa2012}), free boundary
fluids and elastic bodies (\cite{GBMaRa2012}), as well as
charged and Yang-Mills fluids (\cite{GBRa2011a}). There are also
Lagrange-Poincar\'e theorems for field theory
(\cite{CLGPRa2001, CLRa2003, CLRaSh2000, GBRa2010, ElGBHoRa2011}) and 
non-holonomic systems (\cite{CeMaRa2001b}). Lagrange-Poincar\'e equations
also have interesting applications to Riemannian cubics and
splines (\cite{NoRa2015}), the representation of images 
(\cite{BrGBHoRa2011}), certain classes of textures in
condensed matter (\cite{GBMoRa2015}), and some
control (\cite{GBRa2011b}) and optimization (\cite{GBHoRa2013}) problems.

Variational principles play an important role in the design of 
structure preserving numerical algorithms. One discretizes
both spatially and temporally such that the symmetry structure
of the problem is preserved. Integrators based on a discrete
version of Hamilton's principle are called variational integrators
(\cite{MaWe2001}). The resulting equations of motion are the
discrete Euler-Lagrange equations and the associated algorithm
for classical conservative systems is both symplectic as well as
momentum-preserving and manifests very good long time energy
behavior; see \cite{LeMaOrWe2004a, LeMaOrWe2004b} for additional
information. There are versions of such variational integrators
for certain forced (\cite{KaMaOrWe2000}), controlled
(\cite{OBJuMa2011}), constrained holonomic
(\cite{LeMaOr2008, LeOBMaOr2010}), non-holonomic
(\cite{KoMaSu2010}), non-smooth (\cite{FeMaOrWe2003,
DeGBRa2015}), multiscale (\cite{LeOB2011, TaOwMa2010}), and
stochastic (\cite{BROw2008}) systems.
In the presence of symmetry, these systems can be reduced. However,
today a general theory of discrete reduction in all of these cases
is still missing and is currently being developed. If the configuration space is a Lie group, the first
discretization of symmetric Lagrangian systems appears in
\cite{MeVe1991}, motivated by problems in complete integrability;
for an in-depth analysis of such problems see \cite{Suris2003}.

All the above mentioned systems, both in the smooth and discrete
versions, should have various stochastic analogues, depending
on what phenomenon is modeled. The basic idea
is to start with variational principles, motivated by Feynman's
path integral approach to quantum mechanics and also by
stochastic optimal control. The latter has its origins in the
foundational work of Bismut (\cite{Bismut1981, Bismut1982})
in the late seventies and in recent developments by
L\'azaro-Cam\'{\i} and Ortega
(\cite{LCOr2008, LCOr2009a, LCOr2009b, LCOr2009c}). Non-holonomic
systems have been studied in the same spirit (\cite{HoRa2015}).
All of this work investigated mainly stochastic perturbations of
Hamiltonian systems. A very recent approach on the Lagrangian
side, in Euler-Poincar\'e form, has been developed in
\cite{CHR} and \cite{Holm2015}, where both the position and the momentum
of the system are (independently) randomly perturbed, as well as
the Lagrangian.

The stochastic version of Euler-Poincar\'e reduction introduced
in \cite{ACC} is closer in spirit to Feynman's
viewpoint and, particularly, to the approach initiated in the eighties by
Zambrini (c.f. \cite{Z} and references therein as well as
\cite{Y1981,NYZ1981}).
It uses as a main tool the notion of
generalized (or mean-value) derivative  in order to remove the
contribution of the martingale part of the stochastic Lagrangian paths.
This derivative has been introduced in stochastic dynamics by E. Nelson
(\cite{Ne1967}). We also refer to \cite{AC,ACC,CC, KoKo2012} and
references therein for various extensions on infinite dimensional spaces
and applications of this derivative in stochastic Euler-Poincar\'e
reduction.

The crucial idea is that the generalized derivative contains a
contraction term induced by noise (stochastic force) which gives rise
to a second order operator (such as the Laplacian) in the velocity
equation of the stochastic model in continuum mechanics. Then, the
stochastic reduction procedure leads to characterizations of various
partial differential equations whose viscous term only appears in relation
with the Laplacian, such as the incompressible the Navier-Stokes
or the viscous Camassa-Holm equations. This stochastic Euler-Poincar\'e
reduction is formulated on the group of volume preserving diffeomorphisms
and the Lagrangian variables correspond to semimartingales.

The theory of reduction of variational principles of mechanical systems
with advected parameters, leading to Euler-Poincar\'e equations
coupled with advection equations, and hence associated
to semidirect products, has been been developed in \cite{HMR}. For
continuum mechanical models, this method is particularly useful to
characterize several kinds of evolutionary partial differential equations
arising in conservative compressible fluids, such as the compressible
Euler and ideal MHD equations (see, e.g., \cite[Section 7]{HMR}).
Therefore, a first natural question arises whether it is possible to find a
stochastic Euler-Poincar\'e reduction method that would characterize
equations with viscous terms in compressible fluids, such as the
compressible Navier-Stokes equation or the viscous compressible MHD
equation. The main difficulty is that the generalized derivative, alluded
to above, is not capable by itself to generate these viscosity terms since
they do not appear only in connection with the generators of the underlying
stochastic Lagrangian  paths as in the case of  incompressible fluids.
The second natural question, amplifying the first one, is whether
one can formulate a stochastic reduction procedure that would lead to
interesting stochastic partial differential equations, appropriate
for applications to continuum mechanics.

It is well known (\cite{HMR}) that the Euler-Poincar\'e formulation
naturally leads to Kelvin circulation theorems. The classical Kelvin
Circulation Theorem for barotropic ideal fluids
states that the circulation of the velocity around a closed loop
moving with the fluid is constant in time. This statement is intimately
connected to Poisson geometric properties of Euler's ideal
fluid equations (it characterizes the symplectic leaves in the phase space
of Euler's equations; see \cite{MaWe1983}) and has important applications,
for example, in the Lyapunov stability analysis of stationary solutions
(see, e.g., \cite{Arnold1965, Arnold1969, HoMaRaWe1983, HoMaRa2002}).
For more general fluids, this theorem fails; instead of the vanishing of
the time derivative of the circulation around a closed loop moving with
the fluid, there is an explicit right hand side, responsible for generating
circulation, involving advected quantities and the potential energy of
the material. These identities are also known under the same name. For
a general abstract formulation and a large class of examples of such
Kelvin Circulation Theorems, see \cite{HMR}, \cite{HoMaRa2002}. In
addition, these Kelvin Circulation identities are equivalent to
reformulations of the equations of motion that turn out to be convenient
for the qualitative study of the fluid. It is natural hence to seek for
a counterpart of such Kelvin-Noether identities appearing in stochastic
Euler-Poincar\'e reduction.

The main purpose in the paper is to solve the questions mentioned above.
We summarize now the main achievements of  the paper.
\begin{itemize}
\item [(1)] We introduce a contraction matrix for the stochastic
Lagrangian paths, which is different from the generalized derivative
described above. This contraction matrix gives rise to a contraction
force term in the action functional, capable to access separately,
via reduction, each viscosity term, introduced usually by physical
considerations, in the continuum mechanical model. In particular, we
deduce the compressible Navier-Stokes and
the viscous compressible MHD equations (Section \ref{sec_5}) only from
our stochastic variational principle, without any appeal to thermodynamics.

\item[(2)] We study random action functionals, by introducing an
additional stochastic force. Various stochastic partial differential
equations, such as  stochastic (both compressible and incompressible)
Navier-Stokes or Euler equations and stochastic viscous MHD equations,
are deduced from our stochastic reduction procedure.

\item [(3)] We
derive Euler-Poincar\'e equations for stochastic processes defined on semidirect product Lie algebras and
give the associated deterministic constrained variational principle when the stochastic force (in the action functional)
vanishes. In other words,
we develop the semidirect Euler-Poincar\'e reduction for a large
class of stochastic systems.

\item[(4)] We prove a stochastic version of the Kelvin-Noether Circulation
Theorem for our stochastic reduction procedure. Compared with
the result in \cite{HMR}, our (stochastic) evolution equations also depend
on some martingales and some viscosity terms, in addition to the
usual advected quantities (Section \ref{sec_4}).

\end{itemize}

As discussed
earlier, the generalized derivative only produces a trace term
on the contraction part of the associated stochastic Lagrangian path. In
order to obtain different viscosity terms in the models of continuum
mechanics, we have to investigate in more detail the effect of the
contraction induced by the martingale term. To do this, we introduce a
contraction matrix, which carries much more information, involving
each entry in the matrix, and not just their sum (as is the case for
the generalized derivative).

Moreover, partially inspired by \cite{CHR} and \cite{Holm2015}, we  also
consider random perturbations of the action functionals so that  the corresponding critical
points satisfy a stochastic differential equation (a stochastic partial differential equation in
the infinite dimensional case). Therefore, our action functionals have integrands that consist of three
parts: the Lagrangian, a contraction force, and  a stochastic
force, which model the Lagrangian structure, the viscosity,
and the stochastic (martingale) nature of the action.

\medskip

\medskip
\noindent\textbf{Plan of the paper.} In Section \ref{sec:der_semi},
we recall some basic probability notions necessary for the
rest of the paper and give the crucial definition of
the contraction matrix and martingale part for group valued
semimartingales. Section \ref{sec_3} contains the first main
result of the paper, namely the stochastic semidirect product
Euler-Poincar\'e reduction for finite dimensional Lie groups,
both in left and right-invariant versions. We give the
deterministic variational principle and
the reduced equations of motion as well as their random deformations.
In Section \ref{sec_4} we derive a stochastic Kelvin-Noether theorem.
Section \ref{sec_5} presents the second main result of the paper, the
reduction from the material to the spatial representation in infinite
dimensions, which applies to the compressible Navier-Stokes equation and
to the stochastic compressible magnetohydrodynamics equations.
The stochastic reduction process recovers the standard
deterministic equations in Eulerian representation as well as their
random deformations.

\section{The  derivative for semimartingales}
\label{sec:der_semi}
In \cite{ACC},  we gave the notion of generalized derivative for
semimartingales taking values on some topological groups.
In this section, we  decompose a $G$-valued semimartingale (when the
dimension of $G$ is finite, see, e.g. \cite{E})
into its velocity part, martingale part, and contraction part (matrix),
which is crucial for our stochastic reduction procedure.

\subsection{Some probability notions}
\label{subs_2_1}

We  review in this subsection  some basic notions of stochastic
analysis on Euclidean spaces. We recall the concepts omitting
the proofs, which can be found, for example, in \cite{IW}.
\medskip

We denote $\mathbb{R}^+: =[0, \infty[$. Let $(\Omega , {\mathcal P},
\mathbb{P})$ be a probability space. Suppose
we are given a family $({\mathcal P}_t )_{t\in {\mathbb R}^+}$
of sub-$\sigma$-algebras of $\mathcal P$ which is non-decreasing
(namely, ${\mathcal P}_s\subset {\mathcal P}_t$ for
$0\leq s\leq t$) and right-continuous, i.e.,
$\cap _{\epsilon >0}\mathcal{P}_{t+\epsilon} =
\mathcal{P}_t$ for all $t\in \mathbb{R}^+$. We then say that
the probability space is endowed with a \textit{non-decreasing
filtration} $({\mathcal P}_t )_{t \in \mathbb{R}^+}$. A stochastic
process $X: {\mathbb R}^+  \times \Omega  \rightarrow
\mathbb{R}$ is $(\mathcal{P}_t)$-\textit{adapted} if
$X(t, \cdot ):\Omega \rightarrow\mathbb{R}^+$ is
$\mathcal{P}_t$-measurable for every $t \geq 0$.
Typically,  a filtration describes the past history of a process:
one starts with a process $X$ and defines ${\mathcal P}_t$ to be the
sigma-algebra generated by all sets $X (s, \cdot )^{-1} (B)$, with
$0\leq s\leq t$ and $B$ a Borel subset in $\mathbb R$. Then the process
$X$ is automatically $(\mathcal{P}_t )$-adapted.

\medskip

A stochastic process $M:\ {\mathbb R}^+  \times \Omega \rightarrow
\mathbb{R}$ is a ($\R$-valued) \textit{martingale} with respect to
$(\mathcal{P}_t)_{t\ge 0}$ if

\vskip 2mm

(i) $\e |M_\omega (t)|<\infty $ for all $t\geq 0$;

(ii) $M_{\omega}(t)$ is $({\mathcal P}_t )$-adapted;

(iii) $\e_s (M_\omega(t) ))=M_\omega(s)$ a.s.
for all $0\leq s<t$.
\vskip 2mm

In the above definition, $\mathbb{E}$ denotes the expectation
of the random variable with respect to the probability measure
$\mathbb{P}$;  $\mathbb{E}_s(M_{\omega}(t)):=
\e [M_{\omega}(t)| {\mathcal P}_s ] $,
for each $s\geq 0$, is the conditional expectation of the
random variable $M_{\omega}(t)$, $t>s$, relative to the
$\sigma$-algebra  $(\mathcal{P}_s)$,
i.e., $\Omega \ni \omega \mapsto \mathbb{E}_s[M_{\omega}(t)]
\in \mathbb{R}$ is a $\mathcal{P}_s$-measurable function
satisfying
\[
\e\big[\mathbb{E}_s[M_\omega(t)]\chi_A(\omega)\big] =
\e\big[M_\omega(t)\chi_A(\omega)\big], \quad \forall A \in
\mathcal{P}_s,
\]
where $\chi_A$ is the characteristic
function of the set $A$.
Thus, condition (iii) is equivalent to
$\mathbb{E}[(M_\omega(t)-M_\omega(s))\chi_A(\omega)] = 0$
for all $A \in \mathcal{P}_s$ and all $t, s \in \mathbb{R}$
satisfying $t>s\geq 0$.

In this paper  we shall only consider processes defined on  compact time
intervals $[0,T]$ which have continuous sample paths (i.e., continuous
with respect to the time variable  $t$ for almost all $\omega \in \Omega$).

If a martingale $M_\omega(\cdot):\R^+\rightarrow \R$ is continuous
for a.s. $\omega \in \Omega$ and $\mathbb{E}[M_{\omega} (t)^2] <\infty$
for all $t\geq 0$, we say that $M$ has a  \textit{quadratic variation}
$\{\llbracket M_\omega, M_\omega\rrbracket_t \mid t\in [0,T]\}$ if
$M_\omega^2 (t)-\llbracket M_\omega, M_\omega\rrbracket_t$ is a martingale
with respect to $(\mathcal{P}_t)_{t\ge 0}$, and
$\llbracket M_\omega, M_\omega\rrbracket_{\cdot}:\R^+\rightarrow \R$
is a continuous, non-decreasing process with
$\llbracket M_\omega, M_\omega\rrbracket_0 =0$ for a.s.
$\omega \in \Omega$. Such a process is unique
and coincides with the following limit (convergence in probability),
$$
\lim_{n\to \infty} \sum_{t_i ,t_{i+1} \in
\sigma_n}  (M_\omega(t_{i+1})-M_\omega(t_i ))^2
$$
where $\sigma_n$ is a partition of the interval $[0,t]$ and
the mesh converges to zero as $n\rightarrow \infty$. Actually, the
definition of the quadratic variation requires
only right-continuity of $M$.

Moreover, for two
martingales $M$ and $N$, under the same assumptions and
conventions as given above, one can also define their
\textit{covariation}
$$
\llbracket M_\omega, N_\omega\rrbracket_t:= \lim_{n\to \infty}
\sum_{t_i ,t_{i+1} \in \sigma_n}
(M_\omega(t_{i+1})-M_\omega(t_i ))(N_\omega(t_{i+1})-N_\omega(t_i )),
$$
which extends the notion of quadratic variation. Clearly,
$$
2\llbracket M_\omega, N_\omega\rrbracket_t =\llbracket M_\omega+
N_\omega, M_\omega+N_\omega\rrbracket_t
-\llbracket M_\omega, M_\omega\rrbracket_t
-\llbracket N_\omega, N_\omega\rrbracket_t .
$$

\medskip

More generally, one can consider local martingales. A
\textit{stopping time} is a random variable $\tau :\Omega
\rightarrow \mathbb R^+$ such that for all $t\geq 0$,
$\{\omega \in\Omega \mid \tau (\omega )\leq  t \}\in
\mathcal{P}_t$. Then, a stochastic process $M$ is a
\textit{local martingale} if there exists a sequence
of stopping times $\{ \tau_n \mid n\geq 1 \}$, such that
$\lim_{n\rightarrow \infty}\tau_n(\omega) = \infty$ a.s., and
$M^n_\omega (t):=M_\omega (t\wedge \tau_n(\omega) )$ is a square
integrable martingale for
all $n \geq 1$, where $t\wedge \tau_n(\omega): =
{\rm min}(t, \tau_n(\omega))$.  Thus, for a local martingale $M$, we define
$\llbracket M_\omega,M_\omega\rrbracket_t :=
\llbracket M^n_\omega ,M^n_\omega \rrbracket_t$ if $t\leq \tau_n(\omega)$.

A real-valued \textit{Brownian motion}  is a
martingale $W (t)$  with continuous sample paths, $t\in \R^+$, such that
$W^2 (t)-t$ is a martingale; or, equivalently, such that
$\llbracket W_\omega,W_\omega\rrbracket_t =t$ for a.s. $\omega \in \Omega$.

A stochastic process  $X:\Omega \times [0,T] \rightarrow
\mathbb{R}$ is a  \textit{(local) semimartingale}
with respect to the non-decreasing filtration $(\mathcal{P}_t)_{t\ge 0}$
if, for every $t \geq 0$, it can be decomposed into a sum
$$
X_\omega(t) =X_\omega(0) +M_\omega(t) +A_\omega(t),
$$
where $M$ is a local martingale with respect to $(\mathcal{P}_t)_{t\ge 0}$
such that $M_\omega(0) =0$ and $A$ is a
c\`adl\`ag $(\mathcal{P}_t)_{t\ge 0}$-adapted process
 of locally bounded variation with $A_\omega(0) =0$ a.s.
(c\`adl\`ag = ``continue \`a droite, limite \`a gauche''
means, by definition, that $A$ is
right-continuous with left limits at each
$t\geq 0$;  however, we consider only processes that are
continuous in the time variable $t$, which is a standing
assumption throughout this paper).

For a (local) semimartingale we define
$\llbracket X_\omega,X_\omega\rrbracket_t :=
\llbracket M_\omega,M_\omega\rrbracket_t$.
\medskip

Martingales and, in particular, Brownian motion,  are not
(a.s.) differentiable in time (unless they are constant);
therefore, one cannot integrate  with
respect to martingales as one does with respect to functions of bounded
variation.
We recall the definition of the
two most commonly used stochastic integrals, the It\^o and
the Stratonovich integrals.

If $X$  and $Y$ are  real-valued semimartingales with continuous sample paths
such that for some $T>0$,
\[
\e \left[\int_0^T |X_\omega(t) |^2 dt
+\int_0^T |Y_\omega(t) |^2 dt\right] <\infty,
\]
the \textit{It\^o stochastic integral} in the time interval
$[0,t]$, $0<t\leq T$, with respect to $Y$ is defined as the
limit in probability (if the limit exists) of the sums
$$
\int_0^t X_\omega(s) dY_\omega(s) =\lim_{n\to \infty} \sum_{t_i ,t_{i+1}
\in \sigma_n} X_\omega(t_i ) (Y_\omega (t_{i+1})-Y_\omega(t_i ))
$$
where $\sigma_n$ is a partition of the interval $[0,t]$ with
mesh converging to zero as $n\rightarrow \infty$.

The \textit{Stratonovich stochastic integral} is defined by
$$
\int_0^t X_\omega(s) \delta Y_\omega(s)=\lim_{n\to \infty}
\sum_{t_i ,t_{i+1} \in \sigma_n}
\frac{(X_\omega(t_i)+X_\omega(t_{i+1}))}{2} (Y_\omega(t_{i+1})-
Y_\omega(t_i))
$$
whenever such limit exists.

These integrals do not coincide, in general, even though $X$
is a   process with continuous sample paths (due
to the lack of differentiability of the paths of $Y$). The It\^o and  the
Stratonovich integrals are related by
\begin{equation}
\label{Ito-Strat}
\int_0^t X_\omega(s) \delta Y_\omega(s) =
\int_0^t X_\omega(s) d Y_\omega(s) +
\frac{1}{2}\int_0^t  d\llbracket X_\omega , Y_\omega\rrbracket _s.
\end{equation}

If $f\in C^2 (\mathbb{R})$, It\^o's formula states that
\begin{equation}
\label{ito_formula}
f(X_\omega(t) )=f(X_\omega (0) )+\int_0^t f^{\prime} (X_\omega (s) )
dX_\omega(s)
+\frac{1}{2}\int_0^t f^{\prime \prime} (X_\omega(s))
d\llbracket X_\omega,X_\omega\rrbracket_s
\end{equation}
This formula, for Stratonovich integrals, reads,
$$
f(X_\omega(t) )=f(X_\omega(0) )+\int_0^t f^{\prime} (X_\omega(s)) \delta X_\omega(s)
$$
One advantage of Stratonovich integrals is that they allow the
use of the same rules as those of the standard deterministic
differential calculus. On the other hand  an It\^o integral with
respect to a martingale $M$ is again a martingale (under the
integrability condition
$\e \left[\int_0^T |X_\omega(t)|^2 d \llbracket
M_\omega,M_\omega\rrbracket_t \right]<\infty$),
a very important property. For example, we have, as an
immediate consequence, that
$\mathbb{E}_s \left[\int_s^t X_\omega(r) d M_\omega(r)\right] =0$ for
all $0\leq s < t$. This property does not
hold for Stratonovich integrals.

In higher dimensions,  the difference between
the Stratonovich and the It\^o integral in It\^o's formula is
given in terms of the Hessian  of $f$ (see Subsection
\ref{subs_gen_der}). In fact, suppose that $X$ is an $\mathbb R^d$-valued
semimartingale; then It\^o's formula  in $d$-dimensions (see  also
\eqref{ito_formula}) states
that, for every $f\in C^2 (\mathbb R^d )$,

\begin{align}
\label{hessian_formula}
f(X_\omega (t ))&=f(X_\omega(0) )+\sum_{i=1}^d \int_0^t
\partial_i f (X_\omega(s) ) dX_\omega^i (s)
+\frac{1}{2}\sum_{{i,j}=1}^d\int_0^t \partial^2_{i,j}f(X_\omega (s))
d\llbracket X^i_\omega ,X^j_\omega \rrbracket_s  \nonumber \\
&=f(X_\omega(0) )+\sum_{i=1}^d \int_0^t \partial_i f (X_\omega (s))
\delta X^i_\omega (s)
\end{align}

For independent Brownian motions $W^i$, $i=1,\ldots, k$, we have
\begin{equation}
\label{diff_brownian_quadratic}
d\llbracket W^i_\omega , W^j_\omega \rrbracket_t =\delta_{ij} dt
\end{equation}
where $\delta_{ij}$ denotes the Kronecker delta symbol.
As the  covariation of semimartingales is determined by their
martingale parts, the following identities hold (see, e.g.,
\cite{IW}),
\begin{equation}
\label{diff_quadratic}
d\llbracket W^i_\omega , \iota \rrbracket_t =0 \quad\forall
i=1,\ldots, d , \qquad
d\llbracket \iota ,\iota  \rrbracket_t=0,
\end{equation}
where $\iota (t)=t$ is the identity (deterministic) function.

\subsection{The generalized derivative and martingale part for (topological) group valued semimartingales}
\label{subs_gen_der}

Let $G$ denote a topological group, endowed with a Banach manifold
structure (possibly infinite dimensional) whose underlying topology
is the given one, such that all left (or right) translations $L_g$
(resp. $R_g$) by arbitrary $g \in G$ are smooth maps, where $L_gh:=gh$,
$R_gh:= hg$, for all $g, h \in G$. Given  a vector $v \in T_eG$, we denote by
$v^L$ (resp. $v^R$) the left (resp. right) invariant vector field
whose value at the neutral element $e$ of $G$ is $v$, i.e.,
$v^L(g): = T_eL_g v$ (resp. $v^R(g): = T_eR_g v$), where $T_eL_g:
T_eG \rightarrow T_gG$ is the tangent map (derivative) of $L_g$
(and similarly for $R_g$). The operation
$[v_1, v_2]: = \left[v_1^L, v_2^L\right](e)$, for any $v_1, v_2 \in T_eG$,
defines a (left) Lie bracket on $T_eG$. In this paper, we denote by $\mathfrak{g}$ the Lie algebra of
$G$, which is the set of left invariant vector fields on $G$. When working with right invariant
vector fields, we shall still use, formally, the \textit{left} Lie
bracket defined above, i.e., we shall never work with right Lie algebras;
the bracket defined by right invariant vector fields is equal to the
negative of the left Lie bracket defined above.
Denote, as usual, by $\operatorname{ad}_uv:= [u, v]$ the adjoint
action of $T_eG$ on itself and by $\operatorname{ad}_u^\ast: T^*_eG
\rightarrow T^*_eG$ its dual map (the coadjoint action of $T_eG$ on
its dual $T^*_eG$).

Suppose that $\nabla$ is  a left invariant linear connection on $G$,
i.e., $\nabla_{v_1^L}v_2^L$ is a left invariant vector field, for
any $v_1, v_2 \in  T_eG$. Then we define $\nabla_{v_1} v_2 :=
\nabla_{v_1^L}v_2^L (e)$ for all $v_1 ,v_2 \in T_e G$. If right
translation is smooth, in all the definitions above,
we can replace  left translation by right translation in a similar way.
We also assume that the left invariant connection  $\nabla$ is torsion
free, namely
\[
\nabla_{v_1}v_2  - \nabla_{v_2}v_1
= \left[v_1, v_2 \right], \quad \text{for all} \quad v_1, v_2 \in T_eG.
\]
 For a fixed $g_1 \in G$, let
$T_{g_2} L_{g_1}: T_{g_2} G \rightarrow T_{g_1g_2} G$ be the
\textit{tangent map} (or derivative) of $L_{g_1}$ at the point
$g_2 \in G$.
\medskip

Let $G$ be  endowed with a left invariant linear torsion free connection
$\nabla$. The corresponding \textit{Hessian}
${\rm Hess}f(g): T_g G \times T_g G\rightarrow \mathbb{R}$ of
$f \in C^2(G)$  at $g \in G$ is defined by
\begin{equation}\label{e2-1}
{\rm Hess} f(g)\Big(v_1, v_2\Big):=\tilde{v}_1 \tilde{v}_2 f(g)-
\nabla_{\tilde{v}_1}\tilde{v}_2 f(g), \quad v_1, v_2 \in T_g G,
\end{equation}
where $\tilde{v}_i,\ i=1,2$, are arbitrary smooth vector fields on
$G$ such that $\tilde{v}_i(g)=v_i$. Since the connection
is torsion free, $\operatorname{Hess}f(g)$ is a symmetric
$\mathbb{R}$-bilinear form on each $T_g G$. In addition,
${\rm Hess}f= \nabla^2 f = \nabla \mathbf{d}f$ (see, e.g., \cite{E})
is the covariant derivative associated with $\nabla$ of the one-form
$\mathbf{d}f$, where $\mathbf{d}$ denotes the exterior
differential.
\medskip

Given a probability space $(\Omega , {\mathcal P}, \mathbb{P})$ endowed
with a non-decreasing
filtration $(\p_t)_{t\ge 0}$, a \textit{semimartingale with values in $G$}
(with respect to $(\p_t)_{t\ge 0}$) is a $\mathcal P_t$-adapted stochastic
process $g :\Omega \times \mathbb R^+\rightarrow G$ such that,
for every function $f\in C^2 (G)$, $f\circ g: \Omega \times \mathbb{R}^+
\rightarrow \mathbb{R}$ is a real-valued
semimartingale (on $(\Omega , {\mathcal P}, \mathbb{P})$), as introduced
in subsection \ref{subs_2_1} (see,
e.g., \cite{E} for the case of finite dimensional Lie groups).

A semimartingale with values in $G$ is a
$\nabla$-\textit{(local) martingale} if
$$
t \longmapsto
f(g_\omega(t))-f(g_\omega(0))-\frac{1}{2} \int_0^{t} {\rm Hess}
f (g_\omega(s)) d\llbracket g_\omega ,g_\omega \big\rrbracket_s ds
$$
is a  real-valued (local) martingale  for any $f \in C^2(G)$, where
$\llbracket g_\omega ,g_\omega \rrbracket_t$ is the quadratic variation
of $g_\omega$. If $G$ is a finite dimensional Lie group, then
we have the following expression
$$
d\llbracket g_\omega ,g_\omega \rrbracket_t :=
d\left[\!\!\left[ \int_0^{\cdot}\mathbf{P}_s^{-1} \delta g_\omega  (s),
\int_0^{\cdot}\mathbf{P}_s^{-1} \delta g_\omega  (s)\right]\!\!\right]_t,
$$
where $\mathbf{P}_t : T_{g_\omega (0)} G\rightarrow T_{g_\omega (t)} G$
is the (stochastic) parallel translation along the (stochastic) curve
$t \mapsto g_\omega(t)$ associated with the connection
$\nabla$; see, e.g., \cite{E} or \cite{IW}. Moreover, for some infinite
dimensional groups $G$ (for example the diffeomorphism group on a torus),
the quadratic variation is also well defined; we refer the reader to
\cite{ACC,CC} for details (see also Section \ref{sec_5} of this paper).

\medskip

For a $G$-valued semimartingale $g_\omega(\cdot)$, suppose there exist
an integer $m>0$ and $\mathcal{P}_t$-adapted processes
$\boldsymbol{\mathsf{v}}:\Omega\times \R^+ \rightarrow T_eG$,
$\w^i:\Omega \times \R^+ \rightarrow T_e G$, $M^i:\Omega\times \R^+
\rightarrow \R$, $1\le i \le m$, such that
$M^i$ is a ($\R$-valued) martingale with continuous sample  paths,
and for every $f \in
 C^{2}(G)$,
\begin{equation}\label{e2-0a}
\begin{split}
&f(g_\omega (t))=f(g_\omega (0))+\sum_{i=1}^m\int_0^t \left\langle
\mathbf{d}f(g_\omega (s)), T_eL_{\g(s)}\w^i_\omega(s)\right \rangle
dM^i_\omega(s)\\
&+\frac{1}{2}\sum_{i,j=1}^m\int_0^t
{\rm Hess}f (g_\omega (s)))\big(T_eL_{\g(s)}\w^i_\omega(s),T_eL_{\g(s)}
\w^j_\omega(s)\big)d\llbracket M^i_\omega , M^j_\omega \rrbracket_s\\
&+\frac{1}{2}\sum_{i,j=1}^m\int_0^t \left\langle \mathbf{d}f(g_\omega (s)),
T_eL_{\g(t)}
\left(\nabla_{\w^i_\omega(s)}\w^j_\omega(s)\right)\right \rangle
d\llbracket M^i_\omega , M^j_\omega \rrbracket_s\\
&+\frac{1}{2}\sum_{i=1}^m\int_0^t \left\langle \mathbf{d}f(\g(s)),
d\llbracket \w_\omega^i, M_\omega^i\rrbracket_s\right\rangle
+\int_0^t \left\langle \mathbf{d}f(g_\omega (s)),
T_eL_{\g(s)}\boldsymbol{\mathsf{v}}_\omega (s) \right\rangle ds.
\end{split}
\end{equation}

For such a $G$-valued semimartingale $\g$, having the form \eqref{e2-0a}
above, the following is true,
\begin{equation}\label{e2-0}
d\g(t)=T_e L_{\g(t)}\Big(\sum_{i=1}^m \w^i_\omega(t)\delta M_\omega^i(t)
+\boldsymbol{\mathsf{v}}_\omega (t)dt\Big).
\end{equation}
Here $\delta$ denotes the Stratonovich integral (of the tangent vectors
in $G$).

Note that although for a given left invariant connection $\nabla$, the
choice of $\{(\w^i_\omega, M_\omega^i)\mid 1\le i \le m\}$ in \eqref{e2-0}
may not be unique, the decomposition into the
martingale part (which is $\sum_{i=1}^m \w^i_\omega(t)dM^i_\omega(t)$),
and the drift part without contraction
(which is $T_e L_{\g(t)}\boldsymbol{\mathsf{v}}_\omega (s) dt$) in
\eqref{e2-0} is unique.
Then we define the \textit{velocity derivative}
of $g_\omega (\cdot)$ by
\begin{equation}\label{e2-2}
\frac{\D g_\omega (t)}{dt}:=
T_e L_{\g(t)}\boldsymbol{\mathsf{v}}_\omega (t),
\end{equation}
and the \textit{stochastic differential with respect to the martingale
part} of $g_{\omega}(\cdot)$ by
\begin{equation}\label{e2-2a}
d^{\Delta} g_{\omega}(t):=\sum_{i=1}^m T_e L_{\g(t)}
\left(\w^i_\omega(t)dM_\omega^i(t)\right),
\end{equation}
where $dM^i_\omega(t)$ denotes the It\^o integral with respect to the
martingale $M^i_\omega(t)$. Note that the two terms above do not depend
on the choice of the left invariant connection $\nabla$.

In order to obtain the viscous terms in the associated stochastic
Euler-Poincar\'e equation, we need to make a more detailed analysis of
the contraction part of the semimartingale (or stochastic Lagrangian path)
$\g(\cdot)$. For a given left invariant connection $\nabla$ on $G$  and
some fixed choice $\{(\w^i_\omega,M_\omega^i)\mid
1\le i \le m\}$, where both $\w^i_\omega$ and
$M_\omega^i$ are $\mathcal{P}_t$-adapted processes and $M_\omega^i$ are
real valued martingales with continuous sample paths, we define the
{\it contraction matrix}
$\frac{\mathbf{D}^{\nabla,(\w^i_\omega,M_\omega^i)_{i=1}^m}g_\omega(t)}
{dt}$ as the following $T_{\g(t)}G$-valued $m\times m$ matrix:
\begin{equation}
\label{e2-2b}
\begin{split}
\left(\frac{\mathbf{D}^{\nabla,(\w^i_\omega,M_\omega^i)_{i=1}^m}
g_\omega (t)}{dt}\right)_{i,j}
&:=T_eL_{\g(t)}\left(\nabla_{\w^i_\omega(t)}\w^j_\omega(t)
\frac{d\llbracket M^i_\omega , M^j_\omega \rrbracket_t}{dt} \right.\\
&\quad \left.
+\frac{d\llbracket \w^i_\omega, M_\omega^i\rrbracket_t}{dt}1_{\{i=j\}}
\right),\qquad 1\le i, j\le m.
\end{split}
\end{equation}

Therefore, we can split the differential of a $G$-valued semimartingale
into the velocity part, the Hessian (second order )
term, the martingale part, and the contraction part (more accurately the
contraction matrix). Intuitively,
the velocity part could be seen as the direction where the particles
flow, the martingale part represents their random fluctuations, while the
contraction part describes the contraction effect from the noise.

The term $\left(\frac{\mathbf{D}^{\nabla,(\w^i_\omega,M_\omega^i)_{i=1}^m}
g_\omega (t)}{dt}\right)_{i,j}$ corresponds to the contraction  between
the noises  in vectors
$\w^i_\omega$ and  $\w^j_\omega$. Thus, the
contraction matrix describes explicitly the behavior of the noises
interaction along different  vector fields (directions)
$\{\w^i_\omega\}_{i=1}^m$.

Let
$$
{\bf Sum}\left(\frac{\mathbf{D}^{\nabla,(\w^i_\omega,M_\omega^i)_{i=1}^m}
g_\omega (t)}{dt}\right):=
\sum_{i,j=1}^m \left(\frac{\mathbf{D}^{\nabla,
(\w^i_\omega,M_\omega^i)_{i=1}^m}g_\omega (t)}{dt}\right)_{i,j}
\in T_{\g(t)}G
$$
denote the sum of all entries of the matrix $\frac{\mathbf{D}^{\nabla,
(\w^i_\omega,M_\omega^i)_{i=1}^m}g_\omega (t)}{dt}$; for each
fixed $t$ this is
a $T_{\g(t)}G$-valued random variable.

Then it is easy to verify that for a $G$-valued
semimartingale of the form \eqref{e2-0} and any $f \in C^{2}(G)$, the
process
\begin{equation*}
\begin{split}
N_t^f&:=f(g_\omega (t))-f(g_\omega (0)))-\frac{1}{2}\int_0^t
{\rm Hess}f (g_\omega (s)))d\llbracket g_\omega , g_\omega \rrbracket_s\\
&-\frac{1}{2}\int_0^t \left\langle \mathbf{d}f(\g(s)),
{\bf Sum}\left(\frac{\mathbf{D}^{\nabla,(\w^i_\omega,M_\omega^i)_{i=1}^m}
g_\omega (s)}{ds}\right)\right\rangle
-\int_0^t  \left\langle \mathbf{d}f(\g(s)), \frac{\D \g(s)}{ds}
\right\rangle ds
\end{split}
\end{equation*}
is a real-valued local martingale.

We  remark that by (\ref{e2-2})-\eqref{e2-2b},
the terms $\frac{\D}{dt}$, $d^{\Delta}$,
$\frac{\mathbf{D}^{\nabla, (\w_\omega^i,M_\omega^i)_{i=1}^m}}{dt}$
are well defined for semimartingales with values in  a finite dimensional
Lie group as well as in some  infinite dimensional groups (the
diffeomorphism group on a torus for example); see, e.g., \cite{ACC} or
Section \ref{sec_5} below.

In  the stochastic Euler-Poincar\'e reduction  introduced in Section
\ref{sec_3}, the martingale part and the contraction part generate,
respectively, the martingale term and the viscosity term
in  associated (stochastic) Euler-Poincar\'e equation.

Moreover, when
$G$ is a finite dimensional compact Lie group,  for a $G$-valued
semimartingale $\g(\cdot)$ of the  form
\eqref{e2-0}, we have the
following equalities (see, e.g., \cite{E})
\begin{equation}\label{gen_der}
\begin{split}
\frac{D^{\nabla} g_\omega (t)}{dt}&:=\mathbf{P}_t
\left(\lim_{\epsilon \rightarrow 0} \mathbb{E}_t
\left[\frac{\eta_\omega (t+\epsilon )-
\eta_\omega (t)}{\epsilon}\right]\right)\\
&=
\frac{1}{2}{\bf Sum}\left(\frac{\mathbf{D}^{\nabla,
(\w^i_\omega,M_\omega^i)_{i=1}^m}\g(t)}{dt}\right)
+\frac{\D\g(t)}{dt},
\end{split}
\end{equation}
where $\mathbf{P}_t :T_eG \rightarrow T_{\g(t)}G$ is the stochastic
parallel translation associated  to $\nabla$,
$\mathbb{E}_t[\cdot]=\mathbb{E}[\cdot |\p_t]$  denotes the conditional
expectation, and
$$
\eta_\omega (t)=\int_0^t \mathbf{P}_s^{-1} \delta g_\omega (s) \in T_e G.
$$

Therefore, according to the definition, if  a  $G$-valued semimartingale
$g_\omega (t)$ satisfies  $\frac{D^{\nabla} g_\omega (t)}{dt}=0$, then
$g_\omega (t)$ is a $\nabla$-martingale.

In fact, $\frac{D^{\nabla}}{dt}$ is the generalized
derivative in \cite{ACC}, which is a generalization for group-valued
semimartingales of those in \cite{CC,Ne1967,Y1981,NYZ1981,Z}; it
contains a single term formed by the sum of all elements in the
contraction matrix. The generalized derivative is
sufficient to generate
the viscosity terms (second order differential terms) in some partial
differential equations through the stochastic reduction procedure. This
is the case, for example, for the incompressible Navier-Stokes equation;
see, e.g., \cite{ACC,CC,Ne1967,Y1981,NYZ1981,Z}. However, for a large
class of equations in fluid mechanics,  the viscous terms do not depend
only on such kind of contraction terms; see, e.g., the compressible
Navier-Stokes equation or the viscous MHD equation in Section \ref{sec_5}.
This is one of our motivations to introduce the decomposition
of $\frac{D^{\nabla}}{dt}$ above.

The generalized derivative  coincides with the drift of a diffusion
processes. It  was commonly used since the beginning of  Stochastic
Analysis but was first associated with a dynamical interpretation, as
a mean velocity, in the context of Nelson's Stochastic Mechanics
\cite{Ne1967}.

\medskip

Given  a $\mathbb{R}^m $-valued martingale $M_\omega (t)=
(M_{\omega}^1 (t), ...,M_{\omega}^m (t)),~t\in  [0,T] $,
which has a continuous sample path,  (non-random) vectors
$H_i \in T_eG$, $1\le i \le m$, and a $\p_t$-adapted, $T_e G$-valued
semi-martingale $u_\omega:\Omega \times [0,T]\rightarrow T_eG$,
consider the following Stratonovich SDE on $G$,
\begin{equation}\label{e2-3}
\begin{cases}
& d g_{\omega}(t)=T_eL_{g_{\omega}(t)}\left(\sum_{i=1}^m H_i\delta
M_{\omega}^i (t)+ u_{\omega}(t) dt\right),\\
& g_{\omega}(0)=e.
\end{cases}
\end{equation}

As explained in \cite{E}, given the connection $\nabla$,
the difference (contraction term) between the It\^o and Stratonovich
integrals has the following form
\begin{equation*}
\begin{split}
&\sum_{i=1}^m\Big(T_e L_{g_\omega(t)} H_i \delta M_\omega^i (t)-
T_e L_{g_\omega(t)} H_i d M_\omega^i(t)\Big)\\
&=\frac{1}{2}\sum_{i=1}^m d\llbracket (T_e L_{g_\omega(t)} H_i ),
M_\omega^i (t)\rrbracket_t\\
&=\frac{1}{2}\sum_{i,j=1}^m T_e L_{g_\omega(t)}(\nabla_{H_i} H_j)
d\llbracket  M_\omega^i, M_\omega^j\rrbracket_t.
\end{split}
\end{equation*}
Therefore, equation (\ref{e2-3}) is equivalent to
\begin{equation}
\label{eq_2_5}
\begin{cases}
& d g_{\omega}(t)=T_eL_{g_{\omega}(t)}\left(\sum_{i=1}^m
H_i d M_{\omega}^i (t)
+\frac{1}{2}\sum_{i,j=1}^m \nabla_{H_i}H_jd
\llbracket  M_\omega^i, M_\omega^j\rrbracket_t
+ u_{\omega}(t) dt\right),\\
& g_{\omega}(0)=e.
\end{cases}
\end{equation}

If $G$ is a finite dimensional Lie group, there exists a unique strong
solution for (\ref{e2-3}) (c.f. \cite{IW}, \cite{E}) and hence also for
\eqref{eq_2_5}.   When $G$ is the diffeomorphism group on a torus and $u$
is less regular, a weak solution to \eqref{e2-3} still exists
(\cite{ACC}, \cite{CC}) \
under suitable conditions on $H_i$.

Applying It\^o's formula to the solution $\g(t)$ of \eqref{e2-3} (see
\cite{E} for the case where $G$ is finite dimensional and
\cite[Section 4.2]{ACC} or Section \ref{sec_5} below, for the case
where $G$ is the diffeomorphism group on a torus),
for every $f\in C^2 (G)$ we have,
\begin{equation*}
\begin{split}
f(g_\omega(t))&=f(g_\omega(0))+\sum_{i=1}^m \int_0^t
\left\langle \d f(\g(s)), T_eL_{\g(s)}H_i \right\rangle dM_\omega^i(s)\\
&+\frac{1}{2}\int_0^t
{\rm Hess}f (g_\omega(s))d\llbracket g_\omega, g_\omega \rrbracket_s +
\int_0^t \left\langle \d f(\g(s)), T_e L_{g_\omega(s)}u_{\omega}(s)
\right\rangle ds\\
&+\frac{1}{2}\sum_{i,j=1}^m\int_0^t \left\langle \d f(\g(s)),
T_eL_{\g(s)}\nabla_{H_i}H_j\right\rangle
d\llbracket M_\omega^i, M_\omega^j \rrbracket_s
\end{split}
\end{equation*}
Actually, this last equality, valid for each $f\in C^2 (G)$, is
a characterization of the solution of the stochastic differential
equation (\ref{e2-3}) (or \eqref{eq_2_5}), in a weak sense.

Clearly, by the definition (\ref{e2-2}) and \eqref{e2-2a}, we have
\begin{equation}
\label{e2-4}
\begin{split}
&\frac{\D  g_\omega(t)}{d t}=T_e L_{g_\omega(t)}u_{\omega}(t),\\
& d^{\Delta}\g(t)=\sum_{i=1}^m (T_e L_{\g(t)}H_i) dM_\omega^i(t),\\
&\left(\frac{\mathbf{D}^{\nabla, (H_i,M_\omega^i)_{i=1}^m}\g(t)}{dt}
\right)_{i,j}=
T_e L_{\g(t)}(\nabla_{H_i}H_j)
\frac{d\llbracket M_\omega^i, M_\omega^j \rrbracket_t}{dt}.
\end{split}
\end{equation}

\section{Stochastic semidirect product Euler-Poincar\'e reduction}
\label{sec_3}

In this section, partially inspired by \cite{ACC}, \cite{CHR},
\cite{Holm2015}, we extend the deterministic semidirect product
Euler-Poincar\'e reduction, formulated and developed in \cite{HMR},
to the stochastic setting. By such a reduction, we obtain a large
class of partial differential equations and stochastic partial
differential equations with various viscosity terms; see Section
\ref{sec_5} below.

\subsection{Left invariant version}
\label{subsection_3_1}
Let $U$ be a vector space and $U^*$ its dual, also denote by
$\left\langle\cdot ,\cdot \right\rangle_U: U^* \times U
\rightarrow\mathbb{R}$ the (weak)
duality pairing. Suppose that $G$ is a group endowed with a manifold
structure making it into a topological group whose left translation
is smooth. As discussed in subsection \ref{subs_gen_der}, the tangent
space $T_eG$ to $G$ at the identity element $e \in G$ is (isomorphic to) a
Lie algebra.
Assume that $G$ has a left representation on $U$; therefore, there are
naturally induced left representations of the group $G$ and the Lie
algebra $T_eG$ on $U$ and $U^*$. All actions will be denoted by
concatenation. Let $\left\langle \cdot , \cdot \right\rangle_{T_eG} :
T_e^\ast G \times T_eG \rightarrow \mathbb{R}$
be the (weak) duality pairing between $T_e^\ast G$ and
$T_e G$. Define the operator $\diamond: U \times U^*: \rightarrow
T_e^*G$ by
\begin{equation}
\label{e3-1}
\left\langle a \diamond \alpha , v \right\rangle_{T_eG} : =
- \left\langle v\alpha, a  \right\rangle_U =
\left\langle \alpha, v a \right\rangle_U, \qquad v \in T_eG, \quad
a\in U, \quad \alpha \in U^*.
\end{equation}
In fact, $a \diamond \alpha$ is the value at $(a, \alpha)$ of
the momentum map $U \times U^*\rightarrow T^*_eG$ of the
cotangent lifted action induced by the left representation of
$G$ on $U$.
\medskip

Let $\mathscr{S}(G)$ denote the collection of $G$-valued
semimartingales with smooth coefficients defined on the time interval
$[0,T]$. Let $\scr{M}_m:=\{(a_{i,j})_{i,j=1}^m \mid a_{i,j}\in T_e G\}$
be the vector space of all $m\times m$, $T_e G$-valued
matrices. Define $\scr{M}:=\cup_{m=1}^{\infty}\scr{M}_m$.
In order to define the contraction matrix for  $\g \in \mathscr{S}(G)$
having the form \eqref{e2-0}, we need to fix a pair
$\{(\w^i_\omega,M_\omega^i) \mid 1\le i \le m\}$ in the martingale part
of \eqref{e2-0} (the first term of the right hand side of
\eqref{e2-0}, i.e., the It\^o integral). The hypotheses on this set
of pairs remain the same: $\w^i_\omega$ and $M_\omega^i$ are
$\mathcal{P}_t$-adapted processes and $M_\omega^i$ are real valued
martingales with continuous sample paths, for all $i=1, \ldots, m$.
We denote by  $(\g,\w^i_\omega,M_\omega^i)_{i=1}^m$ an element in
$\mathscr{S}(G)$ with a fixed choice $\{(\w^i_\omega,M_\omega^i) \mid
1\le i \le m\}$  in \eqref{e2-0}. Let $\widetilde{\mathscr{S}(G)}$ be
the collection of all  these triples.

Given a (left invariant) linear connection $\nabla$ on $G$, a
point $\alpha_0 \in U^*$, a random (Lagrangian) function
$l: \Omega\times [0,T]\times T_eG \times U^* \rightarrow \mathbb{R}$
such that $l_\omega(t)$ is $\mathcal{P}_t$-adapted for each $t\in [0,T]$,
a (viscosity force) function $p:\scr{M}\times \scr{M}\times T_e G
\rightarrow \R$, a (stochastic force) function $q: [0,T]\times T_eG \times
U^* \rightarrow T_e^*G$, vectors $V_i \in T_e G$ (which are
non-random), $1\le i \le k$, and an $\R^k$-valued martingale
$N_\omega(t)$, we  define a stochastic \textit{action functional}
$J^{\nabla, \alpha_0, l,p,q, (V_i, N_\omega^i)_{i=1}^k}:
\widetilde{\S(G)}\times \widetilde{\S(G)}\times \S(G) \rightarrow \R$ by
\begin{equation}\label{e3-2}
\begin{split}
& J^{\nabla, \alpha_0, l,p,q,(V_i, N_\omega^i)_{i=1}^k}
\Big(\big(g^1_\omega,\w_\omega^{1,i},M_\omega^{1,i}\big)_{i=1}^{m_1},
\big(g^2_\omega,\w_\omega^{2,i},M_\omega^{2,i}\big)_{i=1}^{m_2},\g^3\Big)\\
&:=
\int_0^T l_\omega\left(t, T_{g^1 _\omega(t)}L_{g^1_\omega(t)^{-1}}
\frac{\D g^1_\omega(t)}{dt},  \alpha_\omega(t)\right)dt\\
&\quad +
\int_0^T p\Bigg(T_{g^1 _\omega(t)}L_{g^1_\omega(t)^{-1}}
\frac{\mathbf{D}^{\nabla, (\w_\omega^{1,i},M_\omega^{i,1})_{i=1}^{m_1}}
\g^1(t)}{dt},
T_{g^2_\omega(t)}L_{g^2_\omega(t)^{-1}}\frac{\mathbf{D}^{\nabla,
(\w_\omega^{2,i},M_\omega^{i,2})_{i=1}^{m_2}}\g^2(t)}{dt},\\
&\qquad \qquad
T_{g^1 _\omega(t)}L_{g^1_\omega(t)^{-1}}\frac{\D \g^1(t)}{dt}\Bigg)dt\\
&\quad +\int_0^T \left\langle q\left(T_{g^1 _\omega(t)}
L_{g^1_\omega(t)^{-1}}
\frac{\D g^1_\omega(t)}{dt},  \alpha_\omega(t)\right),
T_{g^1 _\omega(t)}L_{g^1_\omega(t)^{-1}}d^{\Delta}\g^1(t)\right\rangle\\
&\quad -\sum_{i=1}^k \int_0^T \left\langle  q\left(T_{g^1 _\omega(t)}
L_{g^1_\omega(t)^{-1}}
\frac{\D g^1_\omega(t)}{dt},  \alpha_\omega(t)\right), V_i dN_\omega^i(t)
\right\rangle,
\end{split}
\end{equation}
where $\big(g^1_\omega,\w_\omega^{1,i},M_\omega^{1,i}\big)_{i=1}^{m_1}\in
\widetilde{\S(G)}$,
 $\big(g^2_\omega,\w_\omega^{2,i},M_\omega^{2,i}\big)_{i=1}^{m_2}\in
\widetilde{\S(G)}$, $T_{g^1 _\omega(t)}L_{g^1_\omega(t)^{-1}}d^{\Delta}
\g^1(t)$
corresponds to the It\^o integral on the vector space $T_e G$, and
\begin{equation}
\label{a_t}
\alpha_\omega(t):=g^3_\omega(t)^{-1}\alpha_0 .
\end{equation}

\begin{remark}
{\rm We explain intuitively why we want the action functional
$J^{\nabla, \alpha_0, l,p,q,(V_i ,N_\omega^i)_{i=1}^k}$
to have the form \eqref{e3-2}. When the Lagrangian
$l(v)=\frac{1}{2}\left\langle v, v\right\rangle_{\R^d}$, $v \in \R^d$,
is the kinetic energy, for a stochastic Lagrangian path
$d\g(t)=dM_\omega(t)+\u(t)dt=d^{\Delta}g(t)+\frac{\D \g (t)}{dt} dt$
with $M_\omega(t)$ being a $\R^d$-valued martingale, we can  formally
write the kinetic energy as follows
\begin{equation*}
\begin{split}
\int_0^T l(T_{\g(t)}L_{\g(t)^{-1}}d\g (t))
=&\int_0^T \frac{1}{2}\left\langle T_{\g(t)}L_{\g(t)^{-1}}
\frac{\D \g (t)}{dt} ,
T_{\g(t)}L_{\g(t)^{-1}}\frac{\D \g (t)}{dt} \right\rangle dt\\
&+\int_0^T \left\langle T_{\g(t)}L_{\g(t)^{-1}}\frac{\D \g (t)}{dt} ,
T_{\g(t)}L_{\g(t)^{-1}} d^{\Delta} \g(t) \right\rangle\\
&+\frac{1}{2}\int_0^T  \left\langle T_{\g(t)}L_{\g(t)^{-1}}\frac{d^{\Delta} \g(t)}{dt},
T_{\g(t)}L_{\g(t)^{-1}} \frac{d^{\Delta} \g(t)}{dt} \right\rangle\\
&:=I_1+I_2+I_3.
\end{split}
\end{equation*}
Here  $I_1$ represents the kinetic energy of the velocity: it is  the
action functional in the deterministic case, based on which a standard
Euler-Poincar\'e equation is obtained via the reduction procedure. The
summand $I_2$ contains a stochastic differential for the martingale part
of $d\g(t)$ and we can interpret it as the It\^o integral with respect  to
this martingale. Concerning $I_3$, since  it is not well-defined (it is
almost-everywhere infinite), we  drop this term in the
action functional.

Besides the kinetic energy, we could also add some extra terms of the
form
$$
\sum_{i=1}^k\int_0^T
\left\langle T_{\g(t)}L_{\g(t)^{-1}}\frac{\D  \g (t)}{dt},
V_i  dN_\omega^i(t) \right\rangle
$$
which represents
the external stochastic fluctuation for the velocity.

Therefore, we define an action functional as follows
\begin{equation*}
\begin{split}
J(\g(\cdot))=&\int_0^T \frac{1}{2}\left\langle T_{\g(t)}L_{\g(t)^{-1}}
\frac{\D  \g (t)}{dt} ,
T_{\g(t)}L_{\g(t)^{-1}}\frac{\D  \g (t)}{dt} \right\rangle dt\\
&+\int_0^T \left\langle T_{\g(t)}L_{\g(t)^{-1}}\frac{\D  \g (t)}{dt} ,
T_{\g(t)}L_{\g(t)^{-1}} d^{\Delta} \g(t) \right\rangle\\
&-\sum_{i=1}^k\int_0^T\left\langle T_{\g(t)}L_{\g(t)^{-1}}
\frac{\D\g(t)}{dt} ,
V_i dN_\omega^i(t) \right\rangle,
\end{split}
\end{equation*}
which, when we add the viscous term (defined by a viscosity force $q$
and the contraction matrix for $\g$), is a particular case of
\eqref{e3-2} for $q(v,a)=v$, $\forall v \in \R^d$, $a \in U^*$
(we use here the identification of $T_e^* G$ with
$T_e G$)}. 
\hfill $\lozenge$

\end{remark}

From now on, we write $J^{\nabla,(V_i,N_\omega^i)_{i=1}^k}$ for
$J^{\nabla, \alpha_0, l,p,q,(V_i,N_\omega^i)_{i=1}^k}$ for simplicity.
In order to characterize  the critical points of the action functional
and to derive the corresponding Euler-Poincar\'e equation, it is
necessary to consider a variation for
$\big(g^1_\omega,\w_\omega^{1,i},M_\omega^{1,i}\big)_{i=1}^{m_1}\in
\widetilde{\S(G)}$ and
$\big(g^2_\omega,\w_\omega^{2,i},M_\omega^{2,i}\big)_{i=1}^{m_2}\in
\widetilde{\S(G)}$.

For every $\varepsilon \in [0,1)$ and $\p_t$-adapted  process 
$\mathpzc{g}:\Omega \times [0,T]\rightarrow T_e G$
satisfying $\mathpzc{g_{\omega}}(0)=\mathpzc{g_{\omega}}(T)=0$ and 
$\gg(\cdot)\in C^1([0,1];T_e G)$ a.s.,
 let $e_{\omega,\varepsilon,\mathpzc{g}}(\cdot) \in
C^1 ([0,T];G)$ be the unique solution of the (random) time-dependent
ordinary differential equation on $G$
\begin{equation}\label{e3-3}
\begin{cases}
&\frac{d}{dt}e_{\omega,\varepsilon,\mathpzc{g}}(t)=
\varepsilon T_{e}L_{e_{\omega,\varepsilon,\mathpzc{g}}(t)}
\dot{\mathpzc{g_{\omega}}}(t),\\
&e_{\omega,\varepsilon,\mathpzc{g}}(0)=e,
\end{cases}
\end{equation}
where $\dot{\mathpzc{g_\omega}}(t)$ denotes the derivative with
respect to the time variable $t$. Note that this system implies
$e_{\omega,0,\mathpzc{g}}(t) = e$ a.s. for all $t \in [0,T]$.

From now on, in this section, we assume that $G$ is a finite dimensional
Lie group endowed with a left invariant linear connection $\nabla$ and
$U$ is a finite dimensional left $G$-representation space.

We first give the following lemma concerning the variations induced by
$e_{\omega,\varepsilon,\mathpzc{g}}$ on a semimartingale $\g \in \S(G)$.

\begin{lem}\label{l3-1}
Suppose $\g \in \S(G)$ has the form \eqref{e2-0} and let
$$
g^i_{\omega, \varepsilon,\mathpzc{g}}(t):=
g^i_\omega(t)e_{\omega,\varepsilon,\mathpzc{g}}(t), \quad
t\in [0,T], \quad \varepsilon \in [0, 1).$$
Then we have
\begin{equation}
\label{l3-1-1}
\begin{split}
dg_{\omega, \varepsilon,\mathpzc{g}}(t)
&=T_e L_{g_{\omega,\varepsilon ,\mathpzc{g}}(t)}\Big(\sum_{i=1}^{m}
\operatorname{Ad}_{e_{\omega,\varepsilon ,\mathpzc{g}}^{-1}(t)}
\w_\omega^i(t)\delta M_\omega^i(t)
+\operatorname{Ad}_{e_{\omega,\varepsilon ,\mathpzc{g}}^{-1}(t)}
\boldsymbol{\mathsf{v}}_\omega (t)dt
+ \varepsilon \dot{\mathpzc{g}_\omega }(t)dt\Big).
\end{split}
\end{equation}
\end{lem}
\begin{proof}
By It\^o's formula and recalling that the Leibniz rule  holds for
Stratonovich integrals, we have
\begin{equation*}
\begin{split}
dg_{\omega, \varepsilon,\mathpzc{g}}(t)
&=T_e L_{g_{\omega,\varepsilon ,\mathpzc{g}}(t)}\left(\sum_{i=1}^{m}
\operatorname{Ad}_{e_{\omega,\varepsilon ,\mathpzc{g}}^{-1}(t)}
\w_\omega^i(t)\delta M_\omega^i(t)
+\operatorname{Ad}_{e_{\omega,\varepsilon ,\mathpzc{g}}^{-1}(t)}
\boldsymbol{\mathsf{v}}_\omega (t)dt \right.\\
&\qquad  \left. \phantom{\sum_{i=1}^{m}}
+T_{e_{\omega,\varepsilon ,\mathpzc{g}}(t)}
L_{e_{\omega,\varepsilon ,\mathpzc{g}}^{-1}(t)}
\dot{e}_{\omega,\varepsilon , \mathpzc{g}}(t)dt\right)\\
&= T_e L_{g_{\omega,\varepsilon ,\mathpzc{g}}(t)}\left(\sum_{i=1}^{m}
\operatorname{Ad}_{e_{\omega,\varepsilon ,\mathpzc{g}}^{-1}(t)}
\w_\omega^i(t)\delta M_\omega^i(t)
+\operatorname{Ad}_{e_{\omega,\varepsilon ,\mathpzc{g}}^{-1}(t)}
\boldsymbol{\mathsf{v}}_\omega (t)dt+
\varepsilon \dot{\mathpzc{g}_\omega }(t)dt\right),
\end{split}
\end{equation*}
where the last equality is due to \eqref{e3-3}.
\end{proof}

Based on \eqref{l3-1-1}, it is natural to consider
$\big(g_{\omega, \varepsilon,\mathpzc{g}},
\operatorname{Ad}_{e_{\omega,\varepsilon ,\mathpzc{g}}^{-1}(t)}\w_\omega^i,
M_\omega^i\big)_{i=1}^m$ as a deformation for
$\big(\g,\w^i_\omega,M_\omega^i\big)_{i=1}^m$ with $\g\in \S(G)$
having the expression \eqref{e2-0}. Meanwhile, using definitions
\eqref{e2-2}--\eqref{e2-2b}, it is easy to verify that
\begin{equation}
\label{e3-6}
\begin{split}
& T_{g_{\omega, \varepsilon,\mathpzc{g}}(t)}L_{g_{\omega, \varepsilon,
\mathpzc{g}}(t)^{-1}}
\frac{\D g_{\omega, \varepsilon,\mathpzc{g}}(t)}{dt} =
\operatorname{Ad}_{e_{\omega,\varepsilon ,\mathpzc{g}}^{-1}(t)}
\boldsymbol{\mathsf{v}}_\omega (t)+
\varepsilon \dot{\mathpzc{g}_\omega }(t)\\
&T_{g_{\omega, \varepsilon,\mathpzc{g}}(t)}L_{g_{\omega, \varepsilon,
\mathpzc{g}}(t)^{-1}}d^{\Delta}g_{\omega, \varepsilon,\mathpzc{g}}(t)
=\sum_{i=1}^m  \big(\operatorname{Ad}_{e_{\omega,\varepsilon ,
\mathpzc{g}}^{-1}(t)} \w^i_\omega(t) \big)dM_\omega^i(t)\\
&\left(T_{g_{\omega, \varepsilon,\mathpzc{g}}(t)}L_{g_{\omega, \varepsilon,
\mathpzc{g}}(t)^{-1}}\frac{\mathbf{D}^{\nabla, (
\operatorname{Ad}_{e_{\omega,\varepsilon ,\mathpzc{g}}^{-1}(t)}\w^i,
M_\omega^i)_{i=1}^m}
g_{\omega, \varepsilon,\mathpzc{g}}(t)}{dt}\right)_{i,j}\\
&=\left(\nabla_{\operatorname{Ad}_{e_{\omega,\varepsilon ,
\mathpzc{g}}^{-1}(t)}\w_\omega^i(t)}
\operatorname{Ad}_{e_{\omega,\varepsilon ,\mathpzc{g}}^{-1}(t)}
\w_\omega^j(t)\right)\frac{d\llbracket M_\omega^i,
M_\omega^j \rrbracket_t}{dt}
+\frac{d\llbracket \operatorname{Ad}_{e_{\omega,\varepsilon ,\mathpzc{g}}^{-1}}\w^i , M_\omega^i\rrbracket_t}{dt}1_{\{i=j\}}.
\end{split}
\end{equation}
\begin{remark}
{\rm Although by now we assume that $G$ is a finite dimensional Lie group,
by the arguments in \cite[Section 4.2]{ACC} we know that \eqref{e3-6}
still holds when $G$ is the diffeomorphism group on torus, see, e.g.,
\eqref{e4-4} below. Hence Theorem \ref{t3-1} stated below
still holds for the diffeomorphism group on the torus (see Section
\ref{sec_5}).
} \hfill $\lozenge$
\end{remark}

Now we  define the critical point for action functional based on
the variations we introduced above. We say that
$\Big(\big(g^1_\omega,
\w_\omega^{1,i},M_\omega^{1,i}\big)_{i=1}^{m_1},
\big(g^2_\omega,\w_\omega^{2,i},M_\omega^{2,i}\big)_{i=1}^{m_2},
g^3_\omega\Big)\in \widetilde{\S(G)}\times \widetilde{\S(G)}\times\S(G)$
is a \textit{critical point} of $J^{\nabla, (V_i ,N_\omega^i)_{i=1}^k}$
if for every $\p_t$-adapted process $\gg$ satisfying
$\mathpzc{g_{\omega}}(\cdot) \in C^1([0,T]; T_eG)$
and $\mathpzc{g_{\omega}}(0)= \mathpzc{g_{\omega}}(T)=0$ a.s., we have
\begin{equation}
\label{e3-4}
\left.\frac{d}{d\varepsilon}\right|_{\varepsilon=0}
J^{\nabla, (V_i ,N_\omega^i)_{i=1}^k}
\left(\big(g_{\omega,
\varepsilon,\mathpzc{g}}^1,
\operatorname{Ad}_{e_{\omega,\varepsilon ,\mathpzc{g}}^{-1}}
\w_\omega^{1,i}, M_\omega^{1,i}\big)_{i=1}^{m_1},
\big(g_{\omega, \varepsilon,\mathpzc{g}}^2,
\operatorname{Ad}_{e_{\omega,\varepsilon ,\mathpzc{g}}^{-1}}
\w_\omega^{2,i}, M_\omega^{2,i}\big)_{i=1}^{m_2},
g_{\omega, \varepsilon,\mathpzc{g}}^3\right) = 0
\end{equation}
where
\begin{equation}
\label{deformations}
g_{\omega, \varepsilon,\mathpzc{g}}^i(t):=
g^i_\omega(t)e_{\omega,\varepsilon,\mathpzc{g}}(t), \quad
t\in [0,T], \quad  i=1,2,3, \quad \varepsilon \in [0, 1).
\end{equation}
We emphasize the particular form of these deformations in the Lie group:
they correspond to  developments along (random)
directions $\mathpzc{g_{\omega}}(t)$.

\begin{remark}
{\rm
As will be seen in the applications presented in Section
\ref{sec_5}, the reason why we choose three different semimartingales in
the variational principle \eqref{e3-4} is that the viscosity constants
in different  equations may be different.}
\hfill $\lozenge$
\end{remark}

Fixing (non-random) $\{H_i^j \}_{i=1}^{m_j} \in T_e G$, $j=1,2,3$,
as well as $\mathbb R^{m_j}$-valued martingales $M^j_{\omega}(t)=
(M^{j,1}_{\omega} (t),..., M^{j,m_j }_{\omega} (t)), ~j=1,2,3$,
we consider $\big(g^j_\omega,H^j_i,M_\omega^{j,i}\big)_{ i=1}^{m_j}
\in \widetilde{\S(G)}$, $j=1,2,3$, where $\g^j$ are the solutions
of the following SDEs  on $G$,
\begin{equation}
\label{e3-5}
\begin{cases}
& d g^j_\omega(t)=T_eL_{g^j_\omega(t)}\left(\sum_{i=1}^{m_j}
H_i^j\delta M^{j,i}_\omega(t)+ u_{\omega}(t) dt\right),\\
& g^j_\omega(0)=e,
\end{cases}
\end{equation}
and where $\u$ is a $\p_t$-adapted, $T_e G$-valued semimartingale. Note
that $\u(\cdot )$ is not given a priori and  is the same for $j=1,2,3$;
we shall see below that it is the solution of a certain (stochastic)
equation when
$\Big(\big(g^1_\omega,H_i^{1},M_\omega^{1,i}\big)_{i=1}^{m_1},
\big(g^2_\omega,H_i^2,M_\omega^{2,i}\big)_{i=1}^{m_2}, g^3_\omega\Big)$
is a critical point for $J^{\nabla,(H_i^1,M_\omega^{1,i})_{i=1}^{m_1}}$.

\subsection{Stochastic variational principle for stochastic differential equations}
\label{sec_{3.1}}

In the theorem below we use the functional derivative notation. Let
$V$ be (a possibly infinite dimensional) vector space and $V^*$ a
space in weak duality $\left\langle \cdot , \cdot \right\rangle: V^*
\times V \rightarrow \mathbb{R}$ with $V$; in finite dimensions,
$V^*$ is the usual dual vector space, but in infinite dimensions it
rarely is the topological dual. If $f:V \rightarrow \mathbb{R}$
is a smooth function, then the \textit{functional derivative}
$\frac{\delta f}{\delta a} \in V^*$, if it
exists, is defined by $\lim_{\varepsilon \rightarrow 0}
\frac{f(a+\varepsilon b)-f(a)}{\varepsilon} =
\left\langle\frac{\delta f}{\delta a}, b \right\rangle$ for all
$a, b \in V$.

In this section, we assume that $l,p,q$ in the action functional
$J^{\nabla,(V_i,N_\omega^i)_{i=1}^k}$ are smooth with respect to all variables, except, of course, $\omega\in \Omega$.

Thus, in the theorem below,
$\frac{\delta l}{\delta u} \in T^*_eG$,
$\frac{\delta l}{\delta \alpha}\in U$, $\frac{\delta p}{\delta \xi_1},
\frac{\delta p}{\delta \xi_2}\in \scr{M}^*$, and
$\frac{\delta p}{\delta u}\in T^*_e G$ are the partial functional
derivatives of $l: \Omega\times [0,T]\times T_eG \times U^* \rightarrow \mathbb{R}$ and
 $p:\scr{M}\times \scr{M}\times T_eG \rightarrow \R$.
Recall that  here $\scr{M}_m:=\{(a_{i,j})_{i,j=1}^m, a_{i,j}\in T_e G\}$
and $\scr{M}:=\cup_{m=1}^{\infty}\scr{M}_m$.

\begin{thm}\label{t3-1}
Let
$l:\Omega\times [0,T]\times T_eG \times U^* \rightarrow \mathbb{R}$,
$p:\scr{M}\times \scr{M}\times T_e G \rightarrow \R$,
$q: [0,T]\times T_e G \times U^* \rightarrow T_e^* G$
such that $\frac{\delta l_\omega}{\delta u}$ is non-random and
$l_\omega(t)$ is ${\mathcal P}_t$-adapted.
Suppose that  the semimartingales $g^j_\omega(\cdot)$,
$j=1,2,3$, have the form \eqref{e3-5}.
\begin{itemize}
\item[{\rm (i)}] Then
$\Big(\big(g^1_\omega,H_i^{1},M_\omega^{1,i}\big)_{i=1}^{m_1},
\big(g^2_\omega,H_i^2,M_\omega^{2,i}\big)_{i=1}^{m_2}, g^3_\omega\Big)$
is a critical point of $J^{\nabla,(H_i^1,M_\omega^{1,i})_{i=1}^{m_1}}$
{\rm(}given in  \eqref{e3-2}{\rm)} if and only if the $\p_t$-adapted
process $u_\omega(t)$ coupled with the $\p_t$-adapted  process
$\alpha_\omega(t)$ {\rm(}which is defined by \eqref{a_t}{\rm)}
satisfies the following \emph{(stochastic) semidirect product
Euler-Poincar\'e equation for stochastic particle paths:}
\begin{equation}\label{t3-1-1}
\begin{cases}
& d\left(\frac{\delta l_\omega}{\delta u}\left(t,\u(t),\a(t)\right)
+\frac{\delta p}{\delta u}\left(
\tilde H_{\omega,1}(t),\tilde H_{\omega,2}(t),\u(t)\right)\right)\\
&=\sum_{i=1}^{m_1}{\rm ad}^*_{H_i}q(\u(t),\a(t))
dM^{1,i}_{\omega}(t)+ {\rm ad}^*_{\u(t)}
\left(\frac{\delta l_\omega}{\delta u}\left(t,\u(t),\a(t)\right)\right)dt\\
&\quad +
\left(\frac{\delta l_\omega}{\delta \alpha}\left(t,\u(t),\a(t)\right)
\right)\diamond \a(t)dt
+{\rm ad}^*_{\u(t)}\left(\frac{\delta p}{\delta u}
\left(\tilde H_{\omega,1}(t),\tilde H_{\omega,2}(t),\u(t)\right)\right)dt\\
&\quad +K_\omega\left( t,\tilde H_{\omega,1}(t),
\tilde H_{\omega,2}(t),\u(t)\right)dt, \\
&d\alpha_\omega (t)=-\sum_{i=1}^{m_3} H_i^3 \a(t)dM_\omega^{3,i}(t)\\
&\quad +
\frac{1}{2}\sum_{i,k=1}^{m_3} H_k^3 \left(H_i^3 \a (t)\right)
d\llbracket M_\omega^{3,i}, M_\omega^{3,k}\rrbracket_t
- \u(t)\a (t)dt.
\end{cases}
\end{equation}
Here
the operation $\diamond$ is given by formula \eqref{e3-1},
$\tilde H_{\omega,j}(t)\in \scr{M}_{m_j}$, $j=1,2$,
is the $ m_j \times m_j$ matrix whose entries are given by
\begin{equation}\label{t3-1-1a}
\begin{split}
&(\tilde H_{\omega,j}(t))_{i,k}=\big(\nabla_{H_i^j}H_k^j\big)
\frac{d\llbracket M_\omega^{j,i}, M_\omega^{j,k}\rrbracket_t}{dt},\quad
 1\le i,k\le m_j,
\end{split}
\end{equation}
the operator $K_\omega: [0,T]\times \scr{M}\times \scr{M}
\times T_eG \rightarrow T^*_eG$ is defined for every $\omega \in \Omega$ by
\begin{equation}
\label{t3-1-3}
\begin{split}
\langle K_\omega(t,A_1,A_2,u),v \rangle=
 -\sum_{j=1}^2
\left\langle\frac{\delta p}{\delta \xi_j}(A_1,A_2,u),
B_{\omega,j}(t,v))\right\rangle,\quad  \forall\ t\in [0,T],
\end{split}
\end{equation}
\footnote{we delete the constant 1/2 in definition of $K_{\omega}$} where
$A_j\in \scr{M}_{m_j}$, $j=1,2$, $u,v\in T_eG$, and
$B_{\omega,j}(t,v)\in \scr{M}_{m_j}$ is  the $ m_j \times  m_j$ matrix whose entries are
\begin{equation}\label{t3-1-2}
\big(B_{\omega,j}(t,v)\big)_{i,k}:=\big(\nabla_{H_i^j}({\rm ad}_{v}H_k^j)+
\nabla_{{\rm ad}_{v}H_i^j}H_k^j\big)\frac{d\llbracket M_\omega^{j,i},
M_\omega^{j,k}\rrbracket_t}{dt},
1\le i,k\le m_j,\ t\in [0,T].
\end{equation}
\item[{\rm (ii)}]
The first equation in \eqref{t3-1-1} is
equivalent to the \emph{stochastic dissipative Euler-Poincar\'e variational
principle}
\begin{equation}
\label{EP_dissipative_vp}
\begin{split}
&\left. \frac{d}{d\ee}\right|_{\ee=0}\Bigg(\int_0^T
l_\omega(t,u_{\omega,\ee}(t), \alpha_{\omega,\ee}(t)) dt
+\int_0^T p(\tilde H_{\omega,1,\ee}(t), \tilde H_{\omega,2,\ee}(t),
u_{\omega,\ee}(t))dt\\
&+\int_0^T \langle q(t,u_{\omega,\ee}(t),\alpha_{\omega,\ee}(t)),
d\beta_{\omega,\ee}(t)\rangle
-\sum_{i=1}^{m_1} \left(
\int_0^T \langle q(t,u_{\omega,\ee}(t),\alpha_{\omega,\ee}(t)),
H_i^1\rangle dM_\omega^{1,i}(t)\right) \Bigg)= 0
\end{split}
\end{equation}
on $T_eG \times U^*$, for variations of the form
\begin{equation}
\label{left_variations}
\left\{
\begin{aligned}
&\frac{d u_{\omega,\ee}(t)}{d\ee}\Big|_{\ee=0} = \dot{v}_\omega(t)
+ ad_{\u(t)} v_\omega(t),\\
& \frac{d \alpha_{\omega,\ee}(t)}{d\ee}\Big|_{\ee=0} =
- v_\omega(t) \alpha_\omega(t),\\
&\frac{d \tilde H_{\omega,j,\ee}(t)}{d\ee}\Big|_{\ee=0}=
-B_{\omega,j}(t,v_\omega(t)),\ j=1,2,\\
& \frac{d \beta_{\omega,\ee}(t)}{d\ee}\Big|_{\ee=0}=
-\sum_{i=1}^{m_1} \int_0^t ad_{v_\omega(s)} H_i^1 dM_\omega^{1,i}(s),\\
& u_{\omega,0}(t)=\u(t), \alpha_{\omega,0}(t)=\alpha_\omega(t),
\beta_{\omega,0}(t)=\sum_{i=1}^{m_1}\int_0^t H_i^1 dM_\omega^{1,i}(s),
\tilde H_{\omega,j,0}(t)=\tilde H_{\omega,j}(t),
\end{aligned}
\right.
\end{equation}
where $v_\omega(t)$ is an $\p_t$-adapted  process such that
$v_\omega \in C^1([0,T];T_e G)$ and
$v_\omega(0)=0$, $v_\omega(T)=0$ a.s..
{\rm(}Note that this variational principle is \emph{constrained} and
\emph{stochastic}.{\rm)}
\end{itemize}
\end{thm}

\begin{proof} (i) \bf Step 1. \rm
We start by proving that $\a (t)=g^3_\omega (t)^{-1} \alpha_0 $
satisfies the second equation in (\ref{t3-1-1}).

 Since $d\Big(\big(g^3(t)\big)^{-1}g^3(t)\Big)=0$, we have
$$
d\left(g^3_\omega (t)\right)^{-1} =
- T_eR_{(g^3_\omega(t))^{-1}}
T_{g^3_\omega(t)}L_{(g^3_\omega(t))^{-1}}dg^3_\omega(t),
$$
so replacing $dg^3_{\omega} (t)$ by its expression in \eqref{e3-5}
we obtain,
\begin{equation}
\label{eq3_7}
\begin{cases}
& d (g^3_\omega(t))^{-1}=
T_eR_{(g^3_\omega(t))^{-1}}\left(\sum_{i=1}^{m_3}
 -H_i^3\delta M^{3,i}_\omega(t)- \u(t) dt\right),\\
& g^3_\omega(0)^{-1}=e.
\end{cases}
\end{equation}

We now derive the stochastic differential equation satisfied by
$\a (t)$:
\begin{align}
\label{first_g_omega_alpha}
&d\a(t)=d\left(g^3_\omega (t)^{-1} \alpha_0\right) =
\left[- T_{g^3_\omega (t)}L_{g^3_\omega (t)^{-1}}
dg^3_\omega (t) \right]g^3_\omega (t)^{-1}\alpha_0 \\
&\qquad = -\sum_{i=1}^{m_3}
H_i^3 \left(g^3_\omega (t)^{-1}\alpha_0\right)
\delta M^{3,i}_\omega(t)-
\u(t) \left(g^3_\omega (t)^{-1}\alpha_0\right) dt \nonumber
\end{align}

Since we assume $U^*$ to be a finite dimensional vector space,
the difference between the Stratonovich and It\^o integrals (see
\eqref{Ito-Strat}) yields
\begin{align*}
&\sum_{i=1}^{m_3}
\left(H_i^3 \left(g^3_\omega (t)^{-1}\alpha_0\right)\right)
\delta M^{3,i}_\omega(t) \\
&\qquad =
\sum_{i=1}^{m_3} \left(
\left(H_i^3 \left(g^3_\omega (t)^{-1}\alpha_0\right)\right)
d M^{3,i}_\omega(t)+\frac{1}{2}d \llbracket H_i^3 \left(g^3_\omega
(\cdot )^{-1}\alpha_0 \right) , M_\omega^{3,i}\rrbracket_t\right).
\end{align*}
By the same procedure as in \eqref{first_g_omega_alpha}, the (local)
martingale part of $H_i^3 (g^3_{\omega} (\cdot )^{-1} \alpha_0 )$
is equal to $-\sum_{k=1}^{m_3} \int_0^{\cdot} H_k^3 H_i^3
(g^3_{\omega} (t)^{-1} \alpha_0 )dM^{3,k}_\omega (t)$.
Therefore, by
\eqref{diff_brownian_quadratic} and \eqref{diff_quadratic} we derive
$$
\sum_{i=1}^{m_3} d \llbracket H_i^3 \left(
g^3_\omega (\cdot )^{-1}\alpha_0 \right) , M_\omega^{3,i}\rrbracket_t
=-\sum_{i,k=1}^{m_3} H_k^3 (H_i^3 (g^3_\omega (t)^{-1}\alpha_0)) d
\llbracket M_\omega^{3,i}, M_\omega^{3,k} \rrbracket_t.
$$
Using \eqref{first_g_omega_alpha} we have,
\begin{align}
\label{ito_version_first}
&d\a(t)=-\sum_{i=1}^{m_3} H_i^3
\a(t)dM_{\omega}^{3,i} (t)\\
&\quad +\frac{1}{2}\sum_{i,k=1}^{m_3} H_k^3\left( H_i^3 \a (t)
\right)d\llbracket M_\omega^{3,i}, M_\omega^{3,k} \rrbracket_t
-\u(t)\a(t)dt,
\nonumber
\end{align}
which is the second equation in \eqref{t3-1-1}.

\bf Step 2. \rm
Now we prove the first equation
in \eqref{t3-1-1}. Recall from \eqref{e3-3} that, for every
$\p_t$-adapted process $\gg$ satisfying $\mathpzc{g_{\omega}}(\cdot)
\in C^1([0,1]; T_eG)$ and $\mathpzc{g_{\omega}}(0)=
\mathpzc{g_{\omega}}(T)=0$ a.s.,
$e_{\omega,\varepsilon,\mathpzc{g}}(\cdot)
\in C^1 ([0,T];G)~ a.s.$ uniquely solves the following
(random) ordinary differential equation on $G$
$$
\frac{d}{dt}e_{\omega,\varepsilon,\mathpzc{g}}(t)=
\varepsilon T_{e}L_{e_{\omega,\varepsilon,\mathpzc{g}}(t)}
\dot{\mathpzc{g_{\omega}}}(t), \qquad
e_{\omega,\varepsilon,\mathpzc{g}}(0)=e.
$$

By \cite[Lemma 3.1]{ACC}, we have
\begin{equation}\label{t3-1-4}
\left.\frac{d}{d\varepsilon}\right|_{\varepsilon=0}
e_{\omega,\varepsilon,\mathpzc{g}}(t) =\mathpzc{g_{\omega}}(t),
\qquad
\left.\frac{d}{d\varepsilon}\right|_{\varepsilon=0}e_{\omega,\varepsilon,
\mathpzc{g}}(t)^{-1}
=-\mathpzc{g_{\omega}}(t), a.s..
\end{equation}
Since this computation is important in the proof, for the sake of
completeness, we recall it below. Denoting by $\frac{D}{Dt}$ and
$\frac{D}{D\epsilon}$ the covariant derivatives, induced by $\nabla$
on $G$, along curves parametrized
by $t$ and $\varepsilon$, respectively. Since the torsion vanishes,
 Gauss Lemma yields
\begin{align}
\frac{D}{Dt}\frac{d}{d \varepsilon}e_{\omega,\varepsilon,\mathpzc{g}}(t)&=
\frac{D}{D\varepsilon}\frac{d}{dt}e_{\omega,\varepsilon,\mathpzc{g}}(t)=
\frac{D}{D\varepsilon}\big(\varepsilon
T_eL_{e_{\omega,\varepsilon,\mathpzc{g}}(t)}\dot{\mathpzc{g_{\omega}}}(t)
\big)\\
&=T_eL_{e_{\omega,\varepsilon,\mathpzc{g}}(t)}\dot{\mathpzc{g_{\omega}}}(t)
+ \varepsilon\frac{D}{D\varepsilon}
\big(T_eL_{e_{\omega,\varepsilon,\mathpzc{g}}(t)}
\dot{\mathpzc{g_{\omega}}}(t)\big)\nonumber
\end{align}
Taking $\varepsilon =0$ and since $e_{\omega,0,\mathpzc{g}} (t)=e$ for all
$t$, we obtain $\frac{D}{Dt}\frac{d}{d\varepsilon}\big|_{\varepsilon=0}
e_{\omega,\varepsilon,\mathpzc{g}}(t)=\dot{\mathpzc{g_{\omega}}}(t)$.
Moreover $t \mapsto
\frac{d}{d\varepsilon}\big|_{\varepsilon=0}
e_{\omega,\varepsilon,\mathpzc{g_{\omega}}}(t)$ is a curve in the vector
space $T_eG$ and hence
$\frac{d}{dt}\frac{d}{d\varepsilon}\big|_{\varepsilon=0}
e_{\omega,\varepsilon,\mathpzc{g}}(t) =
\frac{D}{Dt}\frac{d}{d\varepsilon}\big|_{\varepsilon=0}
e_{\omega,\varepsilon,\mathpzc{g}}(t) = \dot{\mathpzc{g_{\omega}}}(t)$.
The first equality in \eqref{t3-1-4} is then a consequence of
$\mathpzc{g_{\omega}}(0)=0$ and
$\frac{d}{d\varepsilon}\big|_{\varepsilon=0}
e_{\omega,\varepsilon,\mathpzc{g}}(0)= 0$. Finally, since
$$
\frac{d}{d\varepsilon}e_{\omega,\varepsilon,\mathpzc{g}} (t)^{-1}=
-T_e R_{e_{\omega,\varepsilon,\mathpzc{g}}^{-1}(t)}
T_{e_{\omega,\varepsilon,\mathpzc{g}}(t)}
L_{e_{\omega,\varepsilon,\mathpzc{g}}^{-1}(t)}\frac{d}{d\varepsilon}
e_{\omega,\varepsilon,\mathpzc{g}}(t),
$$
the second equality in \eqref{t3-1-4}  follows from the first.

Note that due to \eqref{t3-1-4} we have
$\frac{d}{d \varepsilon}\big|_{\varepsilon=0}
\operatorname{Ad}_{e_{\omega,\varepsilon ,\mathpzc{g}}^{-1}(t)}v=
-{\rm ad}_{\mathpzc{g}_\omega(t)}v$, $v\in T_e G$. Combining this with
\eqref{e3-6} we have
\begin{align}\label{t3-1-5}
&\left.\frac{d}{d\varepsilon}\right|_{\varepsilon=0}
\left(T_{g^1_{\omega ,\ee,\mathpzc{g}}(t)}
L_{g^1_{\omega, \varepsilon,\mathpzc{g}}(t)^{-1}}
\frac{\D g^1_{\omega, \varepsilon,\mathpzc{g}}(t)}{dt}\right)
\\
&\qquad\qquad
 =\dot{\mathpzc{g_{\omega}}}(t)+
{\rm ad}_{u_\omega (t)}\mathpzc{g_{\omega}}(t)
\nonumber
\end{align}
\begin{equation}\label{t3-1-5a}
\begin{split}
& \left.\frac{d}{d\varepsilon}\right|_{\varepsilon=0}
\left(T_{g^1_{\omega ,\ee,\mathpzc{g}}(t)}
L_{g^1_{\omega, \varepsilon,\mathpzc{g}}(t)^{-1}}
d^{\Delta}g^1_{\omega, \varepsilon,\mathpzc{g}}(t)\right)\\
&=-\sum_{i=1}^{m_1}{\rm ad}_{\mathpzc{g_{\omega}}(t)}H_i^1
dM_\omega^{1,i}(t)
\end{split}
\end{equation}
\begin{equation}\label{t3-1-5b}
\begin{split}
& \left.\frac{d}{d\varepsilon}\right|_{\varepsilon=0}
\left(T_{g^j_{\omega ,\ee,\mathpzc{g}}(t)}
L_{g^j_{\omega, \varepsilon,\mathpzc{g}}(t)^{-1}}
\frac{\mathbf{D}^{\nabla, (H_{\omega,i}^{j,\varepsilon},
M_\omega^{j,i})_{i=1}^{m_j}}g^j_{\omega ,\ee,\mathpzc{g}}(t)}{dt}
\right)_{k,m}\\
&=-\big(\nabla_{{\rm ad}_{\mathpzc{g}_\omega(t)}H_m^j}H_k^{j}+
\nabla_{H_m^j}({\rm ad}_{\mathpzc{g}_\omega(t)}H_k^j)\big)
\frac{d\llbracket M_\omega^{j,k}, M_\omega^{j, m}\rrbracket_t}{dt}\\
&=-B_{\omega,j}(t,\mathpzc{g}_\omega (t)),\ \ j=1,2,
\end{split}
\end{equation}
where $g^j_{\omega,\varepsilon,\mathpzc{g}}(t)=\g^j(t)e_{\omega,
\varepsilon ,\mathpzc{g}}(t)$, $H_{\omega,i}^{j,\varepsilon}
(t):=\operatorname{Ad}_{e_{\omega,\varepsilon ,\mathpzc{g}}^{-1}(t)}H_i^j$,
$B_{\omega,j}(t,\cdot)$ is defined by \eqref{t3-1-2} and we have applied
the property $\llbracket H_{\omega,i}^{j,\varepsilon}, M_\omega^{j,i}
\rrbracket \equiv 0$ since $H_{\omega,i}^{j,\varepsilon}(\cdot)$ is a
process with bounded variation.

Since $g^j_{\omega, \varepsilon,\mathpzc{g}} (t):=
g^j_\omega(t)e_{\omega,\varepsilon,\mathpzc{g}}(t)$ and
$e_{\omega,0,\mathpzc{g}}(t) = e$ for all $t \in [0, T]$, we conclude
$g^j_{\omega, 0,\mathpzc{g}} (t)=g^j_\omega(t)$, for all $t \in [0,T]$,
$j=1,2,3$. Therefore,
\begin{equation}\label{t3-1-6}
\begin{split}
\left.\frac{d}{d\varepsilon}\right|_{\varepsilon=0}
g^3_{\omega ,\ee,\mathpzc{g}}(t)^{-1}\alpha_0 &=
-g^3_\omega(t)^{-1} \left(
\left.\frac{d}{d\varepsilon}\right|_{\varepsilon=0}
g^3_{\omega ,\ee,\mathpzc{g}}(t) \right)g^3_\omega(t)^{-1} \alpha_0 \\
& = - \left(\left.\frac{d}{d\varepsilon}\right|_{\varepsilon=0}
e_{\omega,\varepsilon,\mathpzc{g}}(t) \right)g^3_\omega(t)^{-1} \alpha_0
\stackrel{\eqref{t3-1-4}}= - \mathpzc{g_{\omega}}(t)g^3_\omega(t)^{-1}
\alpha_0\\
&=-\gg(t)\a(t).
\end{split}
\end{equation}

Based on \eqref{e3-2}, \eqref{t3-1-5}--\eqref{t3-1-6} and noting that
$d^{\Delta} g^1_{\omega,\ee,\mathpzc{g}}(t)\Big|_{\varepsilon=0}=
\sum_{i=1}^{m_1}H_i^1dM_\omega^{1,i}(t)$, we have
\begin{equation}\label{t3-1-7}
\begin{split}
&\left.\frac{d}{d\varepsilon}\right|_{\varepsilon=0}
J^{\nabla,(H_{i}^1,M_\omega^{1,i})_{i=1}^{m_1}}\left(
\left(g_{\omega ,\varepsilon,\mathpzc{g}}^1,
H_{\omega,i}^{1,\varepsilon},M_\omega^{1,i}\right)_{i=1}^{m_1},
\left(g_{\omega ,\varepsilon,\mathpzc{g}}^2,
H_{\omega,i}^{2,\varepsilon},M_\omega^{2,i}\right)_{i=1}^{m_2},
g_{\omega ,\varepsilon,\mathpzc{g}}^3\right)\\
&=\int_0^T \left\langle \frac{\delta l_\omega}{\delta u}
\left(t,\u(t), \a(t)\right),
\left.\frac{d}{d\varepsilon}\right|_{\varepsilon=0}\left(
T_{g^1_{\omega ,\ee,\mathpzc{g}}(t)}
L_{g^1_{\omega ,\ee,\mathpzc{g}}(t)^{-1}}
\frac{\mathscr{D} g^1_{\omega ,\ee,\mathpzc{g}}(t)}{dt}\right)
\right\rangle dt\\
&\qquad +\int_0^T \left\langle\frac{\delta p}{\delta u}
\left(\tilde H_{\omega,1}(t),\tilde H_{\omega,2}(t),\u(t)\right),
\frac{d}{d\varepsilon}\Big|_{\varepsilon=0}\left(T_{g^1_{\omega ,\ee,
\mathpzc{g}}(t)} L_{g^1_{\omega ,\ee,\mathpzc{g}}(t)^{-1}}
\frac{\mathscr{D} g^1_{\omega ,\ee,\mathpzc{g}}(t)}{dt}\right)\right\rangle
dt\\
&\qquad+\sum_{j=1}^2\left\langle \frac{\delta p}{\delta \xi_j}
\left(\tilde H_{\omega,1}(t),\tilde H_{\omega,2}(t),\u(t)\right),
\left.\frac{d}{d\varepsilon}\right|_{\varepsilon=0}
\left(T_{g^j_{\omega ,\ee,\mathpzc{g}}(t)}
L_{g^j_{\omega, \varepsilon,\mathpzc{g}}(t)^{-1}}
\frac{\mathbf{D}^{\nabla, (H_{\omega,i}^{j,\varepsilon},
M_\omega^{j,i})_{i=1}^{m_j}}}{dt}\right)\right\rangle \\
&\qquad + \int_0^T \left\langle
\left.\frac{d}{d\varepsilon}\right|_{\varepsilon=0}
\left(g^3_{\omega ,\ee,\mathpzc{g}}(t)^{-1}\alpha_0\right),
\frac{\delta l_\omega}{\delta \alpha}\left(t,\u(t),\a(t)\right)
\right\rangle dt\\
&\qquad +\int_0^T \left\langle q\left(t,\u(t),\a(t)\right),
\left.\frac{d}{d \ee}\right|_{\ee=0}\left(T_{g^1_{\omega ,\ee,
\mathpzc{g}}(t)}
L_{g^1_{\omega ,\ee,\mathpzc{g}}(t)^{-1}}
d^{\Delta} g^1_{\omega,\ee,\mathpzc{g}}(t)\right)\right \rangle\\
& =\int_0^T \left\langle \frac{\delta l_\omega}{\delta u}
\left(t,\u(t),\a(t)\right), \,\dot{\mathpzc{g_{\omega}}}(t)+
{\rm ad}_{\u(t)}\mathpzc{g}_\omega(t)
\right\rangle dt \\
&\qquad -\int_0^T \left\langle
\mathpzc{g_{\omega}}(t)\a(t), \frac{\delta l_\omega}{\delta \alpha}
\left(t,\u(t), \a(t)\right) \right\rangle dt \\
&\qquad +\int_0^T \left\langle \frac{\delta p}{\delta u}
\left(\tilde H_{\omega,1}(t),\tilde H_{\omega,2}(t),\u(t)\right),
\,\dot{\mathpzc{g_{\omega}}}(t)+
{\rm ad}_{\u(t)}\mathpzc{g}_\omega(t)
\right\rangle dt\\
&\qquad -\sum_{j=1}^2\left\langle \frac{\delta p}{\delta \xi_j}
\left(\tilde H_{\omega,1}(t),\tilde H_{\omega,2}(t),\u(t)\right),
B_{\omega,j}(t,\mathpzc{g}_\omega(t))\right\rangle dt\\
& \qquad
-\sum_{i=1}^{m_1}\int_0^T \left\langle q\left(t,\u(t),\a(t)\right),
{\rm ad}_{\gg(t)}H_i dM_\omega^{1,i}(t)\right\rangle\\
&=\int_0^T  \Bigg\langle
-d\left(\frac{\delta l_\omega}{\delta u}\left(t,\u(t),\a(t)\right)+
\frac{\delta p}{\delta u}\left(\tilde H_{\omega,1}(t),
\tilde H_{\omega,2}(t),\u(t)\right)\right)\\
&\qquad
+{\rm ad}^*_{\u(t)} \left(\frac{\delta l_\omega}{\delta u}
\left(t,\u(t),\a(t)\right)\right)dt
+ {\rm ad}^*_{\u(t)} \left(\frac{\delta p}{\delta u}
\left(\tilde H_{\omega,1}(t),\tilde H_{\omega,2}(t),\u(t)\right)\right)dt\\
&\qquad
+\left(\frac{\delta l_\omega}{\delta \alpha}\left(t,\u(t),\a(t)\right)
\right)\diamond \a(t)dt+
K_\omega\left(t,\tilde H_{\omega,1}(t),\tilde H_{\omega,2}(t),\u(t)
\right)dt\\
&\qquad  +\sum_{i=1}^{m_1}{\rm ad}^*_{H_i}
\left(q\left(t,\u(t),\a(t)\right)\right)
dM_\omega^{1,i}(t),\,\mathpzc{g_{\omega}}(t)\Bigg\rangle,
\end{split}
\end{equation}
where the first equality follows from the property
$T_{g^1 _\omega(t)}L_{g^1_\omega(t)^{-1}}
d^{\Delta} g^1_\omega(t)=\sum_{i=1}^{m_1}H_i^1 dM_\omega^{1,i}(t)$
(which implies that the term depending on derivatives of
$q$ vanishes), the last equality is obtained by applying the
following equation
\begin{equation*}
\begin{split}
0=&\frac{\delta l_\omega}{\delta u}\left(T,\u(T),\a(T)\right)\gg(T)-
\frac{\delta l_\omega}{\delta u}\left(0,\u(0),\a(0)\right)\gg(0)\\
&=\int_0^T \left\langle d\left(\frac{\delta l_\omega}{\delta u}
\left(t,\u(t),\a(t)\right)\right), \gg(t)\right\rangle
+\int_0^T \left\langle \frac{\delta l_\omega}{\delta u}
\left(t,\u(t),\a(t)\right), \dot{\gg}(t) \right\rangle
\end{split}
\end{equation*}
(note that $\gg(\cdot)$ has bounded variation and
$\frac{\delta l_\omega}{\delta u}\left(t,\u(t),\a(t)\right)$
is a semimartingale  because $\frac{\delta l_\omega}{\delta u}$ is
differentiable with respect to variable $t$ and is $\mathcal{P}_t$-adapted,
so  $d \llbracket \frac{\delta l_\omega}{\delta u}, \gg\rrbracket_t=0$),
and by applying the definitions of ${\rm ad}^*$,
$\diamond$ (see \eqref{e3-1}), and $K$ (see \eqref{t3-1-3}).

Since  $\mathpzc{g}_\omega(\cdot)$ is a $\p_t$-adapted arbitrary process,
$\Big(\big(g^1_\omega,H_i^{1},M_\omega^{1,i}\big)_{i=1}^{m_1},
\big(g^2_\omega,H_i^2,M_\omega^{2,i}\big)_{i=1}^{m_2}, g^3_\omega\Big)$
is a critical point of $J^{\nabla,(H_i^1,M_\omega^{1,i})_{i=1}^{m_1}}$
if and only if $\u$ satisfies  the first equation in (\ref{t3-1-1}).

This proves statement (i).
\smallskip

(ii) The expressions of the variations \eqref{left_variations} have
been already found in the previous computations. Applying the same
methods in the variational procedure
\eqref{t3-1-7} we obtain \eqref{EP_dissipative_vp}.

\end{proof}

\begin{remark}{\rm
As we shall see in Section \ref{sec_5}, the conclusion of Theorem
\ref{t3-1} still holds when $G$ is the diffeomorphism group of a torus and
the action of $G$ on $U^*$ is the pull back map.

If $G$ is a finite dimensional Lie group and $U$ a  finite dimensional
vector space, then \eqref{t3-1-1} is a actually an SDE. However, when
$G$ is the diffeomorphism group, as illustrated in Section \ref{sec_5},
\eqref{t3-1-1} is a system of SPDEs.}\hfill $\lozenge$
\end{remark}

\begin{remark}
{\rm  The variation \eqref{left_variations} is a stochastic version
of Lin's constrained variational principle (see, e.g.,
\cite[Theorem 1.2]{HMR}). In fact, if we take $p=q=0$ and $H_i^j=0$,
\eqref{left_variations} is the deterministic constrained variational
principle in \cite[Theorem 1.2]{HMR}.}\hfill $\lozenge$
\end{remark}

\begin{remark}
\label{r4-2}
{\rm  For simplicity, in Theorem \ref{t3-1} we assume that the contraction
force $p$ and stochastic force $q$ are independent of the advection
space $U^*$. In fact, following the same procedure in the proof of
Theorem \ref{t3-1}, we could also characterize the critical points of an
action functional even if $p$ and $q$ depend on $U^*$.}
\hfill $\lozenge$
\end{remark}

\subsection{Stochastic variational principle for ordinary
differential equations}
\label{sec_{3.2}}

If we take $q=0$,  and take the expectation in \eqref{a_t}, through
the stochastic reduction procedure in Theorem \ref{t3-1}, we obtain
a system of ODEs for the drift of the underlying stochastic paths
(not the SDE in {\eqref{t3-1-1}).

Let $l$, $p$ be the same terms in Theorem \ref{t3-1} such that
$l:[0,T]\times T_e G \times U^*
\rightarrow \R$ is non-random. We define
$\tilde J^{\nabla}: \widetilde{\S(G)}
\times \widetilde{\S(G)}\times\S(G)\rightarrow \R$ by
\begin{equation*}
\begin{split}
&\tilde J^{\nabla}\Big(\big( g^1_\omega,\w_\omega^{1,i},
M_\omega^{1,i}\big)_{i=1}^{m_1},
\big(g^2_\omega,\w_\omega^{2,i},M_\omega^{2,i}\big)_{i=1}^{m_2},
\g^3\Big)\\
&:=\int_0^T l\left(t,T_{\g^1(t)}L_{ \g^1(t)^{-1}}\frac{\D \g^1(t)}{dt},
\tilde \alpha(t)\right)dt
+ \int_0^T p\Bigg(T_{g^1_\omega(t)}L_{g^1_\omega(t)^{-1}}
\frac{\mathbf{D}^{\nabla, (\w_\omega^{1,i},
M_\omega^{i,1})_{i=1}^{m_1}}\g^1(t)}{dt},\\
&\qquad \qquad \qquad T_{g^2_\omega(t)}L_{ g^2_\omega(t)^{-1}}
\frac{\mathbf{D}^{\nabla, (\w_\omega^{2,i},
M_\omega^{i,2})_{i=1}^{m_2}} \g^2(t)}{dt},
T_{g^1_\omega(t)}L_{g^1_\omega(t)^{-1}}\frac{\D \g^1(t)}{dt}\Bigg)dt,
\end{split}
\end{equation*}
where $\g^j$, $j=1,2,3$ are $G$-valued semimartingales with form
\eqref{e2-0}, and $\tilde \alpha(t):=\e\big[\a(t)\big]=
\e\big[\tilde \g^3(t)^{-1}\alpha_0\big]\in U^*$  is
non-random. The action functional
$\tilde J^{\nabla}$  can be viewed as a deterministic counterpart of
\eqref{e3-2}, where $q=0$, $\a(t)$ is replaced by $\tilde \alpha(t)$,
and there is  no external stochastic force term (stochastic integral term).

Suppose also that  deformations are of the form \eqref{e3-4} with
$\gg$ non-random (we write $\gl$ for $\gg$ in this section); then we
can also define the critical point of $\tilde J^{\nabla}$ in the same
way of that in \eqref{e3-4}. To further simplify notation,
we drop the index $ \omega $ on some of the variables in the statement
of the theorem below; for example, we write  $u$ and
$\llbracket M^{j,i}, M^{j,k}\rrbracket_t$  for $\u$
and $\llbracket M_\omega^{j,i}, M_\omega^{j,k}\rrbracket_t$, respectively,
when such functions are deterministic.

\begin{thm}{\rm(}Stochastic reduction with deterministic drift and
deformations{\em)}
\label{t3-2}
Let the semimartingales $g^j_\omega(\cdot)$,
$j=1,2,3$, have the form \eqref{e3-5} with 
$u \in C^1([0,T];T_e G)$ and $\llbracket M^{j,i}, M^{j,k}\rrbracket_t$,
$1\le j \le 3$, $1\le i,k\le m_j$ being non-random.

{\rm (i)} Then
$\Big(\big(g^1_\omega,H_i^{1},M_\omega^{1,i}\big)_{i=1}^{m_1},
\big(g^2_\omega,H_i^2,M_\omega^{2,i}\big)_{i=1}^{m_2}, g^3_\omega\Big)$
is a critical
point of $\tilde J^{\nabla}$ if and only if
$u(t)$ coupled with $\tilde \alpha(t)$
satisfies the following (ordinary differential) equation
\begin{equation}\label{t3-2-1}
\begin{cases}
& d\left(\frac{\delta l}{\delta u}\left(t,u(t),\tilde \alpha(t)\right)+
\frac{\delta p}{\delta u}\left(\tilde H_1(t),\tilde H_2(t), u(t)\right)
\right)\\
&\quad = {\rm ad}^*_{u(t)}
\left(\frac{\delta l}{\delta u}\left(t,u(t),\tilde \alpha(t)\right)
\right)dt + {\rm ad}^*_{u(t)}
\left(\frac{\delta p}{\delta u}\left(\tilde H_1(t),\tilde H_2(t), u(t)
\right)\right)dt\\
&\quad \quad
+\left(\frac{\delta l}{\delta \alpha}\left(t,u(t),\tilde \alpha(t)\right)
\right)\diamond \tilde \alpha(t)dt
+ K\left(t,\tilde H_1(t),\tilde H_2(t), u(t)\right)dt,\\
&d\tilde \alpha (t)=
\frac{1}{2}\sum_{i,k=1}^{m_3} H_k^3 \left(H_i^3 \tilde \alpha (t)\right)
d\llbracket M^{3,i}, M^{3,k}\rrbracket_t-u (t)\tilde \alpha (t)dt,
\end{cases}
\end{equation}
where $\tilde H_1(t)\in \scr{M}_{m_1}$, $\tilde H_2(t)\in \scr{M}_{m_2}$,
$\diamond$, $K$ are the same terms as in Theorem \ref{t3-1}
{\rm(}except that we omit the subscript $\omega$ in
order to emphasize that these terms are non-random here{\rm)}.

{\rm (ii)}
The first equation in \eqref{t3-2-1} is
equivalent to the following stochastic variational principle
\begin{equation}\label{t3-2-2}
\begin{split}
\left. \frac{d}{d\ee}\right|_{\ee=0}\Bigg(&\int_0^T l(t,u_{\ee}(t),
\tilde \alpha_{\ee}(t)) dt
+\int_0^T p(\tilde H_{1,\ee}(t), \tilde H_{2,\ee}(t),u_{\ee}(t))dt\Bigg)= 0
\end{split}
\end{equation}
on $T_eG \times U^*$ for variations of the form
\begin{equation*}
\left\{
\begin{aligned}
&\frac{d u_{\ee}(t)}{d\ee}\Big|_{\ee=0} = \dot{v}(t) +
{\rm ad}_{u(t)}v(t),\\
& \frac{d \tilde \alpha_{\ee}(t)}{d\ee}\Big|_{\ee=0} =
- v(t) \tilde \alpha(t),\\
& \frac{d \tilde H_{j,\ee}(t)}{d \ee}\Big|_{\ee=0}=-B_j(t,v(t)),
\ \ j=1,2,\\
& u_0(t)=u(t), \tilde \alpha_0(t)=\tilde \alpha(t),\ \
\tilde H_{j,0}(t)=\tilde H_j(t).
\end{aligned}
\right.
\end{equation*}
where $v \in C^1([0,T];T_e G)$ with
$v(0)=0$, $v(T)=0$ is non-random and $B_j(t,v)$ is defined by
\eqref{t3-1-2}. Note that this variational principle is
\emph{constrained} and \emph{deterministic}.
\end{thm}
\begin{proof}
(i) Since $H_i^3$, $u$ are non-random and the action of $T_eG$ on
$U^*$ is linear, we have
\begin{equation*}
\begin{split}
&\E\left[H_j^3\left(H_i^3\a(t)\right)\right]
=H_j^3\left(H_i^3\left(\E[\a(t)]\right)\right)=
H_j^3\left(H_i^3\left(\tilde \alpha(t)\right)\right),\\
&\E\left[u(t)\a(t)\right]=u(t)\E\left[\a(t)\right]=
u(t)\tilde \alpha(t)
\end{split}
\end{equation*}
Then taking the expectation on both side of \eqref{ito_version_first},
we arrive to the second equation of \eqref{t3-2-1}.

Note that $e_{\omega,\ee,\gl}$ is non-random since $\gg$ is non-random;
from \eqref{t3-1-6} we obtain
\begin{equation*}
\begin{split}
&\left.\frac{d}{d\varepsilon}\right|_{\varepsilon=0}
\E\left[g^3_{\omega ,\ee,\mathpzc{g}}(t)^{-1}\alpha_0\right]=
\E\left[\left.\frac{d}{d\varepsilon}\right|_{\varepsilon=0}
g^3_{\omega ,\ee,\mathpzc{g}}(t)^{-1}\alpha_0\right]\\
&=-\E\left[\gl(t)\g^3(t)^{-1}\alpha_0\right]=-\gl(t)
\E\left[\g^3(t)^{-1}\alpha_0\right]
=-\gl(t)\tilde \alpha(t).
\end{split}
\end{equation*}
Based on this and following the same procedure of \eqref{t3-1-7}
(note that here $q=0$),  we have the first equation of \eqref{t3-2-1}.

(ii) By the same steps of \eqref{left_variations} we derive \eqref{t3-2-2}.
\end{proof}

\begin{remark}{\rm
As we will see in Section \ref{sec_5}, for the case that $G$ is a
diffeomorphism group, the system of
\eqref{t3-2-1} is a PDE with viscosity term.
\hfill $\lozenge$}
\end{remark}

\subsection{Right invariant version}

Due to relative sign changes
in the equations of motion and the dissipative constrained
variational principle, with a view to applications for the spatial
representation in continuum mechanics, we give below the right
invariant version of Theorem \ref{t3-1}.

Suppose that $G$ acts on the right on a vector space
$U$ (we will write the action of $g \in G$ on $u \in U$ by $ug$ and
similarly for the induced infinitesimal $\mathfrak{g}$-representation).

Thus, let $\g$ be a $G$-valued semimartingale of the form
\begin{equation}\label{e3-0a}
d\g(t)=T_e R_{\g(t)}\Big(\sum_{i=1}^m \w^i_\omega(t)\delta M_\omega^i(t)
+\boldsymbol{\mathsf{v}}_\omega (t)dt\Big),
\end{equation}
where $T_e R_{g_\omega(t)}$ denotes the differential
of  the right translation
$R_{g_\omega(t)}$ at the point $e$.
For a fixed right invariant connection $\nabla$ on $G$
and  $(\w^i,M_\omega^{i})_{i=1}^m$, where $\w^i_\omega$ and
$M_\omega^i$ are $\mathcal{P}_t$-adapted processes and $M_\omega^i$
are real valued martingales  with continuous sample paths for all
$i=1, \ldots, m$, we  define
\begin{equation*}
\begin{split}
\left(\frac{\mathbf{D}^{\nabla,(\w^i_\omega,M_\omega^i)_{i=1}^m}
g_\omega (t)}{dt}\right)_{i,j}:=&
T_eR_{\g(t)}\Big(\nabla_{\w^i_\omega(t)}\w^j_\omega(t)
\frac{d\llbracket M^i_\omega , M^j_\omega \rrbracket_t}{dt}\\
& \quad
+\frac{d\llbracket \w^i_\omega, M_\omega^i\rrbracket_t}{dt}1_{\{i=j\}}
\Big),\quad 1\le i, j\le m.
\end{split}
\end{equation*}
The terms $\frac{\D \g(t)}{dt}$ and $d^{\Delta}\g(t)$
are defined
similarly as in the left invariant case (see the defining formulas
\eqref{e2-4}).

With $l:\Omega\times [0,T]\times T_e G \times U^*\rightarrow \R$,
$p:\scr{M}\times \scr{M}\times T_e G \rightarrow \R$,
$q:[0,T]\times T_e G \times U^*\rightarrow T_e^* G$,
$V_i\in T_e G, N_\omega^i$, $1\le i \le k$,
satisfying the same conditions as those in subsections
\ref{subsection_3_1} and \ref{sec_{3.1}} (for the  left invariant case),
the action functional $J^{\nabla, (V_i,N_\omega^i)_{i=1}^k}$ is defined
for the right invariant case by
\begin{equation}\label{right_action}
\begin{split}
& J^{\nabla, (V_i,N_\omega^i)_{i=1}^k}
\Big(\big(g^1_\omega,\w_\omega^{1,i},M_\omega^{1,i}\big)_{i=1}^{m_1},
\big(g^2_\omega,\w_\omega^{2,i},M_\omega^{2,i}\big)_{i=1}^{m_2},\g^3\Big)\\
&:=
\int_0^T l_\omega\left(t,T_{g^1 _\omega(t)}R_{g^1_\omega(t)^{-1}}
\frac{\D g^1_\omega(t)}{dt},  \alpha_\omega(t)\right)dt\\
&\quad +
\int_0^T p\Bigg(T_{g^1 _\omega(t)}R_{g^1_\omega(t)^{-1}}
\frac{\mathbf{D}^{\nabla, (\w_\omega^{1,i},M_\omega^{i,1})_{i=1}^{m_1}}
\g^1(t)}{dt}, T_{g^2_\omega(t)}R_{g^2_\omega(t)^{-1}}
\frac{\mathbf{D}^{\nabla, (\w_\omega^{2,i},M_\omega^{i,2})_{i=1}^{m_2}}
\g^2(t)}{dt},\\
&\qquad \qquad \qquad \qquad
T_{g^1 _\omega(t)}R_{g^1_\omega(t)^{-1}}\frac{\D \g^1(t)}{dt}\Bigg)dt\\
&\quad
+\int_0^T \left\langle q\left(t,T_{g^1 _\omega(t)}R_{g^1_\omega(t)^{-1}}
\frac{\D g^1_\omega(t)}{dt},  \alpha_\omega(t)\right),
T_{g^1 _\omega(t)}R_{g^1_\omega(t)^{-1}}d^{\Delta}\g^1(t)\right\rangle\\
&\quad -\sum_{i=1}^k \int_0^T \left\langle
q\left(t,T_{g^1 _\omega(t)}R_{g^1_\omega(t)^{-1}}
\frac{\D g^1_\omega(t)}{dt},  \alpha_\omega(t)\right),
V_i dN_\omega^i(t)\right\rangle
\end{split}
\end{equation}
 and
\begin{equation}
\label{right_version_of_alpha}
\quad \alpha_{\omega}(t):=\alpha_0g_{\omega}^3(t)^{-1}.
\end{equation}

As for the left invariant case, for every (random) $\p_t$-adapted process
$\mathpzc{g}_\omega(\cdot)$ such that  $\gg \in C^1([0,T]; T_eG)$,
$\mathpzc{g_\omega}(0)=\mathpzc{g_\omega}(T)=0$ a.s., and
$\varepsilon \in [0,1)$, let $e_{\omega, \varepsilon,\mathpzc{g}}(\cdot)$
 be the unique solution of the (random) time-dependent
ordinary differential equation on $G$
\begin{equation}\label{e3-3-right}
\begin{cases}
&\frac{d}{dt}e_{\omega,\varepsilon,\mathpzc{g}}(t)=
\varepsilon T_{e}R_{e_{\omega,\varepsilon,\mathpzc{g}}(t)}
\dot{\mathpzc{g_\omega}}(t),\\
&e_{\omega,\varepsilon,\mathpzc{g}}(0)=e.
\end{cases}
\end{equation}
Define
\begin{equation}\label{e3-4-right}
g^j_{\omega, \varepsilon,\mathpzc{g}} (t):=
e_{\omega,\varepsilon,\mathpzc{g}}(t)g^j_\omega(t),\ \ j=1,2,3.
\end{equation}
With such deformations of $g^j_\omega(\cdot)$, we can  consider (right
invariant) critical points of $J^{\nabla, (V_i,N_\omega^i)_{i=1}^k}$
as in \eqref{e3-4}.

In the procedure leading to Theorem \ref{t3-1} and its proof,
we can interchange all left actions
and left translation operators by their right counterparts to obtain
Theorem \ref{t3_2} and \ref{t3_3} below, so we omit the proof here.

\begin{thm}\label{t3_2}
Assume that $G$ is a finite dimensional Lie group endowed with a right
invariant linear connection  $\nabla$, $H_i^j \in T_e G$,
$1\le i \le m_j$, and
$M_\omega^j=\{M_\omega(t)^{j,i}\}_{i=1}^{m_j}$, $j=1,2,3$ is an
$\R^{m_j}$-valued martingale.
Suppose that the semimartingales $g^j_\omega(\cdot )\in \S(G)$, $j=1,2,3$,
have the following form,
\begin{equation}
\label{e3-5-right_new}
\begin{cases}
& d g_{\omega}^j(t)=T_eR_{g_{\omega}^j(t)}\Big(\sum_{i=1}^{m_j}
H_i^j \delta M_{\omega}^{j,i}(t) +\u(t) dt\Big),\\
& g_{\omega}^j(0)=e,
\end{cases}
\end{equation}
where $\u$ is a $\p_t$-adapted, $T_e G$-valued semimartingale.
\item[{\rm (i)}]
If $\frac{\delta l_\omega}{\delta u}$ is non-random and
$l_\omega(t)$ is ${\mathcal P}_t$-adapted,
then  $\Big(\big(g^1_\omega,H_i^{1},
M_\omega^{1,i}\big)_{i=1}^{m_1},
\big(g^2_\omega,H_i^2,M_\omega^{2,i}\big)_{i=1}^{m_2}, g^3_\omega\Big)$ is a critical point
of $J^{\nabla,(H_i^1,M_\omega^{1,i})_{i=1}^{m_1}}$
defined by
\eqref{right_action} if and
only if the $\p_t$-adapted, $T_e G$-valued semimartingale  $\u(\cdot)$
coupled with $\p_t$-adapted, $U^*$-valued semimartingale $\a(\cdot)$
satisfies the following equation
\begin{equation}\label{t3-4-1-right}
\begin{cases}
& d\left(\frac{\delta l_\omega}{\delta u}\left(t,\u(t),\a(t)\right)
+\frac{\delta p}{\delta u}\left(\tilde H_{\omega,1}(t),
\tilde H_{\omega,2}(t),\u(t)\right)\right)\\
&=-\sum_{i=1}^{m_1}{\rm ad}^*_{H_i}q(t,\u(t),\a(t))
dM^{1,i}_{\omega}(t)
- {\rm ad}^*_{\u(t)}\left(\frac{\delta l_\omega}{\delta u}
\left(t,\u(t),\a(t)\right)\right)dt\\
&\quad -{\rm ad}^*_{\u(t)}
\left(\frac{\delta p}{\delta u}\left(\tilde H_{\omega,1}(t),
\tilde H_{\omega,2}(t),\u(t)\right)\right)dt+
\left(\frac{\delta l_\omega}{\delta \alpha}\left(t,\u(t),\a(t)\right)
\right)\diamond \a(t)dt\\
&\quad -K_\omega\left(t,\tilde H_{\omega,1}(t),\tilde H_{\omega,2}(t),
\u(t)\right)dt,   \\
&d\alpha_\omega (t)=-\sum_{i=1}^{m_3} \a(t) H_i^3 dM_\omega^{3,i}(t)+
\frac{1}{2}\sum_{i,k=1}^{m_3} \left(\a(t) H_i^3 \right)H_k^3
d\llbracket M_\omega^{3,i}, M_\omega^{3,k}\rrbracket_t
-\a (t)\u (t)dt,
\end{cases}
\end{equation}
where $\tilde H_{\omega,1}(t)$, $\tilde H_{\omega,2}(t)$, and $K_\omega$
are defined by \eqref{t3-1-1a}, and \eqref{t3-1-3} respectively.
\item[{\rm (ii)}]
The first equation in \eqref{t3-4-1-right} is
equivalent to the \emph{stochastic dissipative Euler-Poincar\'e variational
principle}
\begin{equation}
\label{EP_dissipative_vp-right}
\begin{split}
&\left. \frac{d}{d\ee}\right|_{\ee=0}\Bigg(\int_0^T
l_\omega(t,u_{\omega ,\ee}(t), \alpha_{\omega,\ee}(t)) dt
+\int_0^T p(\tilde H_{\omega,1,\ee}(t), \tilde H_{\omega,2,\ee}(t),
u_{\omega,\ee}(t))dt\\
&+\int_0^T \langle q(t,u_{\omega,\ee}(t),\alpha_{\omega,\ee}(t)),
d\beta_{\omega,\ee}(t)\rangle
-\sum_{i=1}^{m_1}\left(
\int_0^T \langle q(t,u_{\omega,\ee}(t),\alpha_{\omega,\ee}(t)),
H_i^1\rangle  dM_\omega^{1,i}(t)\right) \Bigg)= 0
\end{split}
\end{equation}
on $T_eG \times U^*$, for variations of the form
\begin{equation}
\label{right_variations}
\left\{
\begin{aligned}
&\frac{d u_{\omega,\ee}(t)}{d\ee}\Big|_{\ee=0} =
\dot{v}_\omega(t) - ad_{\u(t)} v_\omega(t),\\
& \frac{d \alpha_{\omega,\ee}(t)}{d\ee}\Big|_{\ee=0} =
- \alpha_\omega(t)v_\omega(t),\\
&\frac{d \tilde H_{\omega,j,\ee}(t)}{d\ee}\Big|_{\ee=0}=
B_{\omega,j}(t,v_\omega(t)),\ j=1,2,\\
& \frac{d \beta_{\omega,\ee}(t)}{d\ee}\Big|_{\ee=0}=
-\sum_{i=1}^{m_1} \int_0^t ad_{v_\omega(s)} H_i^1 dM_\omega^{1,i}(s),\\
& u_{\omega,0}(t)=\u(t), \alpha_{\omega,0}(t)=
\alpha_\omega(t), \beta_{\omega,0}(t)=\sum_{i=1}^{m_1}\int_0^t H_i^1
dM_\omega^{1,i}(s),
\tilde H_{\omega,j,0}(t)=\tilde H_{\omega,j}(t),
\end{aligned}
\right.
\end{equation}
where $v_\omega(t)$ is an $\p_t$-adapted  process such that $v \in C^1([0,T];T_e G)$ and
$v(0)=0$, $v(T)=0$ a.s..
\end{thm}

The right invariant version for the deterministic action functional
in Theorem \ref{t3-2} is the following:
\begin{equation*}
\begin{split}
&\tilde J^{\nabla}\Big(\big( g^1_\omega,\w_\omega^{1,i},M_\omega^{1,i}
\big)_{i=1}^{m_1}, \big(g^2_\omega,\w_\omega^{2,i},M_\omega^{2,i}
\big)_{i=1}^{m_2}, \g^3\Big)\\
&:=\int_0^T l\left(t,T_{ \g^1(t)}R_{ \g^1(t)^{-1}}
\frac{\D \g^1(t)}{dt}, \tilde \alpha(t)\right)dt
+\int_0^T p\Bigg(T_{g^1_\omega(t)}R_{g^1_\omega(t)^{-1}}
\frac{\mathbf{D}^{\nabla, (\w_\omega^{1,i},M_\omega^{i,1})_{i=1}^{m_1}}
\g^1(t)}{dt},\\
& \qquad \qquad\qquad
T_{g^2_\omega(t)}R_{g^2_\omega(t)^{-1}}
\frac{\mathbf{D}^{\nabla, (\w_\omega^{2,i},M_\omega^{i,2})_{i=1}^{m_2}}
\g^2(t)}{dt}, T_{g^1 _\omega(t)}R_{g^1_\omega(t)^{-1}}
\frac{\D \g^1(t)}{dt}\Bigg)dt.
\end{split}
\end{equation*}

Here $l$ is non-random and $\tilde \alpha(t):=\e\big[\alpha_\omega(t)\big]
\in U^*$ and $\a(t):=\alpha_0 \g^3(t)^{-1}$.

\begin{thm}\label{t3_3}
Suppose that  the semimartingales $g^j_\omega(\cdot)$,
$j=1,2,3$, have the form \eqref{e3-5-right_new} with 
$u \in C^1([0,T];T_e G)$ and $\llbracket M^{j,i},
M^{j,k}\rrbracket_t$, $1\le j \le 3$, $1\le i,k\le m_j$ being
non-random {\rm(}we write $u$, $\llbracket M^{j,i}, M^{j,k}\rrbracket_t$
for $\u$ and $\llbracket M_\omega^{j,i}, M_\omega^{j,k}\rrbracket_t$
in this theorem{\rm)}. Consider deformations of the form
\eqref{e3-4-right} with $\gl$ non-random
{\rm(}we write $\gl$ for $\gg$ in this theorem because
it is non-random{\rm)}.

{\rm (i)} Then $\Big(\big(g^1_\omega,H_i^{1},
M_\omega^{1,i}\big)_{i=1}^{m_1},
\big(g^2_\omega,H_i^2,M_\omega^{2,i}\big)_{i=1}^{m_2}, g^3_\omega\Big)$
is a critical
point of $\tilde J^{\nabla}$ if and only if $u(t)$ coupled with
$\tilde \alpha(t)$ satisfies the following (ordinary differential)
equation
\begin{equation}\label{t3-4-1}
\begin{cases}
& d\left(\frac{\delta l}{\delta u}\left(t,u(t),\tilde \alpha(t)\right)
+\frac{\delta p}{\delta u}
\left(\tilde H_1(t),\tilde H_2(t),u(t)\right)\right)\\
&= -{\rm ad}^*_{u(t)}
\left(\frac{\delta l}{\delta u}\left(t,u(t),\tilde \alpha(t)\right)
\right) dt
-{\rm ad}^*_{u(t)}\left(\frac{\delta p}{\delta u}
\left(\tilde H_1(t),\tilde H_2(t),u(t)\right)\right)dt\\
&\quad +\left(\frac{\delta l}{\delta \alpha}\left(t,u(t),\tilde \alpha(t)
\right)\right)\diamond \tilde \alpha(t)dt
-K\left(t,\tilde H_1(t),\tilde H_2(t),u(t)\right)dt,\\
&d\tilde \alpha (t)=
\frac{1}{2}\sum_{i,k=1}^{m_3} \left( \tilde \alpha (t)H_i^3\right)H_k^3
d\llbracket M^{3,i}_\omega, M_\omega^{3,k}\rrbracket_t-
\tilde \alpha(t)u (t)dt,
\end{cases}
\end{equation}
where $\tilde H_1(t)$, $\tilde H_2(t)$, $\diamond$, $K$ are the same terms
as in Theorem \ref{t3-2}.

{\rm (ii)}
The first equation in \eqref{t3-4-1} is
equivalent to the following stochastic variational principle
\begin{equation}\label{t3-4-2}
\begin{split}
\left. \frac{d}{d\ee}\right|_{\ee=0}\Bigg(&\int_0^T l(t,u_{\ee}(t),
\tilde \alpha_{\ee}(t)) dt
+\int_0^T p(\tilde H_{1,\ee}(t), \tilde H_{2,\ee}(t),u_{\ee}(t))dt\Bigg)= 0
\end{split}
\end{equation}
on $T_eG \times U^*$ for variations of the form
\begin{equation*}
\left\{
\begin{aligned}
&\frac{d u_{\ee}(t)}{d\ee}\Big|_{\ee=0} = \dot{v}(t) -
{\rm ad}_{u(t)}v(t),\\
& \frac{d \tilde \alpha_{\ee}(t)}{d\ee}\Big|_{\ee=0} =
- v(t) \tilde \alpha(t),\\
& \frac{d \tilde H_{j,\ee}(t)}{d \ee}\Big|_{\ee=0}=B_j(t,v(t)),\ j=1,2,\\
& u_0(t)=u(t), \tilde \alpha_0(t)=\tilde \alpha(t),\ \
\tilde H_{j,0}(t)=\tilde H_j(t),
\end{aligned}
\right.
\end{equation*}
where $v \in C^1([0,T];T_e G)$ with
$v(0)=0$, $v(T)=0$ is non-random.
\end{thm}

\section{Stochastic Kelvin-Noether theorem}
\label{sec_4}
In this section we study a (stochastic) version of the
Kelvin-Noether Theorem which holds for solutions of stochastic
Euler-Poincar\'e equations with advection terms (see \eqref{t3-1-1}).

Let $G$, $U^*$, $l$, $q$ be as in Section \ref{sec_3}.
Suppose $\c$ is a manifold and $G$ acts on the left on $\c$. Let
$\mathscr{K}: \c \times U^* \rightarrow T_e^{**}G$ be an
equivariant map, i.e.,
\begin{equation}\label{e5-1}
\left\langle \scr{K}\left(g^{-1}c,g^{-1}\alpha\right), \mu\right\rangle=
\left\langle \scr{K}\left(c,\alpha\right),
{\rm Ad}_{g^{-1}}^*\mu\right\rangle,\ \ c\in \c,\ \alpha\in U^*,\
g\in G, \mu \in T_e^*G,
\end{equation}
where $\langle\,, \rangle$ denotes the weak pairing
between $T_e^{**}G$ and $T_e^*G$.

As explained before, we identify the Lie algebra $\G$ with
the tangent space $T_e G$ at the unit element. As usual (see, e.g.,
\cite[Section 4]{HMR}), we define the Kelvin-Noether
quantity $I: \c\times T_eG \times U^*\rightarrow \R$ by
\begin{equation}\label{e5-2}
I\left(c,u,\alpha\right):=\left\langle \scr{K}\left(c,\alpha\right),
\frac{\delta l}{\delta u}
\left(u,\alpha\right)\right\rangle,\ \ c\in \c,\ u \in T_e G,\
\alpha\in U^*.
\end{equation}

Now we are ready to state the Kelvin-Noether Theorem for the solutions of
\eqref{t3-1-1}.
\begin{thm}\label{t5-1}
Suppose $\u(t)$ satisfies the first equation  in \eqref{t3-1-1} with
$\frac{\delta p}{\delta u}
\big(\tilde H_{\omega,1}(t),$ $\tilde H_{\omega,2}(t),\u(t)\big)\equiv 0$,
$\a(t)=g_{\omega}^3(t)^{-1}\alpha_0$ and
$\frac{\delta l}{\delta u}$ non-random.
Let  the semimartingales $g^j_\omega(\cdot)\in \S(G)$,
$j=1,2,3$, be defined by \eqref{e3-5} using this $\u(t)$. For a fixed
$c_0 \in \c$, set $c_\omega(t):=\g^3(t)^{-1}c_0$,
$I_\omega(t):=I\left(c_\omega(t),\u(t),\a(t)\right)$. We have
\begin{equation}
\label{t5-1-1}
\begin{split}
d I_\omega(t)&=\Bigg\langle \scr{K}\left(c_\omega(t),\a(t)\right),
\left(\sum_{i=1}^{m_1}{\rm ad}^*_{H_i^1}q_\omega(t) dM_\omega^{1,i}(t)-
\sum_{i=1}^{m_3}{\rm ad}^*_{H_i^3}\frac{\delta l_\omega}{\delta u}(t)
dM_\omega^{3,i}(t)\right)\\
&\qquad +\Bigg(\frac{\delta l_\omega}{\delta \alpha}(t)\diamond \a(t)+
K_\omega\left(t,\tilde H_{\omega,1}(t),\tilde H_{\omega,2}(t),\u(t)
\right)\\
&\qquad +\frac{1}{2}\sum_{i,k=1}^{m_3}{\rm ad}^*_{H_i^3}
{\rm ad}^*_{H_k^3}\left(\frac{\delta l_\omega}{\delta u}(t)\right)
 \frac{d\llbracket M_\omega^{3,i} , M_\omega^{3,k}\rrbracket_t}{dt}\\
& \qquad -\sum_{i=1}^{m_1}\sum_{k=1}^{m_3}{\rm ad}^*_{H_k^3}
{\rm ad}^*_{H_i^1}q_\omega(t)
\frac{d\llbracket M_\omega^{1,i} , M_\omega^{3,k}\rrbracket_t}{dt}\Bigg)dt
\Bigg\rangle \Bigg],
\end{split}
\end{equation}
where $\frac{\delta l_\omega}{\delta u}(t):=\frac{\delta l}{\delta u}
\left(t,\u(t),\a(t)\right)$,
$\frac{\delta l_\omega}{\delta \alpha}(t):=
\frac{\delta l_\omega}{\delta \alpha}\left(t,\u(t),\a(t)\right)$,
$q_\omega(t):=q\left(t,\u(t),\a(t)\right)$, and
$\tilde H_{\omega,1}(t)$, $\tilde H_{\omega,2}(t)\in \scr{M}$,
$K_\omega:[0,T]\times \scr{M}\times \scr{M}\times T_e G \rightarrow T_e^*G$ are
defined by \eqref{t3-1-1a} and \eqref{t3-1-3}, respectively.

In particular, if $\frac{\delta l_\omega}{\delta u}(t)=q_\omega(t)$,
$m_1=m_3=m$, and $H_i^1=H_i^3=H_i$, $M_\omega^{1,i}(t)=
M_\omega^{3,i}(t)$, $1\le i \le m$, we have
\begin{equation}\label{t5-1-1a}
\begin{split}
dI_\omega(t)&=\Bigg\langle \scr{K}\left(c_\omega(t),\a(t)\right),
\frac{\delta l_\omega}{\delta \alpha}(t)\diamond \a(t)dt+
K_\omega\left(t,\tilde H_{\omega,1}(t),\tilde H_{\omega,2}(t),\u(t)
\right)dt\\
&\qquad -\frac{1}{2}\sum_{i,k=1}^{m}{\rm ad}^*_{H_i}
{\rm ad}^*_{H_k}q_\omega(t)
 d\llbracket M_\omega^{3,i} , M_\omega^{3,k}\rrbracket_t
\Bigg\rangle.
\end{split}
\end{equation}
\end{thm}
\begin{proof}
Due to \eqref{e5-1} we have
\begin{equation}\label{t5-1-2}
\begin{split}
I_\omega(t)&=\left\langle
\scr{K}\left(\g^3(t)^{-1}c_0,\g^3(t)^{-1}\alpha_0\right),
\frac{\delta l}{\delta u}\left(t,\u(t),\a(t)\right)\right\rangle\\
&=\left\langle \scr{K}\left(c_0,\alpha_0\right),{\rm Ad}^*_{\g^3(t)^{-1}}
\frac{\delta l}{\delta u}\left(t,\u(t),\a(t)\right)\right\rangle.
\end{split}
\end{equation}
By It\^o's formula, for any $T_e^*G$-valued semimartingale
$\beta_{\omega}(\cdot)$, we have
\begin{equation}\label{t5-1-2a}
\begin{split}
&d{\rm Ad}^*_{\g^3(t)^{-1}}\beta_{\omega}(t)\\
&\quad ={\rm Ad}^*_{\g^3(t)^{-1}}\big(-{\rm ad}^*_{\g^3(t)^{-1}
\delta \g^3(t)}\beta_{\omega}(t)
+\delta \beta_{\omega}(t)\big)\\
&\quad ={\rm Ad}^*_{\g^3(t)^{-1}}\Bigg(-\sum_{i=1}^{m_3}
{\rm ad}^*_{H_i^3}\beta_{\omega}(t)dM_{\omega}^{3,i}(t)
-{\rm ad}^*_{\u(t)}\beta_{\omega}(t)dt\\
&\qquad +d\beta_{\omega}(t)+\frac{1}{2}\sum_{i,k=1}^{m_3}
{\rm ad}_{H_i^3}^*{\rm ad}_{H_k^3}^* \beta_{\omega}(t)d
\llbracket M_\omega^{3,i}, M_\omega^{3,k}\rrbracket_t-
\sum_{i=1}^{m_3} {\rm ad}_{H_i^3}^*
d\llbracket M_\omega^{3,i} , \beta_{\omega}\rrbracket_t \Bigg).
\end{split}
\end{equation}
Combining \eqref{t5-1-2a} with \eqref{e3-5} and \eqref{t3-1-1} yields
\begin{equation*}
\begin{split}
&d\left({\rm Ad}^*_{\g^3(t)^{-1}}
\frac{\delta l_\omega}{\delta u}(t)\right)\\
&={\rm Ad}^*_{\g^3(t)^{-1}}\Bigg(-\sum_{i=1}^{m_3}
{\rm ad}^*_{H_i^3}\frac{\delta l_\omega}{\delta u}(t)
dM_\omega^{3,i}(t)-{\rm ad}^*_{\u(t)}
\frac{\delta l_\omega}{\delta u}(t)dt+\sum_{i=1}^{m_1}
{\rm ad}^*_{H_i^1}q_\omega(t) dM_\omega^{1,i}(t)\\
&\quad +{\rm ad}^*_{\u(t)}\frac{\delta l_\omega}{\delta u}(t)dt+
\frac{\delta l_\omega}{\delta \alpha}(t)\diamond \a(t)dt
+K_\omega\left(t,\tilde H_{\omega,1}(t),\tilde H_{\omega,2}(t),\u(t)
\right)dt\\
&\quad +\frac{1}{2}\sum_{i,k=1}^{m_3}
{\rm ad}_{H_i^3}^*{\rm ad}_{H_k^3}^*
\left(\frac{\delta l_\omega}{\delta u}(t)\right)d
\llbracket M_\omega^{3,i}, M_\omega^{3,k}\rrbracket_t
-\sum_{i=1}^{m_1}\sum_{k=1}^{m_3}{\rm ad}^*_{H_k^3}
{\rm ad}^*_{H_i^1}q_\omega(t)
d\llbracket M_\omega^{1,i} , M_\omega^{3,k}\rrbracket_t\Bigg).
\end{split}
\end{equation*}
Here we also apply the assumption $\frac{\delta p}{\delta u}
\left(\tilde H_{\omega,1}(t),\tilde H_{\omega,2}(t),\u(t)\right)\equiv 0$.
Putting this into \eqref{t5-1-2} and applying \eqref{e5-2}, we arrive at
\begin{equation*}
\begin{split}
dI_\omega(t)&=
\left\langle \scr{K}\left(c_0,\alpha_0\right),
d\left({\rm Ad}^*_{\g^3(t)^{-1}}
\frac{\delta l_\omega}{\delta u}(t)\right)\right\rangle\\
&=\Bigg\langle \scr{K}\left(c_0,\alpha_0\right), {\rm Ad}^*_{\g^3(t)^{-1}}
\Bigg(\sum_{i=1}^{m_1}{\rm ad}^*_{H_i^1}q_\omega(t) dM_\omega^{1,i}(t)-
\sum_{i=1}^{m_3}{\rm ad}^*_{H_i^3}\frac{\delta l_\omega}{\delta u}(t)
dM_\omega^{3,i}(t)\\
&
+\frac{\delta l_\omega}{\delta a}(t)\diamond \a(t)dt
+K_\omega\left(t,\tilde H_{\omega,1}(t),\tilde H_{\omega,2}(t),\u(t)
\right)dt\\
&+\frac{1}{2}\sum_{i,k=1}^{m_3}
{\rm ad}_{H_i^3}^*{\rm ad}_{H_k^3}^*
\left(\frac{\delta l_\omega}{\delta u}(t)\right)d
\llbracket M_\omega^{3,i}, M_\omega^{3,k}\rrbracket_t-
\sum_{i=1}^{m_1}\sum_{k=1}^{m_3}{\rm ad}^*_{H_k^3}
{\rm ad}^*_{H_i^1}q_\omega(t)
d\llbracket M_\omega^{1,i} , M_\omega^{3,k}\rrbracket_t\Bigg)\Bigg\rangle\\
&=\Bigg\langle \scr{K}\left(c_\omega(t),\a(t)\right),
\left(\sum_{i=1}^{m_1}{\rm ad}^*_{H_i^1}q_\omega(t) dM_\omega^{1,i}(t)-
\sum_{i=1}^{m_3}{\rm ad}^*_{H_i^3}\frac{\delta l_\omega}{\delta u}(t)
dM_\omega^{3,i}(t)\right)\\
&
+\frac{\delta l_\omega}{\delta a}(t)\diamond \a(t)dt+
K_\omega\left(t,\tilde H_{\omega,1}(t),\tilde H_{\omega,2}(t),\u(t)
\right)dt\\
& +\frac{1}{2}\sum_{i,k=1}^{m_3}
{\rm ad}_{H_i^3}^*{\rm ad}_{H_k^3}^*
\left(\frac{\delta l_\omega}{\delta u}(t)\right)d
\llbracket M_\omega^{3,i}, M_\omega^{3,k}\rrbracket_t-
\sum_{i=1}^{m_1}\sum_{k=1}^{m_3}
{\rm ad}^*_{H_k^3}{\rm ad}^*_{H_i^1}q_\omega(t)
d\llbracket M_\omega^{1,i} , M_\omega^{3,k}\rrbracket_t \Bigg\rangle,
\end{split}
\end{equation*}
which proves \eqref{t5-1-1}.

If $\frac{\delta l_\omega}{\delta u}(t)=q_\omega(t)$, $m_1=m_3=m$,
and $H_i^1=H_i^3=H_i$, $M_\omega^{1,i}(t)=M_\omega^{3,i}(t)$,
$1\le i \le m$, then
${\rm ad}^*_{H_i}\frac{\delta l_\omega}{\delta u}(t)=
{\rm ad}^*_{H_i}q_\omega(t)$
which, combined with \eqref{t5-1-1}, yields \eqref{t5-1-1a}.
\end{proof}

\begin{remark}
{\rm If $H_i^j=0$, $1\le i \le m_j$, $j=1,2,3$,
then \eqref{t5-1-1} becomes
\begin{equation*}
dI_\omega(t)=\left\langle \scr{K}\left(c_\omega(t),\a(t)\right),
\frac{\delta l_\omega}{\delta a}(t)\diamond \a(t)dt\right\rangle,
\end{equation*}
thus recovering \cite[Theorem 4.1]{HMR} for the deterministic
Euler-Poincar\'e equation with advection term.
\hfill $\lozenge$}
\end{remark}

\begin{remark}
{\rm As we will see in Section \ref{sec_5},
the term $K_\omega\left(t,\tilde H_{\omega,1}(t),\tilde H_{\omega,2}(t),
\u(t) \right)$ usually corresponds to some viscosity  terms of the system.
\hfill $\lozenge$}
\end{remark}

\section{Application to PDEs and SPDEs in fluid mechanics}
\label{sec_5}

We begin by recalling, from \cite{EM} and \cite{MEF}, the
necessary standard facts about the group of diffeomorphisms
on a smooth compact boundaryless $n$-dimensional manifold $M$.
Then, when we present the compressible Navier-Stokes and MHD equations
in the periodic case,  we shall take $M= \mathbb{T}^3$, the usual three
dimensional flat torus.

\medskip

Let $M$ be a smooth compact boundaryless $n$-dimensional manifold. Define
\begin{equation*}
G^s:=\left\{g: M \rightarrow M {\rm\ is\ a\ bijection} \mid
g,g^{-1}\in H^s(M, M)\right\},\\
\end{equation*}
where $H^s(M, M)$ denotes the manifold of
Sobolev maps of class $s> 1 + \frac{n}{2}$ from $M$ to itself.
The condition $s>\frac{n}{2}$ suffices to ensure the manifold structure
of $H^s(M, M)$;
only for such regularity class does the notion of an $H^s$-map
from $M$ to itself make intrinsic sense. If $s> 1 + \frac{n}{2}$
(the additional regularity is needed in order to
ensure that all elements of $G^s$ are $C^1$ and hence the inverse
function theorem is applicable), then $G^s$ is an open subset in
$H^s(M, M)$, so it is a $C^{\infty}$ Hilbert manifold. Moreover,
it is a group under composition of diffeomorphisms maps, right translation
by any element is smooth, left
translation and inversion are only continuous, and $G^s$ is a
topological group (relative to the underlying manifold topology)
(see \cite{EM}, \cite{MEF}); thus, $G^s$ is not a Lie group.
Since $G^s$ is an open subset of $H^s(M, M)$,
the tangent space $T_eG^s$ to the identity $e: M \rightarrow M$
coincides with the Hilbert space
$\mathfrak{X}^s(M)$ of $H^s$ vector fields on $M$. Denote
by $\mathfrak{X}(M)$ the Lie algebra of $C^{\infty}$ vector
fields on $M$. The
failure of $G^s$ to be a Lie group is mirrored by the fact that
$\mathfrak{X}^s(M)$ is not a Lie algebra: the usual Jacobi-Lie
bracket of vector fields, i.e.,
$[X,Y][f] = X[Y[f]] - Y[X[f]]$ for any $X, Y \in
\mathfrak{X}(M)$ and $f \in C^{\infty}(M)$, where $X[f] :=
\mathbf{d}f(X)$ is the differential of $f$ in the direction $X$,
loses a derivative for finite differentiability class of vector
fields and thus $[\cdot ,\cdot ] : \mathfrak{X}^s(M) \times
\mathfrak{X}^s(M) \rightarrow \mathfrak{X}^{s-1}(M)$ is not an
operation on $\mathfrak{X}^s(M)$. In general, the tangent
space $T_\eta G^s$ at an
arbitrary $\eta\in G^s$ is $T_\eta G^s = \left\{U:M \rightarrow TM
\text{ of class } H^s\mid U(m) \in T_{\eta(m)} M \right\}$.
If $s > 1+\frac{n}{2}$ and $X \in \mathfrak{X}^s(M)$, then its
global (since $M$ is compact) flow $\mathbb{R}\ni t
\mapsto F_t \in G^s$ exists and is a $C^1$-curve in $G^s$
(see, e.g., \cite[Theorem 2.4.2]{MEF}). The candidate of what should
have been the Lie group exponential map is
$\exp: T_eG^s = \mathfrak{X}^s(M) \ni X \mapsto F_1 \in G^s$,
where $F_t$ is the flow of $X$; however, $\exp$ does not cover a
neighborhood of the identity and it is not $C^1$. Therefore, all
classical proofs in the theory of finite dimensional Lie groups
based on the exponential map, break down for $G^s$. From now on,
we shall always assume $s> 1+ \frac{n}{2}$.

Since right translation is smooth, each $X \in
\mathfrak{X}(M)$ induces a $C^{\infty}$ right invariant
vector field $X^R\in \mathfrak{X}(G^s)$ on $G^s$, defined by
$X^R(\eta): = X \circ \eta$. With this notation, we have the
identity $\left[X^R, Y^R \right](e) = [X, Y]$, for any
$X, Y \in \mathfrak{X}(M)$. This is the analogue of saying
that $\mathfrak{X}(M)$ is the ``right Lie algebra'' of $G^s$.

Assume, in addition, that $M$ is connected, oriented, and
Riemannian; denote by $\left\langle \cdot , \cdot \right\rangle$
the Riemannian metric. Let $\mu_g$ be the Riemannian volume form
on $M$, whose expression in local coordinates $(x^1, \ldots, x^n)$
is $\mu_g = \sqrt{\det(g_{ij})} dx^1 \wedge \cdots \wedge dx^n$,
where $g_{ij}: = \left\langle \partial/\partial x^i,
\partial/ \partial x^j \right\rangle$ for all $i,j=1, \ldots n$.
Let $K: T(TM) \rightarrow TM$ be the connector of the Levi-Civita
connection (with Christoffel symbols $\Gamma^i_{jk}$) defined by
the Riemannian metric on $M$; in local coordinates, this intrinsic
object, which is a vector bundle map $K:T(TM) \rightarrow TM$
covering the canonical vector bundle projection
$\tau_M: TM\rightarrow M$, has the expression
$K(x^i,u^i,v^i,w^i)=\left(x^i, w^i + \Gamma^i_{jk} u^jv^k\right)$.
The connector $K$ satisfies hence the identity
$\tau_M \circ K = \tau_M \circ \tau_{TM}$ and has the property that
the vector bundles $\tau_{TM}: T(TM) \rightarrow TM$ and
$\ker K\oplus \ker T \tau_M \rightarrow TM$
are isomorphic. These two properties characterize the connector.
Conversely, the connector $K$ determines $\nabla$ by the formula:
$\nabla_XY := K \circ TY \circ X$ for any $X,Y \in \mathfrak{X}(M)$.

The Riemannian structure on $M$ induces the \textit{weak} $L^2$, or
\textit{hydrodynamic}, \textit{metric}
$\left\langle \! \left\langle\cdot , \cdot
\right\rangle \! \right\rangle_\eta$
on $G^s$ given by
\[
\left\langle \! \left\langle U_\eta, V_\eta
\right\rangle \! \right\rangle_\eta : =
\int_{M} \left\langle U_\eta(m), V_\eta(m)
\right\rangle_{\eta(m)} d\mu_g(m),
\]
for any $\eta \in G^s$, $U_\eta, V_\eta \in T_\eta G^s$. This means
that the association $\eta \mapsto
\left\langle \! \left\langle\cdot , \cdot
\right\rangle \! \right\rangle_\eta$
from $G^s$ to the vector bundle of symmetric covariant two-tensors
on $G^s$ is smooth but that for every $\eta \in G^s$, the map
$T_\eta G^s \ni U_\eta \mapsto \left\langle \! \left\langle U_\eta,
\cdot \right\rangle \! \right\rangle \in T^\ast_\eta G^s$, where
$T^\ast_\eta G^s$ denotes the linear continuous functionals on
$TG^s$, is only injective and not, in general, surjective. This weak
metric is not right invariant (because of the Jacobian appearing
in the change of variables formula in the integral).

The usual proof for finite dimensional Lie groups showing the existence
of a unique Levi-Civita connection associated to a Riemannian metric
breaks down, because
$\left\langle \! \left\langle\cdot , \cdot \right\rangle \! \right\rangle$
is weak; the proof would only show uniqueness.
However,  $K^0:T(TG^s) \rightarrow TG^s$ given by
$K^0(\mathcal{Z}_{U_\eta}) := K \circ \mathcal{Z}_{U_\eta}$, where
$\mathcal{Z}_{U_\eta} \in T_{U_\eta}(TG^s)$, is a connector for
the vector bundle $\tau_{G^s}:TG^s \rightarrow G^s$ (since
$\tau_{G^s} \circ K^0 = \tau_{G^s} \circ \tau_{TG^s}$ and the
vector bundles $\tau_{TG^s}: T(TG^s) \rightarrow TG^s$ and
$\ker K^0\oplus \ker T\tau_{G^s} \rightarrow TG^s$ are isomorphic).
Here, $\mathcal{Z}_{U_\eta} \in T_{U_\eta}(TG^s)$ means that
$\mathcal{Z}_{U_\eta}: M \rightarrow T(TM)$ satisfies
$\tau_{TM} \left(\mathcal{Z}_{U_\eta}(m)\right) \in T_{\eta(m)}M$.
The covariant derivative $\nabla^0$ on $G^s$ is defined by
$\nabla^0_\mathcal{X} \mathcal{Y}: = K^0 \circ T \mathcal{Y}
\circ\mathcal{X}$, for any $\mathcal{X}, \mathcal{Y}\in
\mathfrak{X}(G^s)$. This is the Levi-Civita
connection associated to the weak metric
$\left\langle\!\left\langle\cdot, \cdot\right\rangle\!\right\rangle$
since it is torsion free ($\nabla^0_\mathcal{X} \mathcal{Y} -
\nabla^0_\mathcal{Y} \mathcal{X} = [\mathcal{X}, \mathcal{Y}]$, for
all $\mathcal{X},\mathcal{Y} \in \mathfrak{X}^s (G^s)$, where
$[\mathcal{X}, \mathcal{Y}]$ is the Jacobi-Lie bracket of vector fields
on $G^s$) and $\left\langle \! \left\langle\cdot , \cdot
\right\rangle \! \right\rangle$-compatible
($\mathcal{Z}[\left\langle \! \left\langle \mathcal{X},
\mathcal{Y}\right\rangle \! \right\rangle] =
\left\langle \! \left\langle \nabla_\mathcal{Z} \mathcal{X},
\mathcal{Y} \right\rangle \! \right\rangle
+ \left\langle \! \left\langle \mathcal{X},
\nabla_\mathcal{Z} \mathcal{Y}\right\rangle\!\right\rangle$,
for all $\mathcal{X},\mathcal{Y},\mathcal{Z}\in \mathfrak{X}^s (G^s)$);
see \cite[Theorem 9.1]{EM}. Uniqueness of such a
connection follows from the weak non-degeneracy of the metric
$\left\langle\!\left\langle\cdot,\cdot\right\rangle\!\right\rangle$.
There is an explicit formula for right-invariant covariant
derivatives on diffeomorphism groups (see, e.g., \cite[page 6]{GB}).
For $\nabla^0$ this formula is
\begin{equation}\label{e5-0}
\left(\nabla^0_{\mathcal{X}} \mathcal{Y} \right)(\eta) :=
\frac{\partial}{\partial t}\left(\mathcal{Y}(\eta_t)
\circ \eta_t^{-1}\right) \circ \eta +
(\nabla_{X^ \eta} Y^ \eta) \circ \eta,
\end{equation}
where  $\nabla$ denotes the Levi-Civita connection on $M$,
$\mathcal{X}, \mathcal{Y}\in \mathfrak{X}(G^s)$,
$X^\eta : = \mathcal{X} \circ \eta^{-1}, Y^\eta : =
\mathcal{Y} \circ \eta^{-1} \in \mathfrak{X}^s(M)$, and
$t \mapsto \eta_t$ is a $C^1$ curve in $G^s$ such that
$\eta_0 = \eta$ and $\left.\frac{d}{dt}\right|_{t=0}\eta_t =
\mathcal{X}(\eta)$; this formula is identical to
\cite[(3.1)]{Misiolek1993}. Note that each term on the
right hand side of this formula is only of class $H^{s-1}$ and,
nevertheless, their sum is of class $H^s$ because of the abstract
definition of the covariant derivative on $G^s$. A similar
phenomenon occurs with the geodesic spray
$TG^s \rightarrow T(TG^s)$ of the weak Riemannian
metric; its existence and smoothness for boundaryless $M$ was proved
in \cite{EM}, i.e., the geodesic spray is in
$\mathfrak{X}(TG^s)$. If $M$ has a boundary, this statement is false.

The discussion above shows that one cannot apply the theorems
of Section \ref{sec_3} to the infinite dimensional group $G^s$ directly.
For infinite dimensional problems,
they serve only as a guideline and direct proofs are needed, which
is what we do below. However, for each important
formula, we shall point out the analogue in the finite dimensional
abstract setting of Section \ref{sec_3} which inspired the result, that
still holds for the model presented here.

\subsection{Stochastic semidirect product Euler-Poincar\'e reduction
for $G^s$.}
\label{sec_5_0}
We formulate now the theory presented in Section \ref{sec_3} for the
infinite dimensional group $G^s$.  From now on we consider the case
$M=\mathbb T^3$. We focus on the following type of SDEs on $G^s$,
\begin{equation*}
\begin{cases}
& dg_\omega(t,\theta)=\sum_{i=1}^m H_i(g_\omega(t,\t))
\delta M^i_\omega(t)
+ u_\omega(t,g_\omega(t,\t))dt\\
& g_\omega(0,\theta)=\t,\ \ \t \in \T^3,
\end{cases}
\end{equation*}
where $H_j \in \mathfrak{X}^s(\T^3)$ is non-random,
 $\{M_\omega^i(t)\}_{i=1}^m$ is a
$\R^m$-valued martingale with continuous sample paths on a probability
space $(\Omega,\p,\P)$ with respect to the filtration $\p_t$,
$u:\Omega\times [0,T]\rightarrow \mathfrak{X}^s(\T^3)$ is such that
$\u(t,x)$ is a ($\p_t$-adapted) semimartingale for every $x \in \T^3$.

In particular, here we take the constant vector fields
$H_1=H_{1,\nu}=\sqrt{2\nu}(1,0,0)$,  $H_2=H_{2,\nu}=\sqrt{2\nu}(0,1,0)$,
$H_3=H_{3,\nu}=\sqrt{2\nu}(0,0,1)$ on $\mathbb{T}^3$, where $\nu\ge 0$
is a  (viscosity) constant. This is understood in the trivialization
$T \mathbb{T}^3 = \mathbb{T}^3 \times \mathbb{R}^3$, so
$H_1, H_2, H_3: \mathbb{T}^3 \rightarrow \mathbb{R}^3$ are
constant maps. Let $g_\omega^{\nu,M}(t,\t)$ be the solution to the
following SDE,
\begin{equation}
\label{e4-1}
\begin{cases}
& dg_\omega^{\nu,M}(t,\t)=\sum_{i=1}^3 H_{i,\nu} dM_\omega^i(t)
+\u(t,\g^{\nu,M}(t,\t))dt\\
&\quad \quad \quad \quad\;\;\; =\sqrt{2\nu}dM_\omega (t)+
\u(t,g_\omega^{\nu,M}(t,\t))dt\\
& g_\omega^{\nu,M}(0,\t)=\t,
\end{cases}
\end{equation}
where $M_\omega=\big(M_\omega^1, M_\omega^2, M_\omega^3\big)$
is an $\R^3$-valued martingale with continuous sample paths.

By the standard theory of  stochastic flows (see, e.g., \cite{KU} and
standard embedding theorems),
if $\u$ is regular enough (with respect to the space variable), i.e.,
$\u \in C([0,T];\x^{s'}(\T^3)) $ for some $s'>s$ large enough, then
$g_\omega^{\nu,M}(t,\cdot) \in G^s$ for every $t \in [0,T]$.
From now on, for simplicity, we always assume
$\u$ to be regular enough.

As in \cite[Section 6]{HMR}, let $U^{*}$
be some linear space which can be a space  of functions, densities,
or differential forms on $\T^3$. The action of $G^s$ on $U^{*}$
is the pull back map and the action of the ``Lie algebra'' $T_eG^s$ on
$U^*$ is the Lie derivative.

If we take $\alpha_0=A_0(\t)\cdot\mathbf{d}\theta:=
\sum_{i=1}^3 A_{0,i}(\t)\mathbf{d}\theta_i$
to be a $C^2$ one-form on $\T^3$, we derive the following
result (see also an equivalent expression in
\cite{Ey}, equations (32)-(34) for the deterministic case).
Formula \eqref{p4-1-2} below is the analogue of the second equation
in \eqref{t3-4-1-right}, derived here
by hand for the infinite dimensional group $G^s$.

\begin{prp}\label{p4-1}
Let $g_{\omega}^{\nu}(t)$ be given by \eqref{e4-1} with
$M_\omega=W_\omega$, where $W_\omega$ is a standard
$\R^3$-valued Brownian motion {\rm(}i.e.,
$W_\omega=(W_\omega^1,W_\omega^2,W_\omega^3)$ with $W_\omega^i$,
$1\le i \le 3$, independent $\R$-valued Brownian motions{\rm)}. Define
\begin{equation*}
\alpha_{\omega}(t,\t):=
\left(\alpha_0g_{\omega}^{\nu}(t,\cdot )^{-1}\right)(\theta)=
\left(\left(g_{\omega}^{\nu}(t,\cdot )^{-1}\right)^{*}\alpha_0
\right)(\theta):=
A_{\omega}(t,\t)\cdot \mathbf{d}\theta:=
\sum_{i=1}^3 A_{\omega,i}(t,\t)\mathbf{d}\theta_i,
\end{equation*}
where  $\left(g_{\omega}^{\nu}(t,\cdot )^{-1}\right)^{*}$ denotes the
pull back map by $g_{\omega}^{\nu}(t,\cdot )^{-1}$, and
\begin{equation*}
\tilde \alpha (t,\t):=\E[ \alpha_{\omega}(t,\t)]:=
\A(t,\t)\cdot \mathbf{d}\theta:=
\sum_{i=1}^3\A_i(t,\t )\mathbf{d}\theta_i.
\end{equation*}
Then $A_{\omega}$ satisfies the following SPDE,
\begin{equation}
\label{p4-1-1}
\begin{split}
dA_{\omega,i}(t,\t)=&
-\sum_{j=1}^3\sqrt{2\nu}\partial_j A_{\omega,i}(t,\t,\omega)
dW_{\omega}^j(t)  \\
& -\sum_{j=1}^3\Big(u_{\omega,j}(t,\t) \partial_j A_{\omega,i}(t,\t)
+ A_{\omega,j}(t,\t)\partial_i u_{\omega,j}(t,\t)\Big)dt \\
& +\nu\Delta A_{\omega,i}(t,\t)dt,\ i=1,2,3,
\end{split}
\end{equation}
where we use the notation $u_\omega(t):=(u_{\omega,1}(t),u_{\omega,2}(t),
u_{\omega,3}(t))$ and $\partial_j$ and $\Delta$ stand for the partial
derivative and the Laplacian with respect to the space variable
$\theta$ of $A_\omega(t,\theta)$, respectively.
Equation \eqref{p4-1-1} can also be expressed as
\begin{equation}\label{p4-1-2}
\begin{split}
dA_{\omega}(t, \theta)=
&-\sqrt{2\nu}\nabla A_{\omega}(t,\t )\cdot dW_\omega(t) \\
&- \left(\u(t, \theta) \times \cu A_{\omega}(t, \theta)-
\nabla(\u(t,\theta)\cdot A_{\omega}(t,\theta)) \right)dt
+\nu\Delta A_{\omega}(t)dt
\end{split}
\end{equation}
{\rm(}the term $dA_\omega(t, \theta)$ above denotes the It\^o
differential of $A_\omega(t, \theta)$ with respect to the
time variable{\rm)}.

Moreover, if $\u$ is non-random {\rm(}in this case we write $u$ for
$\u${\rm)}, we have
\begin{equation}
\label{p4-1-3}
\begin{cases}
&\partial_t \A(t,\t)=
\big(u(t,\t) \times \cu \A(t,\t)-\nabla(u(t,\t)\cdot \A(t,\t)) \big)
+\nu\Delta \A(t,\t),\\
& \A(0,\t)=A_0(\t).
\end{cases}
\end{equation}
\end{prp}

\begin{proof}
We use the methods in \cite[Lemma 4.1]{CI} and
\cite[Proposition 4.2]{CI}. It is not hard to see that, for
the $C^2$ (note that we assume $u$ to be regular)
spatial process $A_{\omega,i}(t,\t)$, there exist adapted
spatial processes $h_{\omega,ij}(t,\t)$ and
$z_{\omega,i}(t,\t)$, $1\leqslant i,j \leqslant 3$, such that,
\begin{equation}\label{p4-1-4}
dA_{\omega,i}(t,\t)=\sum_{j=1}^3
h_{\omega ,ij}(t,\t  )dW_{\omega}^j (t)+z_{\omega,i}(t,\t)dt, \quad
i=1,2,3.
\end{equation}
We compute below the expressions of $h_{\omega,ij}(t,\t)$ and
$z_{\omega,i}(t,\t)$.

Notice that by the definition of
$\alpha_{\omega} (t,\theta)$,
$\left(g_{\omega}^{\nu}(t,\theta)\right)^{*}
\alpha_{\omega} (t, \theta)=
\alpha (0,\theta)$ is a constant with respect to the time variable, and
\begin{equation*}
\left(g_{\omega}^{\nu}(t, \theta)\right)^{*}
\alpha_{\omega} (t,\theta)
=\sum_{j=1}^3\left(\sum_{i=1}^3 A_{\omega,i}(t,g_{\omega}^{\nu}(t,\t),
\omega)V_{\omega,ij}(t,\t)\right) d\t_j,
\end{equation*}
where the process $V_{\omega,ij}(t,\t):=
\partial_j g^{\nu}_{\omega,i}(t,\t)$ (here we use the notation
$g^{\nu}_\omega(t)=\big(g^{\nu}_{\omega,1}(t),$ $g^{\nu}_{\omega,2}(t),
g^{\nu}_{\omega,3}(t)\big)$). We get for each $1\le j\le 3$,
\begin{equation}
\label{p4-1-5}
d\Big(\sum_{i=1}^3 A_{\omega,i}(t,g_{\omega}^{\nu}(t,\theta))
V_{\omega,ij}(t, \theta)\Big)=0.
\end{equation}
By \eqref{p4-1-4} and the generalized It\^o formula for spatial
processes (see \cite[Theorem 3.3.1.]{KU}), we have
\begin{equation}\label{p4-1-5a}
\begin{split}
&dA_{\omega,i}(t,g_{\omega}^{\nu}(t, \theta))=
\sum_{j=1}^3\left(h_{\omega,ij}(t,g_{\omega}^{\nu}(t,\theta))
+\sqrt{2\nu}\partial_j A_{\omega,i}(t,g_{\omega}^{\nu}(t,\theta))
\right)dW_{\omega}^j (t)\\
&\quad +\left(z_{\omega,i}(t,g_{\omega}^{\nu}(t,\theta))
+\nu \Delta A_{\omega,i}(t,g_{\omega}^{\nu}(t,\theta))
\phantom{\sum_{j=1}^3}\right. \\
&\qquad\quad  \left.+\sum_{j=1}^3
\left(u_{\omega,j}\left(t,g_{\omega}^{\nu}(t,\theta)\right)
\partial_j A_{\omega,i}(t,g_{\omega}^{\nu}(t, \theta))
+\sqrt{2\nu}\partial_j h_{\omega,ij}\left(t,g_{\omega}^{\nu}(t, \theta )
\right)\right)\right)dt.
\end{split}
\end{equation}
By the theory of  the stochastic flows in \cite[Theorem 3.3.3.]{KU}, from
\eqref{e4-1} we obtain
\begin{equation}\label{p4-1-6a}
dV_{\omega,ij}(t,\theta)=\
\sum_{k=1}^3 \partial_k u_{\omega,i}(t,g_{\omega}^{\nu}(t, \theta))
V_{\omega,kj}(t, \theta)dt.
\end{equation}
In particular, the martingale part of the above equality vanishes due
to the fact that the diffusion coefficients in (\ref{e4-1}) are constant.

According to \eqref{p4-1-5a} and \eqref{p4-1-6a}, for each $1\le j \le 3$,
the It\^o differential  with respect to the time variable is as follows
\begin{equation*}
\begin{split}
&d\left(\sum_{i=1}^3 A_{\omega,i}(t,g_{\omega}^{\nu}(t,\theta))
V_{\omega,ij}(t,\theta)\right)\\
& \qquad =
\sum_{i,k=1}^3\left(h_{\omega,ik}(t,g_{\omega}^{\nu}(t, \theta)) +
\sqrt{2\nu} \partial_k A_{\omega,i}(t,g_{\omega}^{\nu}(t, \theta))
\right)V_{\omega,ij}(t,\theta)dW_{\omega}^k (t)\\
&\qquad \quad +\sum_{i=1}^3\left(\sum_{k=1}^3 \left(
u_{\omega,k}(t,g_{\omega}^{\nu}(t, \theta))
\partial_k A_{\omega,i}(t,g_{\omega}^{\nu}(t,\theta) )
+\sqrt{2\nu}\partial_k h_{\omega,ik}(t,g_{\omega}^{\nu}(t, \theta))
\right) \right.\\
&\qquad\qquad\qquad  \left.\phantom{\sum_{i=1}^3}
+z_{\omega,i}(t,g_{\omega}^{\nu}(t, \theta))+
\nu \Delta A_{\omega,i}(t,g_{\omega}^{\nu}(t,\theta))\right)
V_{\omega,ij}(t,\theta)dt\\
& \qquad \quad
+\sum_{i,k=1}^3 A_{\omega,k}(t, g_{\omega}^{\nu}(t, \theta))
\partial_i u_{\omega,k}(t,g_{\omega}^{\nu}(t, \theta))
V_{\omega,ij}(t,\theta)dt
\end{split}
\end{equation*}

Hence from (\ref{p4-1-5}), we derive for each $1\le j,m \le 3$,
\begin{equation}\label{p4-1-6}
\sum_{i=1}^3\left(h_{\omega,im}(t,g_{\omega}^{\nu}(t, \theta))+
\sqrt{2\nu}\partial_m A_{\omega,i}(t,g_{\omega}^{\nu}(t, \theta))
\right)V_{\omega,ij}(t,\theta)=0
\end{equation}
and
\begin{equation}\label{p4-1-7}
\begin{split}
&\sum_{i=1}^3 \left(z_{\omega,i}(t,g_{\omega}^{\nu}(t, \theta))+
\nu \Delta A_{\omega,i}(t,g_{\omega}^{\nu}(t,\theta))
\phantom{\sum_{i=1}^3} \right.\\
& \qquad \left. +
\sum_{k=1}^3 \Big(u_{\omega,k}(t,g_{\omega}^{\nu}(t,\theta))
\partial_k A_{\omega,i}(t,g_{\omega}^{\nu}(t,\theta))
+\sqrt{2\nu}\partial_k
h_{\omega,ik}(t,g_{\omega}^{\nu}(t,\t))
 \right.\\
&\qquad  \qquad \left. \phantom{\sum_{i=1}^3}
+A_{\omega,k}(t, g_{\omega}^{\nu}(t, \theta))
\partial_i u_{\omega,k}(t,g_{\omega}^{\nu}(t,\theta))\Big)\right)
V_{\omega,ij}(t, \theta)=0.
\end{split}
\end{equation}
As $\{V_{ij}(t,\theta, \omega )\}_{1\leqslant i,j \leqslant 3}$ is a
non-degenerate matrix-valued process  (see, e.g., \cite{KU}), from
\eqref{p4-1-6} we deduce that, for each $1\le i,j\le 3$,
\begin{equation*}
h_{\omega,ij}(t,g_{\omega}^{\nu}(t,\theta))
=-\sqrt{2\nu}\partial_j A_{\omega,i}(t,g_{\omega}^{\nu}(t, \theta)).
\end{equation*}

Since $g_{\omega}^{\nu}(t, \theta)$ is invertible in
$\theta$ we can derive the expression for $h_{\omega,ij}$ not only
as a function of $g_{\omega}^{\nu}(t, \theta)$ but also as a function
(at the origin) of $(t, \theta)$. Indeed, noticing that,
$\omega$-almost surely, $\theta \mapsto g_\omega^{\nu}(t, \theta)$
is a diffeomorphism for each fixed $t$, we get
\begin{equation}\label{p4-1-8}
h_{\omega,ij}(t,\t)=-\sqrt{2\nu}\partial_j A_{\omega,i}(t,\t),\quad
\forall \theta \in \T^3,
\end{equation}
which is the expression for $h_{\omega,ij}(t,\t)$.

Since $\{V_{\omega,ij}(t, \theta)\}_{1\leqslant i,j \leqslant 3}$ is
non-degenerate, by \eqref{p4-1-7}, for each $1\le i \le 3$,
\begin{equation*}
\begin{split}
z_{\omega,i}(t,g_{\omega}^{\nu}(t,\theta))=
&-\nu \Delta A_{\omega,i}(t,g_{\omega}^{\nu}(t,\theta))
-\sum_{k=1}^3 \Big(
u_{\omega,k}(t,g_{\omega}^{\nu}(t,\theta))
\partial_k A_{\omega,i}(t,g_{\omega}^{\nu}(t,\theta))) \\
& -\sqrt{2\nu}\partial_k
h_{\omega,ik}(t,g_{\omega}^{\nu}(t, \theta))+
A_{\omega,k}(t, g_{\omega}^{\nu}(t,\theta) )
\partial_i u_{\omega,k}(t,g_{\omega}^{\nu}(t,\theta))\Big).
\end{split}
\end{equation*}
We use \eqref{p4-1-8} in the above equation and the fact that
$\theta\mapsto g_{\omega}(t, \theta)$ is a
diffeomorphism for each fixed $t$, $\omega$-almost surely and we
obtain the expression for $z_{\omega,i}(t,\t)$, namely,
\begin{equation}\label{p4-1-9}
\begin{split}
z_{\omega,i}(t,\t)=&\nu \Delta A_{\omega,i}(t,\t) \\
&-\sum_{k=1}^3\left(u_{\omega,k}(t,\t) \partial_k A_{\omega,i}(t,\t) +
A_{\omega,k}(t,\t) \partial_i u_{\omega,k}(t,\t)\right),\quad
\forall \t \in \T^3.
\end{split}
\end{equation}
Combining \eqref{p4-1-4}, \eqref{p4-1-8}, and \eqref{p4-1-9}
proves \eqref{p4-1-1}. We can check that \eqref{p4-1-1} is
equivalent to \eqref{p4-1-2} by direct computation.

If $\u(t,\cdot)=u(t,\cdot)$ is non-random, then it is easy to verify that
\begin{equation*}
\begin{split}
\mathbb{E}\left[u(t, \theta) \times \cu A_{\omega}(t, \theta)\right]&=
u(t,\t)\times  \E\left[\cu A_{\omega}(t, \theta)\right]\\
&=u(t,\t)\times \cu \E\left[A_{\omega}(t, \theta)\right]
=u(t,\t)\times \cu \A_\omega(t,\t).\\
\E\left[\nabla\left(\u(t,\theta)\cdot A_{\omega}(t,\theta)\right)\right]&=
\nabla \left(u(t,\t)\cdot \E\left[A_\omega(t,\t)\right]\right)\\
&=\nabla \left(u(t,\t)\cdot \A(t,\t)\right).
\end{split}
\end{equation*}
So taking the expectation of the two sides of equation \eqref{p4-1-2},
\eqref{p4-1-3} follows.
\end{proof}

\begin{remark}\label{r4-0}
{\rm
In Proposition \ref{p4-1}, $U^*$ is taken to be a space of differential
forms on $\T^3$. Note that the action of $G^s$ on $U^{*}$
is the pull back map and the action of the ``Lie algebra'' $T_eG^s$ on
$U^*$ is the Lie derivative. Then, for $H_{1,\nu}=\sqrt{2\nu}(1,0,0)$,
$H_{2,\nu}=\sqrt{2\nu}(0,1,0)$, $H_{3,\nu}=\sqrt{2\nu}(0,0,1)$, and
$\alpha_\omega (t,\t)=A_\omega(t,\t)\cdot \mathbf{d}\theta$, we have
\begin{equation*}
\begin{split}
&\sum_{i=1}^3\a(t)H_{i,\nu} dW_\omega^i(t)=\sqrt{2\nu}
\left(\nabla A_\omega(t,\t)\cdot dW_\omega(t)\right)\cdot
\mathbf{d}\theta,\\
&\sum_{i=1}^3 \alpha_\omega(t)H_{i,\nu}H_{i,\nu}=
\nu\Delta A_\omega(t,\t)\cdot \mathbf{d}\theta,\\
&  \alpha_\omega(t)u_\omega(t)=
\big(u_\omega(t,\t) \times \cu A_\omega(t,\t)-
\nabla(u_\omega(t,\t)\cdot A_\omega(t,\t)) \big)\cdot \mathbf{d}\theta,
\end{split}
\end{equation*}
which implies that \eqref{p4-1-2} is just the second equation of
\eqref{t3-4-1-right}.

In the same way, we can verify that the second equation of \eqref{t3-4-1}
is \eqref{p4-1-3}.} \hfill $\lozenge$
\end{remark}

\begin{remark}
\label{r4-1}
{\rm
By the same procedure as in the proof of Proposition \ref{p4-1},
if  $\alpha_0$ is replaced by another term, such as
a function or a density, we can still prove the corresponding
evolution equation for $\alpha_\omega (t):=
\alpha_0 g_{\omega}^{\nu}(t, \cdot )^{-1}=
\left(g_{\omega}^{\nu}(t, \cdot )^{-1}\right)^*\alpha_0$.

For example, if $\alpha_0=\beta_0: \T^3 \rightarrow \R$ is a
$C^2$ function, then $\alpha_\omega (t, \theta)$ satisfies the following
stochastic transport equation,
\begin{equation*}
\begin{cases}
&d\alpha_\omega (t,\t)=-\sqrt{2\nu}\nabla \a(t,\t)\cdot dW_\omega(t)-
u_\omega(t,\t)\cdot \nabla \alpha_\omega (t,\t)dt
+\nu\Delta \alpha_\omega (t,\t)dt,\\
&\alpha_\omega (0,\t)=\beta_0(x).
\end{cases}
\end{equation*}
This equation has been studied in \cite{FGP} which illustrates that the
added stochastic force (noise) can turn an ill-posed ODE into a
well-posed one.

If $\alpha_0=D_0(\t)d^3 \t$ is a density (volume form), write
$\alpha_\omega (t,\theta)=D_\omega(t,\t)d^3 \t$. Then
$D_\omega(t, \theta)$ satisfies the following equation
\begin{equation}\label{r4-1-2}
\begin{cases}
&d D_\omega(t,\t)= -\sqrt{2\nu}\nabla D_\omega(t,\t)\cdot dW_\omega(t)-
{\rm div}(D_\omega u_\omega)(t,\t)dt+\nu\Delta D_\omega(t,\t)dt,\\
&D_\omega(0,\t)=D_0(\t).
\end{cases}
\end{equation}

Assume that $\u(\cdot)=u(\cdot)$ is non-random and
$\alpha_0=D_0(\t)d^3 \t$ is a probability measure, let
$\tilde \alpha(t):=\E\left[\a(t)\right]:=\tilde D(t,\t)d^3 \t$. Then
$\tilde D(t,\t)$ satisfies the following forward Kolmogorov equation (or
Fokker-Planck equation),
\begin{equation}\label{r4-1-1}
\begin{cases}
&d \tilde D(t,\t)= -
{\rm div}(\tilde D u )(t,\t)dt+\nu\Delta \tilde D(t,\t)dt,\\
&\tilde D(0,\t)=D_0(\t).
\end{cases}
\end{equation}
Moreover, let  $\hat g_{\omega}^{\nu}(t, \theta)$ be the process
satisfying
\begin{equation*}
d\hat g_{\omega}^{\nu}(t,\theta)=
\sqrt{2\nu}dW_{\omega} (t)+u(t,\hat g_{\omega}^{\nu}(t,\theta))dt
\end{equation*}
whose initial distribution is $D_0(\t)d^3 \t$. Then
for every $t \in [0,T]$, the distribution of
$\hat g_{\omega}^{\nu}(t, \theta )$ is of the form $\tilde D(t,\t)d^3 \t$,
where $\tilde D(t,\t)$ satisfies \eqref{r4-1-1}.} \hfill $\lozenge$
\end{remark}

\begin{remark}
{\rm
By carefully tracking the  proof of Proposition \ref{p4-1}, if we take
$M_\omega$ in \eqref{e4-1} to be a general $\R^3$-valued martingale,
equation \eqref{p4-1-2} becomes}
\begin{equation*}
\begin{split}
\qquad\quad \;\; dA_{\omega}(t, \theta)=
&-\sqrt{2\nu}\nabla A_{\omega}(t,\t )\cdot dM_\omega(t)-
\u(t, \theta) \times \cu A_{\omega}(t, \theta)dt\\
&+\nabla(\u(t,\theta)\cdot A_{\omega}(t,\theta))dt +
\nu\sum_{i,j=1}^3\partial_i \partial_j
A_{\omega}(t)d\llbracket M_\omega^{i}, M_\omega^{j}\rrbracket_t.
\qquad \qquad \lozenge
\end{split}
\end{equation*}
\end{remark}

For each $\p_t$-adapted process $v$ such that $v_\omega(\cdot,\cdot)
\in C^1([0,1];\x^{s}(\T^3))$ (for $s$ large enough) with
$v_\omega(0,\t)=v_\omega(T,\t)=0$ a.s., the deformation \eqref{e3-3-right}
(for right invariant systems) in the
formulation here is determined by the following stochastic
flows $e_{\omega,\ee,v}(t,\cdot) \in G^s$ (see e.g., \cite{ACC}, \cite{CC}
for the deterministic counterpart)
\begin{equation}\label{e4-4a}
\begin{cases}
&\frac{d e_{\omega,\ee,v}(t,\t)}{dt}=
\ee \dot{v}_\omega(t, e_{\omega,\ee,v}(t,\t))\\
& e_{\omega,\ee,v}(0,\t)=\t
\end{cases}
\end{equation}
Setting $g_{\omega,\ee,v}^{\nu,M}(t,\t):=
e_{\omega,\ee,v}(t,g_\omega^{\nu,M}(t,\t))$,  where $g_\omega^{\nu,M}$
is the solution to \eqref{e4-1}, and using such deformations,
we can also define the critical point for an action functional in the
same way as in \eqref{e3-4}, Section \ref{sec_3}.

By the analysis in \cite[Section 4.2]{ACC} (especially (4.5)-(4.6)
in \cite{ACC}) for the (infinite dimensional group) $G^s$, we have
\begin{equation*}
\begin{split}
&dg_{\omega,\ee,v}^{\nu,M}(t)=T_e R_{g_{\omega,\ee,v}^{\nu,M}(t)}
\left(\sum_{i=1}^3 H_{\omega,i,\nu,\ee}(t)\delta M_\omega^i(t)
+{\rm Ad}_{e_{\omega,\ee,v}(t)}\u(t)dt+\ee\dot{v}_\omega(t)dt\right),
\end{split}
\end{equation*}
where $H_{\omega,i,\nu,\ee}(t)={\rm Ad}_{e_{\omega,\ee,v}(t)}H_{i,\nu}$.
Based on the above equation for $g_{\omega,\ee,v}^{\nu,M}$ and
according to the definition of $\frac{\D}{dt}$ and
$\frac{\mathbf{D}^{\nabla_0,(H_{i,\nu,\ee}^{\nu},M_\omega^i)_{i=1}^3}}
{dt}$ (especially using  the right-invariant version of \eqref{e2-4} and
\eqref{e3-6}), it is easy to verify that
\begin{equation}\label{e4-4}
\begin{split}
&T_{g_{\omega}^{\nu,M}(t, \theta)}R_{g_{\omega}^{\nu,M}(t,\theta)^{-1}}
\frac{\D g_{\omega}^{\nu,M}(t, \theta)}{dt}
=\u(t,\theta),\\
&\frac{d}{d\ee}\Bigg|_{\ee=0}\left(T_{g_{\omega,\ee,v}^{\nu,M}(t, \theta)}
R_{g_{\omega,\ee,v}^{\nu,M}(t,\theta)^{-1}}
\frac{\D g_{\omega,\ee,v}^{\nu,M}(t, \theta)}{dt}\right)\\
&\quad \quad =\left({\rm ad}_{v_\omega(t)}\u(t)\right)(\t)=
-[v_\omega(t,\cdot),\u(t,\cdot)](\t),
\end{split}
\end{equation}
\begin{equation}\label{e4-4b}
\begin{split}
&T_{g_{\omega}^{\nu,M}(t, \theta)}R_{g_{\omega}^{\nu,M}(t,\theta)^{-1}}
\left(d^{\Delta}g_{\omega}^{\nu,M}(t, \theta) \right)
=\sqrt{2\nu}dM_\omega(t),\\
&\frac{d}{d\ee}\Bigg|_{\ee=0}\left(T_{g_{\omega,\ee,v}^{\nu,M}(t, \theta)}
R_{g_{\omega,\ee,v}^{\nu,M}(t,\theta)^{-1}}
d^{\Delta}g_{\omega,\ee,v}^{\nu,M}(t, \theta)\right)\\
&\quad \quad =\sum_{i=1}^3 \left({\rm ad}_{v_\omega(t)}H_{i,\nu}\right)(\t)
dM_\omega^i(t)=\sum_{i=1}^3
\sqrt{2\nu}\partial_i v_{\omega}(t,\t)dM_\omega^i(t),
\end{split}
\end{equation}
\begin{equation}\label{e4-4c}
\begin{split}
&T_{g_{\omega}^{\nu,M}(t, \theta)}R_{g_{\omega}^{\nu,M}(t,\theta)^{-1}}
\left(\frac{\mathbf{D}^{\nabla^0,(H_{i,\nu},M_\omega^i)_{i=1}^3}}{dt}
\right)_{i,j}=\nabla^0_{H_{i,\nu}} H_{j,\nu}\frac{d\llbracket M_\omega^i,
M_\omega^j\rrbracket_t}{dt}=0,\\
& \frac{d}{d\ee}\Bigg|_{\ee=0}T_{g_{\omega,\ee,v}^{\nu,M}(t, \theta)}
R_{g_{\omega,\ee,v}^{\nu,M}(t,\theta)^{-1}}
\left(\frac{\mathbf{D}^{\nabla^0,(H_{i,\nu,\ee}^{\nu},M_\omega^i)_{i=1}^3}}
{dt}\right)_{i,j}\\
&\quad \quad =\big(\nabla^0_{{\rm ad}_{v_\omega(t)}H_{i,\nu}}H_{j,\nu}
+\nabla_{H_{i,\nu}}^0({\rm ad}_{v_\omega(t)}H_{j,\nu})\big)
\frac{d\llbracket M_\omega^i, M_\omega^j\rrbracket_t}{dt}\\
&\quad \quad =
2\nu\partial_i\partial_j v_\omega(t,\t)\frac{d\llbracket M_\omega^i,
M_\omega^j\rrbracket_t}{dt},\\
\end{split}
\end{equation}
 where $\nabla^0$ denotes the connection on $\mathfrak{X}(G^s)$
 defined by \eqref{e5-0}
(in particular, we apply the property that $\nabla^0_X Y(\t)=
\sum_{i,j=1}^3 X_i(\t)\partial_i Y_j(\t)\partial_j$
for every $X=\sum_{i=1}^3 X_i(\t)\partial_i$,
$Y=\sum_{i=1}^3 Y_i(\t)\partial_i$ $\in \mathfrak{X}(\T^3)$ because the Christoffel symbols are zero, $\T^3$ being the flat torus).

Based on these formulas,  the procedure on the variational principle
in the proof of Theorem \ref{t3_2} and \ref{t3_3}  also holds for
the infinite dimensional group $G^s$ needed here. Hence the first equation
of \eqref{t3-4-1-right} (also the first equation of \eqref{t3-4-1})
remains  true for $G^s$.

\medskip

Combining all the conclusions above, we deduce that
\textit{Theorem \ref{t3_2} and \ref{t3_3} still hold for
the infinite dimensional  group $G^s$.}

\subsection{Compressible Navier-Stokes equation}\label{sec5-1}

Suppose $\nabla^0$ is the connection on $\mathfrak{X}(G^s)$ defined by
\eqref{e5-0}. Let $U^*$ denote the vector space of all probability
densities on $\T^3$  and define $\alpha_0:=D_0(\t) d^3 \t \in U^*$.
Let $\scr{M}_m(G^s)$ be the collection of all $m\times m$
$\x^s(\T^3)$-valued matrices and define $\scr{M}(G^s):=
\cup_{m=1}^{\infty}\scr{M}_m(G^s)$.

As in \cite{HMR}, we take the dual space $(\x^s(\T^3))^*$ of $\x^s(\T^3)$
to be the vector space $\Omega^1(\T^3)$ of all differential one-forms on
$\T^3$ (here we fix the volume measure  to be the
Lebesgue measure on $\T^3$).

We define the Lagrangian $l:\Omega\times [0,T]\times \x^s(\T^3) \times U^*
\rightarrow \R$ by
\begin{equation*}
\begin{split}
&l_\omega\left(t,u,\alpha\right):=
\int_{\T^3}\left( \frac{D(\t)}{2}|u(\t)|^2
-D(\t)e_\omega(t,D(\t))\right)d^3 \t,\\
&\qquad\qquad
\forall \ u\in \x^s(\T^3), \quad \forall\alpha=D(\t)d^3\t \in U^*,
\end{split}
\end{equation*}
where $e_\omega(t,D)$ is the fluid's specific
internal energy, and the $\p_t$-adapted pressure $p_\omega(t)$ is given by
$\mathbf{d}e_\omega(t)=-p_\omega(t) \mathbf{d}\big(\frac{1}{D}\big)$
($\mathbf{d}$ denotes the space differential). See
\cite[Section 7]{HMR} for more details on such Lagrangians. Here we use
a random version, since the pressure may depend on the randomness of the
system.

Then $\frac{\delta l}{\delta u}\left(t,u,\alpha\right)
=uD \in\Omega_1(\T^3)$
is non-random, independent of $t$, and
\begin{equation*}
\left\langle\frac{\delta l}{\delta u}\left(t,u,\alpha\right), 
v\right\rangle
=\int_{\T^3}\langle u(\t),v(\t)\rangle D(\t)d^3 \t,\ \ \forall
u,v \in \x^s(\T^3),\ \alpha=D(\t)d^3\t \in U^*.
\end{equation*}
Define the contraction force $\tilde p:\scr{M}(G^s)\times \scr{M}(G^s)
\times \x^s(\T^3)\rightarrow \R$ by
\begin{equation}\label{e5-3}
\begin{split}
&\tilde p(A,B,u):=\frac{1}{2}\int_{\T^3}u(\t)\cdot{\bf Tr}(A)(\t)d^3\t
+\frac{1}{2}\sum_{i,j=1}^m \int_{\T^3}\mathbf{P}_i(u(\t))
\mathbf{P}_j\big((B)_{i,j}(\t)\big)d^3\t,\\
&\qquad \qquad\forall A\in \scr{M}_n(G^s),\ B\in \scr{M}_m(G^s),
\quad \forall u\in \x^s(\T^3),
\end{split}
\end{equation}
where ${\bf Tr}: \scr{M}(G^s)\rightarrow \x^s(\T^3)$ is the trace
operator and $\mathbf{P}_i:\R^3 \rightarrow \R$ is the projection
operator defined by
\begin{equation*}
\mathbf{P}_i(x_1,x_2,x_3):=
\begin{cases}
& x_i,\ \ \text{if}\ 1\le i \le 3,\\
& 0,\ \ \ \text{if}\ i>3.
\end{cases}
\end{equation*}
We take the stochastic force $q:[0,T]\times\x^s(\T^3)\times U^*
\rightarrow (\x^s(\T^3))^*$ to be
$q(t,u,\alpha):=\langle u,\cdot \rangle$.

With $\nabla^0$, $l$, $\tilde p$, $q$, $\alpha_0$
given above, we
define an action functional $\mathbf{J}^{\nu}:=J^{\nabla^0,
(H_{i,\nu},W_\omega^i)_{i=1}^3}$ according to \eqref{right_action}
as follows
\begin{equation}\label{e5-1a}
\begin{split}
&\mathbf{J}^{\nu}\left(\left(g_\omega^{1},\w_i^1,M_\omega^{1,i}
\right)_{i=1}^{m_1},\left(g_\omega^{2},\w_i^1,M_\omega^{2,i}
\right)_{i=1}^{m_1},g_\omega^{3}\right)\\
&\quad :=
\int_0^T\int_{\T^3}\left(\frac{1}{2}\left|w_\omega(t,\t)\right|^2
D_{\omega}(t,\t)-
D_\omega(t,\t)e\left(D_\omega(t,\t)\right)\right) d^3 \t dt\\
&\qquad\quad  +\int_0^T \tilde p\left(\frac{\mathbf{D}^{\nabla^0,
(\w_i^1,M_\omega^{1,i})_{i=1}^{m_1}}\g^1(t)}{dt},
\frac{\mathbf{D}^{\nabla^0,(\w_i^2,M_\omega^{2,i})_{i=1}^{m_2}}\g^2(t)}
{dt},w_\omega(t)\right)dt\\
&\qquad\quad
+\int_0^T \int_{\T^3}\left\langle w_\omega(t,\t),
d\Xi_\omega(t,\t)\right\rangle d^3\t\\
&\qquad\quad -\sum_{i=1}^3\sqrt{2\nu}\int_0^T \int_{\T^3}
 w_{\omega,i}(t,\t) d^3\t dW_\omega^i(t),
\end{split}
\end{equation}
where $w_\omega(t,\cdot):=T_{g_\omega^1(t)}R_{g_\omega^1(t)^{-1}}
\left(\frac{\D g^1_\omega(t)}{dt}\right)=
\big(w_{\omega,1}(t,\cdot), w_{\omega,2}(t,\cdot),
w_{\omega,3}(t,\cdot)\big)$,
$d\Xi_{\omega}(t,\cdot):=T_{g_\omega^1(t)}R_{g_\omega^1(t)^{-1}}
\left(d^{\Delta} g_\omega^1(t)\right)$, $D_\omega(t,\t)d^3\t
=\left(\g^3(t,\cdot)^{-1}\right)^*\alpha_0$, and
$W_\omega$ is a standard $\R^3$-valued Brownian motion.

Let $g_\omega^{\nu}$ be the solution of \eqref{e4-1} with $\nu>0$ and
$M_\omega=W_\omega$  the same $\R^3$-valued Brownian motion as in the
definition of $\mathbf{J}^{\nu}$ above. In particular, $g_\omega^0$ is a
solution of \eqref{e4-1} with the same $\u$ and $\nu=0$, which in fact
is an ODE for each fixed $\omega\in \Omega$.

Suppose also that
$\tilde g_\omega^{\nu}$ is a solution of the following SDE,
\begin{equation}\label{e4-2a}
\begin{cases}
& d\tilde g_\omega^{\nu}(t,\t)=\sum_{i=1}^3H_{i,\nu} d\tilde W_\omega (t)+
\u(t,\tilde g_\omega^{\nu}(t,\t))dt\\
& \tilde g_\omega^{\nu}(0,\t)=\t,
\end{cases}
\end{equation}
where $\tilde W_\omega$ is an  standard $\R$-valued Brownian motion,
$H_{i,\nu}$, $1\le i \le 3$, and $\u$ are the same as in \eqref{e4-1}.
From now on, we use the notation $(\tilde W_\omega^1,
\tilde W_\omega^2,\tilde W_\omega^3)=
(\tilde W_\omega,\tilde W_\omega,\tilde W_\omega)$ to denote an
$\R^3$-valued Brownian motion with three equal components
$\tilde W_\omega^1 = \tilde W_\omega^2=\tilde W_\omega^3$, the same
$\R$-valued Brownian motion.

We can therefore characterize the critical points of $\mathbf{J}^{\nu}$
as follows.

\begin{thm}
\label{t5-4}{\rm(}SPDE case{\rm)}
$\left(\left(g_\omega^{\nu},H_{i,\nu},W_\omega^i\right)_{i=1}^3,
\left(\tilde g_\omega^{\mu},H_{i,\mu},\tilde W_\omega^i\right)_{i=1}^3,
g_\omega^{0}\right)$ is a critical point of $\mathbf{J}^{\nu}$ if and
only if $\left(\u,D_{\omega}\right)$ satisfies the following stochastic
compressible Navier-Stokes equation,
\begin{equation}\label{t4-2-1}
\begin{cases}
& d\u(t)=-\u(t)\cdot \nabla \u(t) dt-\frac{1}{D_\omega(t)}\Big(\sqrt{2\nu}\nabla \u(t)\cdot dW_\omega(t)
-\nu \Delta \u(t) dt\\
&\qquad \qquad \qquad \qquad \qquad \qquad \qquad \qquad
-\mu \nabla{\rm div}\u(t)+\nabla p_\omega(t)dt\Big),\\
& dD_\omega(t)=-
{\rm div}\left(\u(t) D_\omega(t) \right)dt,
\end{cases}
\end{equation}
where $D_\omega(t,\t)d^3 \t:=\left(g_{\omega}^{0}(t,\cdot )^{-1}\right)^{*}
\left(D_0(\t) d^3\t\right)$.

\end{thm}

\begin{thm}{\rm(}PDE case{\rm)}
\label{thm_5_6}
Let
\begin{equation*}
\begin{split}
&\mathbf{J}\left(\left(g_\omega^{1},
\w_i^1,M_\omega^{1,i}\right)_{i=1}^{m_1},
\left(g_\omega^{2},\w_i^1,M_\omega^{2,i}\right)_{i=1}^{m_1},
g_\omega^{3}\right)\\
&\quad :=
\int_0^T\int_{\T^3}\left(\frac{1}{2}\left|w_\omega(t,\t)\right|^2
\tilde D(t,\t)- \tilde D(t,\t)e\left(t,\tilde D(t,\t)\right)\right)
d^3 \t dt\\
&\qquad +\int_0^T \tilde p\left(\frac{\mathbf{D}^{\nabla^0,
(\w_i^1,M_\omega^{1,i})_{i=1}^{m_1}}\g^1(t)}{dt},
\frac{\mathbf{D}^{\nabla^0 ,(\w_i^2,M_\omega^{2,i})_{i=1}^{m_2}}\g^2(t)}
{dt}, w_\omega(t)\right)dt,
\end{split}
\end{equation*}
where $e(t,D)$ is non-random and satisfies
$\mathbf{d}e(t)=-p(t) \mathbf{d}\big(\frac{1}{D}\big)$ {\rm(}with non-
random pressure term $p(t)${\rm)} and
\[
w_\omega(t,\cdot):=T_{g_\omega^1(t)}R_{g_\omega^1(t)^{-1}}
\left(\frac{\D g^1_\omega(t)}{dt}\right),\qquad
\tilde D(t,\cdot)d^3\t=
\E\left[\left(\g^3(t,\cdot)^{-1}\right)^*\alpha_0\right].
\]

Suppose that $\u=u$ in \eqref{e4-1} and \eqref{e4-2a} is non-random and
the deformations \eqref{e4-4a} are defined with $v$ non-random. Then
$\left(\left(g_\omega^{\nu},H_{i,\nu},W_\omega^i\right)_{i=1}^3,
\left(\tilde g_\omega^{\mu},H_{i,\mu},\tilde W_\omega^i\right)_{i=1}^3,
g_\omega^{0}\right)$ is a critical point
of $\mathbf{J}$ if and only if {\rm(}the non-random
variables{\rm)} $\left(u,\tilde D\right)$ satisfy the
following {\rm(}deterministic{\rm)} classical
Navier-Stokes equations for compressible fluid flow
\begin{equation}\label{t4-2-2}
\begin{cases}
& du(t)=-\left(u(t)\cdot \nabla u(t)\right)dt+
\frac{1}{\tilde D(t)}\Big(\nu\Delta u(t) dt+
\mu\nabla {\rm div} u(t)-\nabla p(t)dt\Big),\\
& d\tilde D(t)=-{\rm div}\left(u(t)\tilde D(t)\right)dt.
\end{cases}
\end{equation}
\end{thm}

\begin{proof}(\textit{Theorem \ref{t5-4}}.)

By \eqref{r4-1-2}, we know that  $D_\omega(t,\t)$ satisfies the second
equation of \eqref{t4-2-1}. As explained above, since Theorem \ref{t3_2}
still holds for $G^s$, it suffices to show that the first equation of
\eqref{t4-2-1} is just the first one in \eqref{t3-4-1-right} for our model.

Relations \eqref{e4-4}-\eqref{e4-4c}, combined with the definition \eqref{e4-2a} of
$\tilde{g}_\omega^{\nu}$, yield the identities
\begin{equation}
\label{t4-2-4a}
\begin{split}
&T_{\tilde g_{\omega}^{\nu}(t, \theta)}
R_{\tilde g_{\omega}^{\nu}(t,\theta)^{-1}}
\left(\frac{\mathbf{D}^{\nabla^0 ,(H_{i,\nu},
\tilde W_\omega^i)_{i=1}^3}\tilde g_\omega^{\nu}(t)}{dt}
\right)_{i,j}=
\nabla_{H_{i,\nu}}^0 H_{j,\nu}\frac{d\llbracket \tilde W_\omega^i,
\tilde W_\omega^j\rrbracket_t}{dt}=0,\\
& \frac{d}{d\ee}\Bigg|_{\ee=0}T_{\tilde g_{\omega,\ee,v}^{\nu}(t, \theta)}
R_{\tilde g_{\omega,\ee,v}^{\nu}(t,\theta)^{-1}}
\left(\frac{\mathbf{D}^{\nabla^0,(H_{i,\nu,\ee},
\tilde W_\omega^i)_{i=1}^3}\tilde \g^{\nu}(t)}{dt}\right)_{i,j}\\
&\qquad =\big(\nabla^0_{{\rm ad}_{v_\omega(t)}H_{i,\nu}}H_{j,\nu}+
\nabla^0_{H_{i,\nu}}({\rm ad}_{v_\omega(t)}H_{j,\nu})\big)
\frac{d\llbracket \tilde W_\omega^i, \tilde W_\omega^j\rrbracket_t}{dt}=
2\nu\partial_i\partial_j v_\omega(t,\t).
\end{split}
\end{equation}

For every $u,\tilde u \in \x^s(\T^3)$, $\alpha=D(\t)d^3\t \in U^*$,
we easily get the following formulas:
\begin{equation}\label{t4-2-4}
\begin{split}
&\frac{\delta l}{\delta u}\left(t,u,\alpha\right)=uD,\\
&{\rm ad}^*_{u}\left(uD\right)=  ({\rm div} u) uD+u\cdot \nabla (Du)
+\frac{D}{2}\nabla(|u|^2),\\
&\sum_{i=1}^3{\rm ad}^*_{H_i}q dW_\omega^i(t)
=\sqrt{2\nu}\nabla u\cdot dW_\omega(t),\\
&\left(\frac{\delta l_\omega}{\delta \alpha}(t,u,\alpha)\right)
\diamond \alpha=\frac{D}{2}\nabla(|u|^2)-\nabla p_\omega(t).
\end{split}
\end{equation}
The last equality is obtained by repeating the argument in
\cite[Section 7]{HMR} (especially (7.4), (7.18), and (7.19)),
even though $p_\omega(t)$ is random.

On the other hand, for every $A,\tilde A\in \scr{M}_n(G^s)$
and $B,\tilde B\in \scr{M}_m(G^s)$, we have,
\begin{equation}\label{t4-2-6}
\begin{split}
&\left\langle\frac{\delta \tilde p}{\delta \xi_1}\left(A,B,u\right),
\tilde A\right\rangle=\frac{1}{2}\int_{\T^3}
u(\t)\cdot {\bf Tr }(\tilde A(\t))d^3\t,\\
& \left\langle\frac{\delta \tilde p}{\delta \xi_2}\left(A,B,u\right),
\tilde B\right\rangle
=\frac{1}{2}\sum_{i,j=1}^m\int_{\T^3}
\mathbf{P}_i(u(\t))\mathbf{P}_j((\tilde B)_{i,j}(\t))d^3\t,\\
&\left\langle\frac{\delta \tilde p}{\delta u}\left(A,B,u\right),
\tilde u\right\rangle =\frac{1}{2}\int_{\T^3}
\tilde u(\t)\cdot {\bf Tr }(A(\t))d^3\t
+\frac{1}{2}\sum_{i,j=1}^m\int_{\T^3}
\mathbf{P}_i(\tilde u(\t))\mathbf{P}_j((B)_{i,j}(\t))d^3\t.
\end{split}
\end{equation}
Hence, using $M_\omega^1=W_\omega$, and \eqref{e4-4}, \eqref{t4-2-4a},
\eqref{t4-2-6}, we get the formula for the
operator $K$ defined by \eqref{t3-1-3}:
\begin{align*}
\left\langle K_\omega\left(t,A,B,u\right),\tilde u\right\rangle&=
-\sum_{i=1}^3\int_{\T^3} u(\t)\cdot \left(\nabla^0_{[\tilde u,H_{i,\nu}]}
H_{i,\nu}+\nabla^0_{H_{i,\nu}}[\tilde u,H_{i,\nu}]\right)(\t)d^3\t\\
&\quad -\sum_{i,j=1}^3 \int_{\T^3}u_i(\t)\mathbf{P}_j
\left(\nabla^0_{[\tilde u,H_{i,\mu}]}H_{j,\mu}+
\nabla^0_{H_{i,\mu}}[\tilde u,H_{j,\mu}]\right)(\t)d^3\t \\
&=-\nu\int_{\T^3} u(\t) \cdot \Delta \tilde u(\t) d^3\t
-\mu\int_{\T^3}\tilde{u}(\t)\cdot
\nabla {\rm div} u(\t)d^3\t
\end{align*}
that is,
\begin{equation}\label{t4-2-5a}
\begin{split}
K_\omega\left(t,A,B,u\right)&=-\nu\Delta u-\mu\nabla {\rm div} u,\
\ A,B\in \scr{M}_3(G^s),\ u\in \x^s(\T^3),\ t\in [0,T],\\
\frac{\delta \tilde p}{\delta u}\left(0,0,u\right)&=0.
\end{split}
\end{equation}
Thus, combining the equalities above, the  first equation in
\eqref{t3-4-1-right} becomes
\begin{equation}\label{t4-2-3}
\begin{split}
&d\left(\u(t) D_\omega(t)\right)=-
\sqrt{2\nu}\nabla \u(t)\cdot dW_\omega(t)-
\left({\rm div}\u(t)\right) \u(t) D_\omega(t)dt\\
&-\u(t)\cdot \nabla \left(\u(t)D_\omega(t)\right)dt+\nu\Delta \u(t)dt+
\mu \nabla ({\rm div}\u)(t)dt -\nabla p_\omega(t) dt.
\end{split}
\end{equation}
Using the second equation of \eqref{t4-2-1}, we get
\begin{equation}\label{t4-2-5}
\begin{split}
\u(t)dD_\omega(t)&=-
\left({\rm div} \u(t)\right) \u(t) D_\omega(t)dt
-\left(\u(t)\cdot \nabla D_\omega(t)\right) \u(t)dt
\end{split}
\end{equation}
and together with \eqref{t4-2-3}, we obtain the first equation
of \eqref{t4-2-1}.
\end{proof}

\begin{proof} \it {\rm(}Theorem \ref{thm_5_6}.{\rm)} \rm
This follows carrying out the same computations as in the previous proof
and the one in Theorem \ref{t3_3}.
\end{proof}

\begin{remark}
{\rm We emphasize that the usual Navier-Stokes equations for compressible
fluids \eqref{t4-2-2} were deduced from our stochastic variational
principle, without any appeal to thermodynamic considerations in order
to get the dissipative terms; these terms appear entirely due to the
type of stochastic processes we consider.}
\end{remark}

\begin{remark}
{\rm For the incompressible case, i.e., $D_\omega(t,\t)\equiv 1$,
equation \eqref{t4-2-1} becomes
\begin{equation*}
d\u(t)=-\sqrt{2\nu}\nabla \u\cdot dW_\omega(t)
-\u\cdot \nabla \u dt+\nu \Delta \u dt-\nabla p_\omega(t)dt,
\end{equation*}
which is a stochastic incompressible Navier-Stokes equation.
} \hfill $\lozenge$
\end{remark}

\begin{remark}
{\rm Taking the viscous force $\tilde p=0$ in the definition of $\mathbf{J}^{\nu}$ (formula \eqref{e5-1a}) and following the same steps
as in Theorem \ref{t5-4}, it is easy to verify that the associated
critical point $(\u(t),D_\omega(t))$ of $\mathbf{ J}^{\nu}$
satisfies the following stochastic compressible Euler equation,
\begin{equation*}
\;\qquad \qquad \begin{cases}
& d\u(t)=-\u(t)\cdot \nabla \u(t) dt-
\frac{1}{D_\omega(t)}\Big(\sqrt{2\nu}\nabla \u(t)\cdot dW_\omega(t)
+\nabla p_\omega(t)dt\Big),\\
& dD_\omega(t)=-
{\rm div}\left(\u(t) D_\omega(t) \right)dt.
\:\,\quad \qquad \qquad \qquad \qquad \qquad \qquad \qquad \qquad \lozenge
\end{cases}
\end{equation*}
}
\end{remark}

For existence of solutions of stochastic compressible Navier-Stokes
equations we refer to \cite{BH}, \cite{WW}.

\begin{remark}
{\rm We illustrate here how the contraction force $\tilde{p}$, defined by
\eqref{e5-3}, gives rise to the term modeling viscosity in the
compressible Navier-Stokes equation (and MHD equation later).
Other choices for the contraction force $\tilde{p}$ yield different
dissipative equations.

For example, let $\tilde p: \scr{M}(G^s)\times \x^s(\T^3)\times U^*
\rightarrow \R$ be defined by
\begin{equation}\label{r5-1-1}
\tilde p\left(A,u,D(\t)d^3\t\right)=
\frac{1}{2}\int_{\T^3}{\bf Tr}(A)(\t)\cdot u(\t)D(\t)d^3\t,
\end{equation}
where $A\in \scr{M}(G^s)$, $u \in \x^s(\T^3)$, $\ D(\t)d^3\t\in U^*$.
Define the action functional}$\mathbf{J}^{\nu}$  in the same way as
in Theorem \ref{t5-4} with $\tilde p$ replaced by the expression in
\eqref{r5-1-1}.

As explained in Remark \ref{r4-2}, although $\tilde p$ depends on $U^*$,
we can repeat the procedure in Theorem  \ref{t5-4} to show that
$\left(\left(g_\omega^{\nu},H_{i,\nu},W_\omega^i\right)_{i=1}^3,
\left(\tilde g_\omega^{\mu},H_{i,\mu},\tilde W_\omega^i\right)_{i=1}^3,
g_\omega^{\nu}\right)$ is a critical point of $\tilde{\mathbf{J}}^{\nu}$
if and only if $(\u(t),D_\omega(t))$ satisfies the following
system of equations:
\begin{equation*}
\begin{cases}
d\u(t)&=-\u(t)\cdot \nabla \u(t) dt+2\nu\langle \nabla \u(t),
\nabla \log D_\omega(t)\rangle dt+\nu \Delta \u(t)dt\\
&\quad
-\frac{\sqrt{2\nu}\nabla \u(t)}{D_\omega(t)}\cdot dW_\omega(t)
-\frac{\nabla p_\omega(t)}{D_\omega(t)}dt,\\
dD_\omega(t)&=-\sqrt{2\nu}\nabla D_\omega(t)\cdot dW_\omega(t)-
{\rm div}\left(\u(t) D_\omega(t) \right)dt+\nu\Delta D_\omega(t)dt.
\end{cases}
\end{equation*}
The term $\langle \nabla \u(t), \nabla \log D_\omega(t)\rangle$ in the
equation above is crucial for energy dissipation. This term also appears
in Brenner's model; see, e.g., \cite{Brenner2005a, Brenner2005b, FeVa2010}.
 \hfill $\lozenge$
\end{remark}

\subsection{Compressible MHD equation}

Let $\alpha_0:=(b_0(\cdot),\B_0(\t)\cdot\mathbf{d}\s, D_0(\t)d^3\t)$,
where $b_0 $ is a $C^2$ function on $\T^3$,
$\B_0(\t) \cdot \mathbf{d}\s$ is an exact two-form on
$\mathbb{T}^3$, i.e., there is some one-form
$A_0(\t)\cdot \mathbf{d}\theta$ such that
\begin{equation}\label{e4-2}
\B_0(\t) \cdot \mathbf{d}\s=
\mathbf{d}\Big(A_0(\t) \cdot \mathbf{d}\t\Big)
=\sum_{1\le j<k \le 3,i \neq j, i \neq k}
\big(\cu A_0(\t)\big)_i \mathbf{d}\t_j \wedge \mathbf{d}\t_k,
\end{equation}
and $D_0(\t)d^3 \t$ is a density on $\T^3$. We let $U^*$ denote
the vector space of all such triples
$(b(\cdot),\B(\t)\cdot\mathbf{d}\s, D(\t)d^3\t)$.

As in \cite[equation (7.16)]{HMR}, let
$l: \Omega\times [0,T]\times \mathfrak{X}^s(\mathbb{T}^3) \times U^*
\rightarrow \R$ be defined by
\begin{equation*}
l_\omega(t,u,b,\B,D)=\int_{\mathbb{T}^3} \left(\frac{D(\t)}{2}|u(\t)|^2-
D(\t)e_\omega(t,D(\t),b(\t))-\frac{1}{2}|\B(\t)|^2\right) d^3\t,
\end{equation*}
where $u \in  \mathfrak{X}^s(\mathbb{T}^3)$
is the Eulerian (spatial) velocity
of the fluid, $b \in C^2(\T^3)$ is the entropy function,
$\B(\t)\cdot \mathbf{d}\s$ is an exact differential two-form
as in \eqref{e4-2} representing the magnetic field in the
fluid, $D(\t)d^3 \t$ is a density on $\T^3$
representing the mass density of the fluid,
and the function $e_\omega(t,D,b)$ is the fluid's specific
internal energy. The pressure $p_\omega(t )$ is  $\mathcal{P}_t$-measurable
for all $t$ and the temperature
$T_\omega(t)$ of the fluid (also  $\mathcal{P}_t$-measurable for all $t$)
are given in terms of a thermodynamic equation of state for the specific internal energy $e$,
namely $\mathbf{d}e_\omega(t)=-p_\omega(t) \mathbf{d}\big(\frac{1}{D}\big)
+T_\omega(t) \mathbf{d}b$. As explained in \cite[Section 7]{HMR}
it is assumed that $c^2_\omega: = \frac{\partial p_\omega}{\partial D} >0$,
where $c_\omega$ is the adiabatic sound speed.

As in subsection \ref{sec5-1}, we work with a contraction force
$\tilde p:\scr{M}(G^s)\times \scr{M}(G^s)\times \x^s(\T^3)\rightarrow \R$,
defined by \eqref{e5-3}, and
a stochastic force $q:\x^s(\T^3)\times U^* \rightarrow (\x^s(\T^3))^*$,
defined by
$$
q(u,\alpha):=\langle u,\cdot\rangle,\ \ \forall u \in \x^s(\T^3).
$$

With $\nabla^0$, $l$, $\tilde p$, $q$, $\alpha_0$, $(H_{i,\nu},
W_\omega^i)_{i=1}^3$ as in subsection \ref{sec5-1}, we
define the action functional $\mathbf{J}_1^{\nu}:=
J^{\nabla^0,(H_{i,\nu},W_\omega^i)_{i=1}^3}$ according to
\eqref{right_action}, which in this particular case becomes
\begin{equation}\label{e4-3}
\begin{split}
&\mathbf{J}_1^{\nu}\left(\left(g_\omega^{1},\w_i^1,M_\omega^{1,i}
\right)_{i=1}^{m_1},
\left(g_\omega^{2},\w_i^1,M_\omega^{2,i}\right)_{i=1}^{m_1},
g_\omega^{3},\g^4,\g^5\right)\\
&\quad :=
\int_0^T l_\omega\left(t,w_\omega(t,\cdot),\B_\omega(t,\cdot),
b_\omega(t,\cdot),D_\omega(t,\cdot)\right) dt\\
&\qquad +\int_0^T \tilde p\left(\frac{\mathbf{D}^{\nabla^0,
(\w_i^1,M_\omega^{1,i})_{i=1}^{m_1}}\g^1(t)}{dt},
\frac{\mathbf{D}^{\nabla^0,(\w_i^2,M_\omega^{2,i})_{i=1}^{m_2}}
\g^2(t)}{dt},w_\omega(t)\right)dt\\
&\qquad +\int_0^T \int_{\T^3}\left\langle w_\omega(t,\t),
d\Xi_\omega(t,\t)\right\rangle d^3\t\\
&\qquad -\sum_{i=1}^3\sqrt{2\nu}\int_0^T \int_{\T^3}
w_{\omega,i}(t,\t) d^3\t dW_\omega^i(t),
\end{split}
\end{equation}
where
\begin{align*}
w_\omega(t,\cdot)&:=T_{g_\omega^1(t)}R_{g_\omega^1(t)^{-1}}
\left(\frac{\D g^1_\omega(t)}{dt}\right)=
\big(w_{\omega,1}(t,\cdot), w_{\omega,2}(t,\cdot),
w_{\omega,3}(t,\cdot)\big)\\
d\Xi_{\omega}(t,\cdot)& :=T_{g_\omega^1(t)}R_{g_\omega^1(t)^{-1}}
\left(d^{\Delta} g_\omega^1(t)\right)\\
D_\omega(t,\cdot)d^3\t&:=\left(\g^3(t,\cdot)^{-1}\right)^*
\left(D_0(\t)d^3 \t\right) \\
\B_\omega(t,\t)\cdot \mathbf{d}\s&:=
\left(g_{\omega}^{4}(t,\cdot)^{-1}\right)^{*}
\left(\B_0(\t)\cdot \mathbf{d}\s\right)\\
b_\omega(t,\t)&:=
\left(g_{\omega}^{5}(t,\cdot )^{-1}\right)^{*}b_0,
\end{align*}
and $W_\omega(t)$ is a standard $\R^3$-valued Brownian motion.

Let $g_\omega^{\nu}$ be the solution of \eqref{e4-1} with $\nu>0$ and
$M_\omega=W_\omega$ the same $\R^3$-valued Brownian motion as in the
definition of $\mathbf{J}_1^{\nu}$ above. Although in the definition
of $\mathbf{J}_1^{\nu}$, five semimartingales are needed, we can
define its critical points in the same way as in \eqref{e3-4} along
deformations \eqref{e4-4a}. Moreover, the critical points of
$\mathbf{J}_1^{\nu}$ are characterized as follows.

\begin{thm}
\label{t5-10}{\rm(}SPDE case{\rm)}
$\left(\left(g_\omega^{\nu},H_{i,\nu},W_\omega^i\right)_{i=1}^3,
\left(\tilde g_\omega^{\mu},H_{i,\mu},\tilde W_\omega^i\right)_{i=1}^3,
g_\omega^{0},\g^{\nu_1},\g^{\nu_2}\right)$
is a critical point of $\mathbf{J}_1^{\nu}$ if and only if the following
stochastic compressible MHD equations hold for
$(u_\omega (t),b_\omega (t),\B_\omega (t),D_\omega (t))$,
\begin{equation}\label{t4-3-1}
\begin{cases}
&d \u(t)=-\u(t)\cdot \nabla \u(t) dt-\frac{1}{D_\omega(t)}\Big(\sqrt{2\nu}
\nabla \u(t) \cdot dW_\omega(t)-
\nu \Delta \u(t) dt-\mu\nabla {\rm div}\u(t) dt\\
&\qquad\qquad\qquad\qquad\qquad\qquad\qquad\qquad
-\B_\omega(t) \times \cu \B_\omega(t)dt+\nabla p_\omega(t)dt\Big),\\
&d D_\omega(t)= -{\rm div}(\u(t) D_\omega(t)) dt,\\
&d \B_\omega(t)=-\sqrt{2\nu_1}\nabla \B_\omega(t) \cdot dW_\omega(t)+
\cu(\u(t)\times \B_\omega(t))dt+\nu_1 \Delta \B_\omega(t)dt,\\
& db_\omega(t)=-\sqrt{2\nu_2}\nabla b_\omega(t) \cdot dW_\omega(t)-\u(t)
\cdot \nabla b_\omega(t) dt+\nu_2 \Delta b_\omega(t)dt,
\end{cases}
\end{equation}
where $\g^{\nu}$ and $\tilde{g}_\omega^{\mu}$ are the solution of the
SDE \eqref{e4-1} and \eqref{e4-2a} respectively, $D_\omega(t,\t)d^3\t
=\left(\g^{0}(t,\cdot)^{-1}\right)^*\left(D_0(\t)d^3 \t\right)$,
$\B_\omega(t,\t)\cdot \mathbf{d}\s:=
\left(g_{\omega}^{\nu_1}(t,\cdot)^{-1}\right)^{*}$
$\left(\B_0(\t)\cdot \mathbf{d}\s\right)$,
$ b_\omega(t,\t):=
\left(g_{\omega}^{\nu_2}(t,\cdot )^{-1}\right)^{*}
b_0$.
\end{thm}

\begin{thm}
\label{t_5_11}{\rm(}PDE case{\rm)}
Set
\begin{equation*}
\begin{split}
&\mathbf{J}_1\left(\left(g_\omega^{1},\w_i^1,M_\omega^{1,i}
\right)_{i=1}^{m_1},\left(g_\omega^{2},\w_i^1,
M_\omega^{2,i}\right)_{i=1}^{m_1},g_\omega^{3},\g^4,\g^5\right)\\
&\quad:=
\int_0^T l\left(t,w_\omega(t,\cdot),\tilde \B_\omega(t,\cdot),
\tilde b_\omega(t,\cdot),\tilde D_\omega(t,\cdot)\right) dt\\
&\qquad +\int_0^T \tilde p\left(\frac{\mathbf{D}^{\nabla^0,
(\w_i^1,M_\omega^{1,i})_{i=1}^{m_1}}\g^1(t)}{dt},
\frac{\mathbf{D}^{\nabla^0,(\w_i^2,M_\omega^{2,i})_{i=1}^{m_2}}\g^2(t)}
{dt},w_\omega(t)\right)dt,
\end{split}
\end{equation*}
where $l$ is non-random {\rm(}hence the pressure $p(t)$ and temperature
$T(t)$ are non-random{\rm)}, $w_\omega(t,\cdot):=T_{g_\omega^1(t)}
R_{g_\omega^1(t)^{-1}}
\left(\frac{\D g^1_\omega(t)}{dt}\right)$,
$\tilde \B(t,\t)\cdot \mathbf{d}\s:=
\E\left[\left(g_{\omega}^{\nu_1}(t,\cdot)^{-1}\right)^{*}
\left(\B_0(\t)\cdot \mathbf{d}\s\right)\right]$,
$\tilde  b(t,\t):=
\E\left[\left(g_{\omega}^{\nu_2}(t,\cdot )^{-1}\right)^{*}
b_0\right]$, $\tilde D(t,\cdot)d^3\t=\E\left[
\left(\g^0(t,\cdot)^{-1}\right)^*\left(D_0(\t)d^3 \t\right)\right]$.

Suppose $\u=u$ is non-random in \eqref{e4-1}, \eqref{e4-2a}, and
that in the deformation $v$ in \eqref{e4-4a} is also non-random. Then
$\left(\left(g_\omega^{\nu},H_{i,\nu},W_\omega^i\right)_{i=1}^3,
\left(\tilde g_\omega^{\mu},H_{i,\mu},\tilde W_\omega^i\right)_{i=1}^3,
g_\omega^{0},\g^{\nu_1},\g^{\nu_2}\right)$
is a critical point of $\mathbf{J}_1$ if and only if
$\left(u,\tilde \B,\tilde D,\tilde b\right)$ satisfies the following
compressible MHD equations
\begin{equation}\label{t4-3-2}
\begin{cases}
&du(t)=-u(t)\cdot \nabla u(t)dt +\frac{1}{\tilde D(t)}\
\Big(\nu \Delta u(t)+\mu \nabla {\rm div}u(t)
-\tilde \B(t) \times \cu \tilde \B(t)dt-\nabla p(t)\Big)dt,\\
& d\tilde D(t)=-{\rm div}(u(t)\tilde D(t))dt,\\
& d\tilde \B(t)=\cu(u(t)\times\tilde \B(t))dt+\nu_1\Delta \tilde \B(t)dt,\\
&d\tilde b(t)=-u(t)\cdot\nabla\tilde b(t) dt+\nu_2 \Delta \tilde b(t) dt.
\end{cases}
\end{equation}
\end{thm}

\begin{proof} \it {\rm(}Theorem \ref{t5-10}.{\rm)}\rm

Equation \eqref{r4-1-2} implies that $D_\omega(t,\t)$ satisfies the
second equation of \eqref{t4-3-1}. Since $\B_0(\t)\cdot \mathbf{d}\s=
\mathbf{d}\big(A_0(\t) \cdot \mathbf{d}\t\big)$ for some
one-form $A_0(\t) \cdot \mathbf{d}\t$, it follows that
\begin{equation*}
\begin{split}
\B_\omega(t,\t)\cdot \mathbf{d}\s&=
\big(g_{\omega}^{\nu_1}(t,.)^{-1}\big)^{*}
\big(\B_0(\t)\cdot b\mathbf{d}\s\big)\\
&=\big(g_{\omega}^{\nu_1}(t,.)^{-1}\big)^{*}
\mathbf{d}\big(A_0(\t) \cdot \mathbf{d}\t\big)\\
&=\mathbf{d}\big(\big(g_{\omega}^{\nu_1}(t,.)^{-1}\big)^{*}
\big(A_0(\t) \cdot \mathbf{d}\t\big)\big)(\t)\\
&= \mathbf{d}\big(A_\omega(t,\t)\cdot \mathbf{d}\t\big),
\end{split}
\end{equation*}
where
\begin{equation*}
A_\omega(t,\t)\cdot\mathbf{d}\t:=\big(g^{\nu_1}(t,.)^{-1}\big)^{*}
\big(A_0(\t)\cdot \mathbf{d}\t \big),\quad
\cu A_\omega(t):=\B_\omega(t).
\end{equation*}
By Proposition \ref{p4-1},  equation (\ref{p4-1-2}) holds for
$A_\omega(t)$ with viscosity constant $\nu=\nu_3$, and hence
$\B_\omega(t)=\cu A_\omega(t)$ satisfies the third equation of
\eqref{t4-3-1}. We also have $\nabla \cdot \B_\omega(t)=
\nabla \cdot (\cu A_\omega(t))=0$.

In the same way, we verify that the fourth equation in \eqref{t4-3-1}
holds for $b_\omega(t)$.

According to Theorem \ref{t3_2}, \eqref{e4-4c},
\eqref{t4-2-5a} (which implies that
$\frac{\delta \tilde p}{\delta u}
\Big(\tilde H_{\omega,1},\tilde H_{\omega,2},u_\omega(t)\Big)\equiv 0$
here), we conclude that
$\left(\left(g_\omega^{\nu},H_{i,\nu},W_\omega^i\right)_{i=1}^3,
\left(\tilde g_\omega^{\mu},H_{i,\mu},\tilde W_\omega^i\right)_{i=1}^3,
g_\omega^{0},\g^{\nu_1},\g^{\nu_2}\right)$
is a critical point of $\mathbf{J}_1$
if and only if the following equation holds
\begin{equation}\label{t4-3-3}
\begin{split}
& d\left(\frac{\delta l_\omega}{\delta u}\right)(t)=
-\sum_{i=1}^3{\rm ad}^*_{H_{i,\nu}}q(t,\u,b_\omega,
\B_\omega,D_\omega)dW_\omega^i(t)
-{\rm ad}^*_{\u(t)} \frac{\delta l_\omega}{\delta u}dt+
\frac{\delta l_\omega}{\delta b}\diamond b_\omega dt\\
&\qquad\qquad\qquad +\frac{\delta l_\omega}{\delta \B}\diamond \B_\omega dt
+\frac{\delta l_\omega}{\delta D}\diamond D_\omega dt
-K_\omega\left(t,\tilde H_{\omega,1}(t),\tilde H_{\omega,2}(t),\u(t)
\right)dt,
\end{split}
\end{equation}
where $K_\omega$, $\tilde H_{\omega,1}(t)$, $\tilde H_{\omega,2}(t)$ are
the same terms as in \eqref{t3-4-1-right}.

From the computation in \cite[Section 7]{HMR}, particularly
(7.4), (7.18), and (7.19), we get,
\begin{equation*}
\begin{split}
& \frac{\delta l_\omega}{\delta b}\diamond b+
\frac{\delta l_\omega}{\delta \B}\diamond \B
+\frac{\delta l_\omega}{\delta D}\diamond D=
\frac{D}{2}\nabla \left(|u|^2\right)+\B \times \cu \B-\nabla p_\omega.
\end{split}
\end{equation*}
Combining all of the above with \eqref{e4-4}-\eqref{e4-4c},
\eqref{t4-2-4}-\eqref{t4-2-5a}, into \eqref{t4-3-3} yields,
\begin{equation}\label{t4-3-4}
\begin{split}
d\left(\u(t) D_\omega(t)\right)=&-
\sqrt{2\nu}\nabla \u(t)\cdot dW_\omega(t)-
\left({\rm div}\u(t)\right) \u(t) D_\omega(t)dt\\
&-\u(t)\cdot \nabla \left(\u(t)D_\omega(t)\right)dt+
\B_\omega(t) \times \cu \B_\omega(t) dt\\
&+\nu \Delta \u(t)dt+\mu\nabla {\rm div}\u(t)dt-\nabla p_\omega(t) dt.
\end{split}
\end{equation}
Putting the second equation of \eqref{t4-3-1} into \eqref{t4-3-4}, we 
derive the first equation in (\ref{t4-3-1}).
\end{proof}

\begin{proof} \it {\rm(}Theorem \ref{t_5_11}.{\rm)}\rm

The proof of \eqref{t4-3-2} follows by repeating the same computations
as above and the ones in the proof of Theorem \ref{t3_3}.
\end{proof}

\begin{remark}{\rm
The reason for choosing processes $g^{\nu_i}$ with different
constants $\nu_i$ is that the viscosity constants in equation 
(\ref{t4-3-1}) are different.

\hfill $\lozenge$}
\end{remark}

\begin{remark}{\rm
In particular, if we take $D(t)=1$, $b(t)=1$ for every $t \in [0,T]$ in
(\ref{t4-3-2}),  we obtain the following incompressible viscous MHD equations (see, e.g., \cite{ST}),
\begin{equation*}
\begin{cases}
&\partial_t u+u\cdot \nabla u+\nabla p+\B \times
\cu \B=\nu \Delta u\\
&\partial_t \B=\cu(u\times \B)+\nu_1 \Delta \B\\
&{\rm div} u=0. \hfill
\lozenge
\end{cases}
\end{equation*}
}
\end{remark}

\subsection{Stochastic Kelvin-Noether Theorem in Continuum Mechanics}

We now apply the results on Section \ref{sec_4} in continuum mechanics.
Following the formulation in \cite[Section 6]{HMR}, we take here
$G=G^s$ (so $T_e G= \mathfrak{X}^s(\T^3)$), $U$ a linear space
whose formal dual $U^*$ represents the advection terms
(such as mass density, entropy, the magnetic field viewed as a
differential two-form, etc), $\c=\{\gamma\in C([0,1];G^s)\mid
\gamma(0)=\gamma(1)\}$ the set of all continuous $G^s$-valued loops.

As explained in \cite[Section 6]{HMR}, the dual
$\big(\mathfrak{X}^s(\T^3)\big)^*=\Omega^s(\T^3)\otimes
{\rm Den}(\T^3)$, where $\Omega^s(\T^3)$ denotes the space of
$H^s$-differential one-forms on $\T^3$, and
${\rm Den}(\T^3)$ is the set of all densities on $\T^3$. Given a
mass density $\rho=\rho(\t)d^3\t$, we define the circulation map
$\scr{K}: \c\times U^* \rightarrow
\big(\mathfrak{X}^s(\T^3)\big)^{**}$ by
\begin{equation*}
\left\langle \scr{K}(\gamma,a),\alpha\right\rangle
=\oint_{\gamma(\cdot)}\frac{\alpha}{\rho},\quad
\gamma\in \c,\quad a\in U^*,\quad \alpha \in
\big(\mathfrak{X}^s(\T^3)\big)^{*}.
\end{equation*}
Since $\alpha\in \big(\mathfrak{X}^s(\T^3)\big)^*=\Omega^s(\T^3)\otimes
{\rm Den}(\T^3)$, $\rho\in {\rm Den}(\T^3)$, and $\frac{\alpha}{\rho}\in
\Omega^s(\T^3)$, the circulation integral above is well-defined.

Let $\scr{L}_u$ denote the Lie derivative in the direction
$u \in \mathfrak{X}^s(\T^3)$. As shown in
\cite[Page 37, formula (6.2)]{HMR}, we have
\begin{equation*}
{\rm ad}_u^* V=\scr{L}_u V,\quad  u \in  \mathfrak{X}^s(\T^3),\quad
V\in \big(\mathfrak{X}^s(\T^3)\big)^{*}.
\end{equation*}

Suppose that $\g^{\nu}(\cdot)$ is the solution of \eqref{e4-1}
with $M_\omega=W_\omega$ being a standard $\R^3$-valued Brownian
motion and $\tilde g_\omega^{\nu}(\cdot)$ is the solution to \eqref{e4-2a}.
As illustrated above, Theorem \ref{t3_2} still
holds for the infinite dimensional group $G^s$.

So, for a given Lagrangian functional
$l: \Omega\times[0,T]\times\mathfrak{X}^s(\T^3)\times U^* \rightarrow \R$
such that $\frac{\delta l_\omega}{\delta u}$ is non-random, and supposing
that $\tilde p:\scr{M}\times \scr{M}\times \mathfrak{X}^s(\T^3)
\rightarrow \R$,
$q:[0,T]\times\mathfrak{X}^s(\T^3)\times U^*\rightarrow \R$ are the
same terms as in Section \ref{sec5-1}, we
define the action functional $J^{\nabla^0,(H_{i,\nu},W_\omega^i)_{i=1}^3}$
 by \eqref{e5-1a}. Hence by Theorem \ref{t3_2} and the
computations in Section \ref{sec_5_0} above,  we conclude that
$\left(\left(g_\omega^{\nu},H_{i,\nu},W_\omega^i\right)_{i=1}^3,
\left(\tilde g_\omega^{\mu},H_{i,\mu},\tilde W_\omega^i\right)_{i=1}^3,
g_\omega^{\nu_1}\right)$ is a critical point of
$J^{\nabla^0,(H_{i,\nu},W_\omega^i)_{i=1}^3}$ if and only if
$(\u(t),\a(t))$ satisfy the following equations
(note that  in the present situation, \eqref{e4-4c} and
\eqref{t4-2-6}  imply
$\tilde H_1(t)=\tilde H_2(t)\equiv 0$ and
$\frac{\delta \tilde p}{\delta u}\left(0,0,\u(t)\right) \equiv 0$),
\begin{equation}
\begin{split}\label{ne5-1}
& d\frac{\delta l_\omega}{\delta u}(t)=-\sqrt{2\nu}\sum_{i=1}^{3}
\partial_i \u(t) dW^{i}_{\omega}(t)
- {\rm ad}^*_{u_\omega(t)}
\left(\frac{\delta l_\omega}{\delta u}(t)\right)dt+
\frac{\delta l_\omega}{\delta \alpha}(t)\diamond \a(t)dt\\
&\qquad \qquad\;+\nu\Delta u_\omega(t)+\mu \nabla {\rm div} \u(t) dt,\\
&d\a(t)=-\sum_{i=1}^3\scr{L}_{H_{i,\nu_1}}\a(t)dW_\omega^i(t)
-\scr{L}_{u_\omega(t)}\a(t)dt+
\frac{1}{2}\sum_{i=1}^3\scr{L}_{H_{i,\nu_1}}\scr{L}_{H_{i,\nu_1}}\a(t)dt,
\end{split}
\end{equation}
where $u_\omega(\cdot)$ denotes the drift in
\eqref{e4-1}, $\frac{\delta l_\omega}{\delta u}(t):=
\frac{\delta l_\omega}{\delta u}\big(t,u_\omega(t),\a(t)\big)$,
$q_\omega(t):=q\big(t,\u(t),\a(t)\big)$, and
$\frac{\delta l_\omega}{\delta \alpha}(t):
=\frac{\delta l_\omega}{\delta \alpha}\big(t,u_\omega(t),\a(t)\big)$.
Here, we have also applied the property that $q(t,u,\alpha)=
\langle u,\cdot\rangle$.

Given  $\gamma_0\in \c$,
the Kelvin-Noether
quantity $I:\c\times \mathfrak{X}^s(\T^3)\times U^* \rightarrow \R$ is
defined by
\begin{equation}\label{ne5-2}
I_\omega(t):=
\oint_{\gamma_\omega(t)(\cdot)}\frac{1}{\rho_\omega(t)}
\frac{\delta l_\omega}{\delta u}(t),
\end{equation}
where $\gamma_\omega(t)(\cdot):=\gamma_0\big(\g^{\nu}(t,\cdot)\big)$,
$(\u(t),\a(t))$ is a solution of
equation \eqref{ne5-1}, and $\rho_\omega(t)=
\rho\big((\g^{\nu})^{-1}(t,\t)\big)d^3\t$.

The stochastic Kelvin-Noether Theorem on $G^s$ takes the following form.
\begin{prp}\label{nt5-1}
Let $I_\omega(t)$ be defined by \eqref{ne5-2} and suppose that
$(\u(t),\a(t))$ satisfies \eqref{ne5-1} with $\nu=\nu_1$ and
$\frac{\delta l}{\delta u}=q=\langle u,\cdot \rangle$. Then we have
\begin{equation}\label{nt5-1-1}
dI_\omega(t)=\oint_{\gamma_\omega(t)(\cdot)}\frac{1}{\rho_\omega(t)}
\left(\frac{\delta l_\omega}{\delta \alpha}(t)\diamond \alpha_\omega(t)
+\mu\nabla {\rm div}\u(t)\right)dt.
\end{equation}
\end{prp}
\begin{proof}
By definition of $\gamma_\omega(t)(\cdot)$ and the change of variables
formula, we obtain
\begin{equation}\label{nt5-1-2}
\begin{split}
I_\omega(t)=\oint_{\gamma_\omega(t)(\cdot)}\frac{1}{\rho_\omega(t)}
\frac{\delta l_\omega}{\delta u}(t)=
\oint_{\gamma_0(\cdot)}\frac{1}{\rho}\left(\g^{\nu}(t)\right)^*\left[
\frac{\delta l_\omega}{\delta u}(t)\right]
\end{split}
\end{equation}

By carefully tracking the proof of Proposition \ref{p4-1}, we know
that the following right invariant version of \eqref{t5-1-2a} still holds
\begin{equation*}
\begin{split}
d(\g^{\nu}(t))^*\left[
\frac{\delta l_\omega}{\delta u}(t)\right]
&=(\g^{\nu}(t))^*\left[\sum_{i=1}^3{\rm ad}^*_{H_{i,\nu}}
\frac{\delta l_\omega}{\delta u}(t)dW_\omega^{i}(t)+
{\rm ad}^*_{\u(t)}\frac{\delta l_\omega}{\delta u}(t)+
d\frac{\delta l_\omega}{\delta u}(t) \right.\\
&\qquad \left.+
\frac{1}{2}\sum_{i=1}^3\left({\rm ad}^*_{H_{i,\nu}}{\rm ad}^*_{H_{i,\nu}}
\frac{\delta l_\omega}{\delta u}(t)+
2{\rm ad}^*_{H_{i,\nu}}d\llbracket W_\omega^{i} ,
\frac{\delta l_\omega}{\delta u}(t)
\rrbracket_t\right)\right].
\end{split}
\end{equation*}
Then, replacing here $d\frac{\delta l_\omega}{\delta u}(t)$ by its
expression given in \eqref{ne5-1} and using the identity
$\langle\u(t),\cdot\rangle=q_\omega(t)=
\frac{\delta l_\omega}{\delta u}(t)$, we get

\begin{equation*}
\begin{split}
&d(\g^{\nu}(t))^*\left[
\frac{\delta l_\omega}{\delta u}(t)\right]\\
&=(\g^{\nu}(t))^*\left[\frac{\delta l_\omega}{\delta \alpha}(t)\diamond \
\a(t) + \nu\Delta \u(t)
+\mu\nabla {\rm div}\u(t)-\frac{1}{2}\sum_{i=1}^3
{\rm ad}^*_{H_{i,\nu}}{\rm ad}^*_{H_{i,\nu}}q_\omega(t)\right]dt\\
&=(\g^{\nu}(t))^*\left[\frac{\delta l_\omega}{\delta \alpha}(t)\diamond
\a(t) +\mu\nabla {\rm div}\u(t)\right]dt.
\end{split}
\end{equation*}

Combining this with \eqref{nt5-1-2} yields
\begin{equation*}
\begin{split}
dI_\omega(t)&=
\oint_{\gamma_0(\cdot)}\frac{1}{\rho}d\left(
(\g^{\nu}(t))^*\left[
\frac{\delta l_\omega}{\delta u}(t)\right]\right)\\
&=\oint_{\gamma_0(\cdot)}\frac{1}{\rho}(\g^{\nu}(t))^*\left[
\frac{\delta l_\omega}{\delta \alpha}(t)\diamond \a(t)+
\mu\nabla {\rm div}\u(t)\right]dt\\
&=\oint_{\gamma_\omega(t)(\cdot)}\frac{1}{\rho_\omega(t)}
\left(\frac{\delta l_\omega}{\delta \alpha}(t)\diamond \a(t)+
\mu\nabla {\rm div}\u(t)\right)dt,
\end{split}
\end{equation*}
which finishes the proof of \eqref{nt5-1-1}.
\end{proof}

\begin{remark} {\rm It is worthwhile to note that the second
summand in the integrand of \eqref{nt5-1-1} is the bulk viscosity
term appearing in the classical Navier-Stokes equations for compressible
fluid flow, except that here, the velocity is random.}
\end{remark}

\paragraph{Acknowledgments.} We thank Darryl Holm for his
interest in our work and the many discussions we had about
the geometric framework of stochastic mechanics. We are grateful
for the hospitality of the Instituto Superior T\'ecnico of the University 
of Lisbon, the Bernoulli Center of the Swiss Federal Institute of 
Technology Lausanne, and the Shanghai  Jiao Tong University, 
which facilitated our collaboration.

\end{document}